\documentclass[12pt,english]{article}
\usepackage[T1]{fontenc}
\usepackage[latin9]{inputenc}
\usepackage[letterpaper]{geometry}
\usepackage{color}
\usepackage{babel}
\usepackage{refstyle}
\usepackage{url}
\usepackage{amsmath}
\usepackage{amsthm}
\usepackage{amssymb}
\usepackage{graphicx}
\usepackage{setspace}
\usepackage[authoryear]{natbib}
\usepackage[unicode=true,
 bookmarks=true,bookmarksnumbered=true,bookmarksopen=true,bookmarksopenlevel=5,
 breaklinks=false,pdfborder={0 0 0},pdfborderstyle={},backref=false,colorlinks=true]
 {hyperref}
\hypersetup{
 citecolor=blue, linkcolor=blue}

\makeatletter

\AtBeginDocument{\providecommand\secref[1]{\ref{sec:#1}}}
\AtBeginDocument{\providecommand\eqref[1]{\ref{eq:#1}}}
\AtBeginDocument{\providecommand\oappxref[1]{\ref{oappx:#1}}}
\AtBeginDocument{\providecommand\propref[1]{\ref{prop:#1}}}
\AtBeginDocument{\providecommand\fnref[1]{\ref{fn:#1}}}
\AtBeginDocument{\providecommand\appxref[1]{\ref{appx:#1}}}
\AtBeginDocument{\providecommand\axmref[1]{\ref{axm:#1}}}
\AtBeginDocument{\providecommand\thmref[1]{\ref{thm:#1}}}
\AtBeginDocument{\providecommand\lemref[1]{\ref{lem:#1}}}
\AtBeginDocument{\providecommand\defref[1]{\ref{def:#1}}}
\AtBeginDocument{\providecommand\factref[1]{\ref{fact:#1}}}
\AtBeginDocument{\providecommand\figref[1]{\ref{fig:#1}}}
\providecommand{\tabularnewline}{\\}
\RS@ifundefined{subsecref}
  {\newref{subsec}{name = \RSsectxt}}
  {}
\RS@ifundefined{thmref}
  {\def\RSthmtxt{theorem~}\newref{thm}{name = \RSthmtxt}}
  {}
\RS@ifundefined{lemref}
  {\def\RSlemtxt{lemma~}\newref{lem}{name = \RSlemtxt}}
  {}

\numberwithin{equation}{section}
\numberwithin{figure}{section}
\theoremstyle{plain}
\newtheorem*{ax*}{\protect\axiomname}
\theoremstyle{definition}
\newtheorem{defn}{\protect\definitionname}
\theoremstyle{plain}
\newtheorem{ax}{\protect\axiomname}[section]
\theoremstyle{plain}
\newtheorem{thm}{\protect\theoremname}
\theoremstyle{plain}
\newtheorem{prop}{\protect\propositionname}
\theoremstyle{plain}
\newtheorem{lem}{\protect\lemmaname}
\theoremstyle{plain}
\newtheorem{fact}{\protect\factname}

\usepackage{amssymb}
\usepackage{refstyle}
\usepackage{accents}
\usepackage{charter}

\def\sloppy{%
  \tolerance 1000%
  \emergencystretch 1em%
  \hfuzz .5\p@
  \vfuzz\hfuzz}
\newref{sec}{name=Section~}
\newref{subsec}{name=Subsection~}
\newref{axm}{name=Axiom~}
\newref{lem}{name=Lemma~}
\newref{def}{name=Definition~}
\newref{prop}{name=Proposition~}
\newref{thm}{name=Theorem~}
\newref{remark}{name=Remark~}
\newref{cor}{name=Corollary~}
\newref{fig}{name=Figure~}
\newref{eq}{name=Equation~}
\newref{appx}{name=Appendix~}
\newref{oappx}{name=Online Appendix~}
\newref{fn}{name=Footnote~}
\newref{fact}{name=Fact~}

\makeatother

\providecommand{\axiomname}{Axiom}
\providecommand{\definitionname}{Definition}
\providecommand{\factname}{Fact}
\providecommand{\lemmaname}{Lemma}
\providecommand{\propositionname}{Proposition}
\providecommand{\theoremname}{Theorem}

\begin{document}
\title{\vskip -2emOrdered Reference Dependent Choice}
\author{Xi Zhi ``RC'' Lim\thanks{Shanghai Jiao Tong University. Email: xzlim@sjtu.edu.cn. I thank co-advisors
Mark Dean and Pietro Ortoleva for invaluable training and guidance,
for and beyond this job market paper. I also thank Anujit Chakraborty,
Yeon-Koo Che, Paul Cheung, Soo Hong Chew, David Dillenberger, Efir
Eliaz, Evan Friedman, Navin Kartik, Matthew Kovach, Qingmin Liu, Shuo
Liu, Elliot Lipnowski, Yusufcan Masatlioglu, Bin Miao, Xiaosheng Mu,
Daniel Rappoport, Collin Raymond, Gil Riella, Jingni Yang, Chen Zhao,
Songfa Zhong, Weijie Zhong, and participants at various seminars/conferences
for useful feedback. This paper is based on Chapter 1 of my 2020 doctoral
dissertation at Columbia University, some results in earlier versions
have since been relegated to \citet{lim2023ordcextra,lim2023avoidable}.}}
\date{2024/02/17\\
Most recent public version: \url{http://s.xzlim.com/ordc}}
\maketitle
\begin{abstract}
This paper studies how violations of structural assumptions like expected
utility and exponential discounting can be connected to basic rationality
violations, even though these assumptions are typically regarded as
independent building blocks in decision theory. A reference-dependent
generalization of behavioral postulates captures preference shifts
in various choice domains. When reference points are fixed, canonical
models hold; otherwise, reference-dependent preference parameters
(e.g., CARA coefficients, discount factors) give rise to \textquotedblleft non-standard\textquotedblright{}
behavior. The framework allows us to study risk, time, and social
preferences collectively, where seemingly independent anomalies are
interconnected through the lens of reference-dependent choice.
~
\end{abstract}
\newpage{}

\section{Introduction}

In various branches of economics, multiple assumptions come together
to form the basis of an economic model, and interesting findings often
emerge from the unforeseen interplay among these assumptions. The
empirical failure of these models, however, need not lie in the substance
of each individual assumption but is rooted in their indiscriminate
applications.

In individual decision-making, the standard model of choice faces
two distinct strands of empirical challenges. First, \emph{structural
assumptions} like the expected utility form and exponential discounting
are violated in simple choice experiments, such as the Allais paradox
and present bias behavior. Second, and separately, studies show that
choices are often affected by reference points, resulting in behavior
that violates \emph{basic rationality assumptions} like the\emph{
}weak axiom of revealed preferences\emph{ }(WARP). With few exceptions,
these two classes of departures have been studied separately, and
independently for each domain of choice, leading to the development
of models that attempt to explain one phenomenon in isolation of the
others.\footnote{\textbf{\label{fn:lit}Risk domain}: rank-dependent utility \citep{Quiggin1982},
quadratic utility \citep{machina1982expected}, disappointment aversion
\citep{gul1991theory}, betweenness preferences \citep{chew1983generalization,fishburn1983transitive,dekel1986axiomatic},
and cautious expected utility \citep{cerreia2015cautious} maintain
basic rationality. \textbf{Time domain}: various models of hyperbolic
discounting \citep{loewenstein1992anomalies,frederick2002time}, quasi-hyperbolic
discounting \citep{phelps1968second,laibson1997golden}, and related
generalizations \citep{chakraborty2021present,chambers2023decreasing}
maintain basic rationality. \textbf{Others}: \citet{kHoszegi2007refence}
and \citet{ortoleva2010status} use reference dependency to explain
structural violations. \citet{hara2015coalitional} maintain structural
assumption but relaxes basic rationality. \textbf{In richer settings}:
\citet{bordalo2012salience,lanzani2022correlation} relax (state-independent)
Independence and (state-independent) Transitivity to study correlated
lotteries and \citet{noor2015menu} relax Independence (for ex-ante
preference) and WARP (for ex-post choices) to study two-stage self-control
problems.}

This paper introduces a unified framework that studies how the two
types of violations may be in part related to one another, stemming
from a common source. The central approach is motivated by a simple
observation: Suppose preference parameters (e.g., utility functions
capturing risk attitude and discount factors representing degree of
patience) are influenced by reference alternatives, then even decision
makers who typically adhere to normative postulates (e.g., maximize
exponentially discounted expected utility) would every so often violate
rationality assumptions and structural assumptions\textemdash when
reference alternatives change. On the other hand, choices made under
the same reference would fully align with both postulates. Thus, while
the two types of assumptions are conventionally treated as separate
building blocks of a choice model\textemdash introduced as independently
motivated axioms\textemdash their deviations are intrinsically connected
by systematic shifts in preferences.

To illustrate, a myriad of documented anomalies, including the Allais
paradox, suggests that decision makers exhibit increased risk aversion
when presented with safer options \citep{allais1990allais,Wakker1996,herne1999effects,bleichrodt2002context,andreoni2011uncertainty}.
While this behavior contradicts the expected utility theory, it aligns
with the expected utility framework when coupled with context-dependent
utility functions that vary in concavity. This observation motivates
the model in the risk domain.
\begin{equation}
c\left(A\right)=\underset{p\in A}{\text{\ensuremath{\arg\max}}}\,\,\sum_{x}p\left(x\right)u_{r}\left(x\right)\label{eq:risk}
\end{equation}
Standard expected utility applies when the safest alternative, which
acts as the reference $r$, is fixed; but when it changes, a safer
reference leads to a more concave utility function $u_{r}$, reflecting
a systematic increase in risk aversion.

This observation is not exclusive to expected utility. In time preferences,
present bias individuals who are less patient in short-term decisions
violate exponential discounting \citep{laibson1997golden,frederick2002time,benhabib2010present,chakraborty2021present}.
However, their behavior could be consistent with the exponential discounting
form when paired with context-dependent discount factors that capture
changing time preferences.
\begin{equation}
c\left(A\right)=\underset{\left(x,t\right)\in A}{\text{\ensuremath{\arg\max}}}\,\,\delta_{r}^{t}u\left(x\right)\label{eq:time}
\end{equation}
When a problem offers sooner payments, it alters the reference point
$r$, prompting the decision maker to use a lower discount factor
$\delta_{r}$ that reflects increased impatience. Again, changes
in preferences are systematic along a certain order, and behavior
is otherwise standard.

For social preferences, it is well-documented in economics and psychology
that the very same individuals display different degree of altruism
in different choice settings, for example when a balanced split of
reward is available than when it is not \citep{ainslie1992picoeconomics,rabin1993incorporating,nelson2002equity,fehr2006economics,sutter2007outcomes}.
These context-dependent preferences could be consistent with
\begin{equation}
c\left(A\right)=\underset{\left(x,y\right)\in A}{\text{\ensuremath{\arg\max}}}\,\,x+v_{r}\left(y\right)\label{eq:social}
\end{equation}
where increased altruism is captured by a utility from sharing, $v_{r}\left(\cdot\right)$,
that systematically increases when more-equitable splits become the
reference.

This paper aims to examine these behaviors as one collective, addressing
three key questions: (1) \emph{Under what conditions do non-standard
behaviors across various choice domains permit such representations?}
(2) \emph{What do they have in common}? and (3)\emph{ How does their
systematic departure from canonical models inform the relationship
between rationality assumptions and structural assumptions?}

It turns out that, although these behavioral anomalies are typically
investigated in largely separate and domain-specific studies, the
behavioral content of the proposed models is underpinned by a ``meta''
axiomatic foundation referred to as \emph{Reference Dependence} (RD).
RD is the key innovation of this paper, introducing a reference-dependent
approach that can generalize a large class of behavioral postulates
or axioms, be it ``rational'' or ``structural''. When applied
to the risk, time, and social domains, it yields three complete characterizations
that resonate with one another.

To illustrate the idea, \secref{ORD} applies RD only to rationality
assumption by requiring that in every (finite) choice set, at least
one alternative would preserve WARP among choice behavior from its
subsets. Intuitively, if the failure of rationality is caused by reference
dependence, then rationality should continue to hold at least for
choice sets that share the same reference. It turns out that this
postulate characterizes a two-step choice process: A reference order
is maximized to identify the reference alternative $r\left(A\right)$
of a choice problem $A$. The reference alternative then determines
a utility function that the decision maker maximizes. Intuitively,
the context of a choice problem is captured by the alternative that
ranks highest in the reference order and the underlying context-dependent
preference is subsequently determined.
\begin{equation}
c\left(A\right)=\underset{z\in A}{\text{\ensuremath{\arg\max}}}\,\,U_{r\left(A\right)}\left(z\right)\label{eq:general}
\end{equation}

It has not gone unnoticed that Equations \ref{eq:risk}, \ref{eq:time},
and \ref{eq:social} are special cases of \eqref{general}, sharing
two essentially components: (i) a reference order and (ii) reference-dependent
preference parameters. It is also apparent that basic rationality\textemdash the
assumption that one persistent utility function is being maximized\textemdash can
fail, which makes the proposed explanations not particularly appealing,
at least until the recent accumulation of theoretical interest and
empirical evidence against basic rationality itself.

It turns out that, barring technical challenges, the axiomatic characterization
of each of these behaviors requires little more than adapting RD to
their domain-specific normative postulates. For risk preference, Risk-RD
preserves both WARP (rationality axiom) and von Neumann-Morgenstern's
Independence (structural axiom) when the safest alternative is maintained.
For time preference, Time-RD preserves normative postulates WARP (rationality
axiom) and Stationarity (structural axiom) when the earliest available
payment is fixed. For social preference, Social-RD calls for consistency
with WARP (rationality axiom) and Quasi-linearity (structural axiom)
when the most-balanced options coincide. The underlying intuition
is universal: Upholding the reference point ensures the validity of
all normative postulates, so that violations of structural assumptions\textemdash whatever
they are and whatever the domain\textemdash are linked to reference
dependent preferences manifested in basic rationality violations.\footnote{The generality of this exercise is demonstrated in \oappxref{tau-and-proofs}.}
A second axiom, which does not involve reference points, captures
systematic changes in preferences by requiring that choices cannot
become more risk loving / more patient / more selfish when a subset
of the original choice set is considered.

Notwithstanding its intuitiveness, this approach does not fully align
with the conventional wisdom in decision theory (and economics in
general) where assumptions are weakened one at a time. Relaxing both
rationality postulates and structural postulates leads to an instinctive
concern about admitting too wide a range of behavior. However, the
\emph{joint }generalization\emph{ }introduced by RD exhibits greater
discipline than an \emph{independent} generalization, and it contributes
to three interrelated insights that span all three domains, forming
the core of this study.

First, the models predict how structural anomalies traditionally detected
in binary comparisons (such as the Allais paradox and present bias
behavior) will manifest as WARP violations when larger choice sets
are considered, providing testable predictions that could bridge the
two largely separate empirical literature. To illustrate, suppose
Option 1 is a later payment and it is chosen over a sooner payment
Option 2, but the opposite decision emerges when both options are
symmetrically advanced, an anomaly known as present bias.\footnote{This behavior violates Stationarity, the axiom responsible for exponential
discounting, which requires a consistent preference between two options
even when the decision is revisited at later point in time.} Then, it is predicted that adding a particular unchosen alternative
Option 3 to the original comparison could switch the choice to Option
2, causing a WARP violation; similar observations link the common
ratio effect in risk preferences to WARP violations. They resonate
with the motivation of the present framework in suggesting that deviations
from standard models may arise from changing preferences rather than
being a mere failure of structural assumptions. This plausible connection,
however, is obscured in studies that assume basic rationality, thereby
missing the opportunity to draw insights from behavior in non-binary
choice sets that could offer a fundamentally different perspective
on traditional anomalies.

Second, the models suggest how basic rationality and structural postulates
can be inextricably linked, even though they are typically regarded
as independent building blocks of individual decision-making. \propref{WARP}
shows that introducing just WARP or just Independence to the risk
model immediately implies standard expected utility behavior, even
though these postulates must be jointly imposed in a general setting.
This means a decision maker who has \emph{any} utility representation
will also have an expected utility representation; similar results
are obtained for time and social domains. The proposed models thus
capture distinct non-standard behavior when contrasted with a substantial
body of the literature that only generalizes structural assumptions.
For example, even though many models can explain the Allais paradox
and present bias behavior, choice behavior from prominent models like
rank-dependent utility, quadratic utility, disappointment aversion,
betweenness, cautious expected utility, hyperbolic discounting, and
quasi-hyperbolic discounting overlap with mine only in the special
case where behavior is fully standard.\footnote{See \fnref{lit} for references.}
That is, for non-standard decision makers, our models provide mutually
exclusive predictions.

Third, the innovation in this exercise is due crucially to an underexplored
generalization of structural assumptions forbidden by traditional
adherence to rationality assumptions. To see this, consider the risk
domain and suppose lottery $p$ is preferred to lottery $q$. The
von Neumann-Morgenstern's Independence condition requires their common
mixtures to have the same preference order, meaning that $p^{\alpha}s$
is preferred to $q^{\alpha}s$.\footnote{Lottery $p^{\alpha}s$ refers to the (compound) lottery generated
by mixing lottery $p$ with probability $\alpha$ and lottery $s$
with probability $\left(1-\alpha\right)$. Independence says the preference
between $p^{\alpha}s$ and $q^{\alpha}s$ should be the same as the
preference between $p$ and $q$, since they differ only by a common
term.} Although a generalization of Independence loosens this requirement,
WARP makes it impossible to discuss how $p^{\alpha}s$ is preferred
to $q^{\alpha}s$ in \emph{some} choice sets while the opposite holds
in others. Relaxing WARP immediately allows for this kind of generalization
and paves the way for a context-dependent implementation of structural
postulates. This paper presents one of many possible demonstrations
and, perhaps counter-intuitively, shows that weakening both kinds
of postulates could bring us ``closer'' to canonical models. Similar
limitations are present when a (complete) preference relation serves
as the primitive, which might explain why this approach has not received
much attention.

While the three observations are formulated within the scope of ordered
reference, they might make a case for the comprehensive examination
of conceptually different behavior that complement foundational groundwork
already established by isolated investigations. Perhaps the examination
of individual decision-making rightly began with a theoretical decomposition
of a complex behavior into key components\textemdash encompassing
conceptually distinct notions like ``rationality axioms'' and ``structural
axioms''\textemdash to focus our studies and interpretations. But
the natural progression now, knowing that each of these components
fails to some extent, entails a joint investigation of these theoretical
constructs. 

Related literature is next. \secref{ORD} introduces the basic framework
and RD. Sections \ref{sec:risk}-\ref{sec:social}, the main parts
of this paper, take the unified framework to risk, time, and social
domains; they introduce axioms, provide representation theorems, study
implications, and discuss evidence. \secref{Conclusion} concludes.
Key proofs are in \appxref{proofs}. Technical results and omitted
proofs are relegated to \oappxref{tau-and-proofs}.

\subsection{Related literature}

The engagement of reference alternatives relates to the extensive
literature on reference-dependent preferences, originating from the
seminal work of \citet{kahneman1979prospect} on loss aversion and
initially explored under the assumption that reference points are
directly observed.\footnote{And relatedly, \citet{tversky1991loss,kahneman1991anomalies}.}
Subsequently, the scope of mechanisms involving exogenous reference
points has broadened beyond the realm of gain-loss utility, e.g.,
general status quo bias \citep{masatlioglu2005rational,masatlioglu2014canonical},
ambiguity aversion \citep{ortoleva2010status}, wishful thinking \citep{kovach2020twisting},
and categorical thinking \citep{ellis2022choice}.\footnote{Other work related to status quo bias includes \citet{rubinstein1999choice,sagi2006anchored,apesteguia2009theory,dean2017limited}.}

A new way to study reference points was popularized by \citet{kHoszegi2006model}
where endogenous reference points capture seemingly reference-dependent
behavior even though reference points are not directly observed. These
studies encompass both objective reference \citep{kivetz2004alternative,orhun2009optimal,bordalo2013salience,tserenjigmid2019}
and subjective reference \citep{kHoszegi2006model,ok2015revealed,freeman2017preferred}.
The present paper falls into this category and explores a novel use
of endogenous reference to proxy for domain-specific contexts and
govern domain-specific preference shifts.

Although the understudied link between rationality assumptions and
structural assumptions forms the core of this paper, the framework
can be applied to the generic choice domain where only rationality
assumptions are considered. In this case, the model and its behavioral
implications coincide with \citet{rubinstein2006two}'s Triggered
Rationality.\footnote{\citet{ravid2021bad} studies the same behavior under a model of bad
temptation.} That same model is also studied in \citet{kibris2018theory} and
\citet{giarlotta2022semantics} using a different axiom that essentially
says ``if dropping $x$ in the presence of $y$ causes a WARP violation,
then dropping $y$ in the presence of $x$ cannot''. Their axiom
is an appealing alternative when applied only to rationality violations,
as it cannot be extended to structural assumptions like Independence
and Stationarity.\footnote{\label{fn:sra}Let $p',q'$ be common mixtures of $p,q$. Using notation
$\left\{ \underline{p},q\right\} $ to denote ``$p$ is chosen from
the choice set $\left\{ p,q\right\} $'', the behavior $\left\{ \underline{p},q,p',q'\right\} $,
$\left\{ \underline{p},q,p'\right\} $, $\left\{ \underline{p},q,q'\right\} $,
$\left\{ \underline{p},p',q'\right\} $, $\left\{ \underline{q},p',q'\right\} $,
$\left\{ p,\underline{q}\right\} $, $\left\{ \underline{p},p'\right\} $,
$\left\{ \underline{p},q'\right\} $, $\left\{ \underline{q},p'\right\} $,
$\left\{ \underline{q},q'\right\} $, $\left\{ p',\underline{q}'\right\} $
satisfies \citet{kibris2018theory}'s Single Reversal even after modifying
it to consider Independence violation as a reversal, but there is
no reference alternative in $\left\{ p,q,p',q'\right\} $ because
$\left\{ \underline{p},q,p',q'\right\} $ and $\left\{ p,\underline{q}\right\} $
violate WARP whereas $\left\{ \underline{p},q,p',q'\right\} $ and
$\left\{ p',\underline{q}'\right\} $ violate Independence. \axmref{risk_axiom1}
rules out this behavior.}

More broadly, theories that systematically apply to different domains
of choice include, among others, loss aversion \citep{kahneman1979prospect,kHoszegi2006model},
attraction effect \citep{HuberPaynePuto1982}, compromise effect \citep{Simonson1989}\emph{,
}salience \citep{bordalo2012salience,bordalo2013salience}, and focusing
\citep{koszegi2013focusing}. These models consider evaluations of
multi-attributes alternatives that are affected by the attributes
of available alternatives, some of which later generalized as categorical
thinking \citep{ellis2022choice}. They provide valuable insights
that help us understand how psychological and attention factors can
influence economic decisions by affecting our perception of an alternative.

To this end, \citet{ellis2022choice}'s study may be the closest to
mine, since they also consider reference points and explore applications
in various choice settings. Their study focuses on the endogenous
formation of categories due to exogeneously given reference points,
and when two alternatives are assigned to different categories, they
are evaluated differently and potentially result in a preference reversal.
The key mechanism that connects reference and preference is thus categorization.
In contrast, the present framework considers endogenous reference
points, uses the functional forms of standard models to evaluate alternatives,
and captures deviations using changes in preference parameters. \citet{kovach2020twisting}'s
wishful thinking lies in between the two approaches; a decision maker's
subjective belief depends on exogeneously given status quo (the reference
point), but she is otherwise standard in maximizing subjective expected
utility.

In terms of characterization, \emph{Reference Dependence} (RD) offers
new tools. First, it is known that the equivalence between canonical
axioms and canonical models breaks down when data is limited or incomplete;
this technical issue emerges as choice problems are partitioned into
reference-dependent subsets.\footnote{See \citet{Samuelson1948} and \citet{aumann1962utility}. For example,
if $\mathcal{B}$ does not contain all doubletons and tripletons,
then a choice correspondence on $\mathcal{B}$ that satisfies WARP
(and Continuity) does not necessarily admit a utility representation.
This challenge extends to richer domains; for example in the risk
domain, if the underlying set of lotteries is not a convex subset
of lotteries, then a choice correspondence that satisfies WARP and
Independence does not necessarily admit an expected utility representation
(even if it admits a utility representation).} Instead of strengthening axioms \citep{Houthakker1950sarp,echenique2020testable,declippel2021limiteddataset}
or embracing more general models \citep{dubra2004expected,manzini2008on,evren2014scalarization,hara2015coalitional},
RD exploits reference formation to impart adequate structure to each
subset of behavior so that a standard representation emerges. The
method bears qualitative similarity to \citet{ke2022local}'s \emph{weak
local independence}, which characterizes local expected utility using
local compliance of canonical Independence. Second, systematic deviations
from structural assumptions are imposed by relating small and large
choice sets, achieving effects similar in spirit to \citet{dillenberger2010preferences,cerreia2015cautious}'s
\emph{negative certainty independence} and \citet{chakraborty2021present}'s
\emph{weak present bias} in more standard settings. These unexpected
connections invite curiosity into the potential role of reference
dependence in studies that do not explicitly consider them.

Finally, the paper aligns with a broader agenda regarding the comprehensive
examination of behaviors conventionally studied in isolation, providing
breadth to the already established depth. This agenda includes, among
others, empirical studies of possible interrelations in behavioral
traits \citep{falk2018global,chapman2023econographics,stango2023we},\footnote{Less representative samples are used in \citet{burks2009cognitive}
(truck drivers) and \citet{dean2019empirical} (university students).} methodological development that separates preference inconsistency
and parametric misspecification \citep{halevy2015,polisson2020revealed,echenique2020testable,echenique2023approximate,deClippel2023relaxed},
experiments that assess a broad spectrum of anomalies as potential
mistakes \citep{nielsen2022choices}, theoretical investigation that
links non-standard risk and time preferences \citep{chakraborty2020relation},
and revealed preference analyses that highlight basic rationality
postulates in rich/different domains \citep{dembo2021ever,halevy2023difficult,chen2023consistency}.

\section{\label{sec:ORD}Basic Framework}

Let $Y$ be a separable metric space, endowed with the standard Euclidean
metric $d_{2}$, that represents the set of all alternatives. Let
$\mathcal{A}$ be the set of all finite and nonempty subsets of $Y$,
also called choice sets. The primitive of this paper is a\emph{ }choice
correspondence $c:\mathcal{A}\rightarrow\mathcal{A}$ where $c\left(B\right)\subseteq B$
for all $B\in\mathcal{\mathcal{A}}$. I assume throughout the paper
that $c$ is continuous:
\begin{ax*}[Continuity]
\label{axm:ord_continuity}$c$ has a closed graph.\footnote{That is, $x_{n}\rightarrow_{d}x$, $A_{n}\rightarrow_{H}A$, and $x_{n}\in c\left(A_{n}\right)$
for every $n=1,2,...$ imply $x\in c\left(A\right)$, where $\rightarrow_{H}$
refers to convergence in the Hausdorff distance, defined by $d_{H}\left(X,Y\right)=\max\left\{ \sup_{x\in X}\inf_{y\in Y}d_{2}\left(x,y\right),\sup_{y\in Y}\inf_{x\in X}d_{2}\left(x,y\right)\right\} $.}
\end{ax*}
The risk, time, and social preferences studied in this paper share
a common starting point: For any given choice set, the decision maker
is seemingly standard by maximizing a single utility function. But
globally, behavior is non-standard because this function depends on
possibly different reference alternatives across choice sets.\footnote{This naturally bounds non-standard behavior: When $|Y|$ is finite,
there are at most $|Y|$ distinct utility functions, but there are
around $2^{|Y|}$ choice sets, and this difference increases exponentially
in $|Y|$.}
\begin{defn}
A choice correspondence $c$ admits an Ordered-Reference Dependent
Utility (ORDU) representation if there exist a linear order $\left(R,Y\right)$
and a set of utility functions $\left\{ U_{y}:Y\rightarrow\mathbb{R}\right\} _{y\in Y}$
such that 
\[
c\left(A\right)=\arg\max_{y\in A}U_{r\left(A\right)}\left(y\right)
\]
 and $r\left(A\right)=\max\left(R,A\right)$ for all $A$, where $c$
has a closed graph.\footnote{A linear order $\left(R,Y\right)$ is a complete, reflexive, transitive,
and antisymmetric binary relation $R$ on $Y$.}
\end{defn}
Existing theories that incorporate a reference order can be traced
back to \citet{rubinstein2006two}'s \emph{Triggered Rationality},
which coincide with ORDU. Restricted to the generic choice domain,
\citet{kibris2018theory,giarlotta2022semantics,kibris2021random}
expand on this trajectory by exploring different axiomatizations,
stochastic choice, and connections to psychological constraints /
limited consideration. The latter suggests how different kinds of
rationality violations may be related, complementing the present framework
that focuses on context-dependent preferences. They also capture interesting
narratives in the generic choice domain. \citet{kibris2018theory}
suggest that the top results when consumers search for a product are
conspicuous, serving as the reference and influencing their final
decisions; \citet{giarlotta2022semantics} propose that the frog's
legs dish in Luce and Raiffa\textquoteright s Dinner is salient, becoming
the reference and increasing a consumer's confidence, thence preference,
for steak; \citet{kibris2021random} consider the case of marketing
campaigns where a consumer is more likely to recall an advertised
product and uses it as benchmark to make consumption decisions.

Despite its simplicity and intuitiveness, focusing on the generic
choice domain is not without caveats. The formation of completely
subjective reference points adds challenges to their identification.
Compounding this issue is the lack of structure in each reference-dependent
utility function and the absence of a systematic relationship between
these utility functions.

Explored in Sections \ref{sec:risk}-\ref{sec:social} (risk, time,
and social preferences), a richer choice domain provides natural remedies
to, and in fact benefit from, the flexibility of this model. First,
it significantly expands the interpretation of a reference order,
where it ranges from being partially subjective (ranking lotteries
by riskiness, \secref{risk}) to fully objective (Gini index, \secref{social}),
so as to capture the relevant domain-specific context. In turn, the
reference order serves as a natural anchor along which domain-specific
preference shifts are manifested, such as increasing risk aversion
or decreasing patience along the established order. The two components\textemdash reference
order and reference effect\textemdash interact with each another,
yielding a framework that captures a highly specific and tractable
form of set-dependent preferences.

Moreover, the models in all three choice domains share a ``meta''
axiomatic framework that in its simplest form characterizes ORDU.
To illustrate the basic idea, consider following definition that maintains
the content of the weak axiom of revealed preferences (WARP) but allows
for selective application.
\begin{defn}
\emph{$c$ satisfies WARP over $\mathcal{S}\subseteq\mathcal{A}$}
if for all $A,B\in\mathcal{S}$, if $B\subset A$ and $c\left(A\right)\cap B\ne\emptyset$,
then $c\left(A\right)\cap B=c\left(B\right)$.
\end{defn}
The classical rationality assumption on choice behavior entails imposing
WARP over $\mathcal{S}=\mathcal{A}$. The following postulate imposes
WARP only locally, using a reference point as the anchor.
\begin{ax}
\label{axm:ord_warp}For every choice set $A\in\mathcal{A}$, $c$
satisfies WARP over $\left\{ B\in\mathcal{\mathcal{A}}:x\in B\subseteq A\right\} $
for some $x\in A$.
\end{ax}
\begin{thm}
\label{thm:ORD}$c$ satisfies \axmref{ord_warp} and Continuity if
and only if it admits an ORDU representation.
\end{thm}
\axmref{ord_warp} captures choice behavior that satisfies WARP in
a reference-dependent manner and coincides with the \emph{reference
point property} in \citet{rubinstein2006two}. To understand this
axiom, suppose choices from $A$ and its subset $B$ constitute a
WARP violation. If this is caused by a change in reference point,
specifically, that the reference alternative of $A$ was removed when
we go to subset $B$, then a natural limitation of WARP violations
would arise: Had we not\emph{ }removed the reference alternative of
$A$, choices must satisfy WARP. To put it differently, suppose that
by preserving some alternative $x$ in $A$, choices from the subsets
of $A$ would comply with WARP, then $x$ is a candidate reference
alternative\emph{ }of\emph{ $A$}. \axmref{ord_warp} demands that
every choice set contains (at least) one candidate reference alternative,
which makes is less demanding than the standard postulate that imposes
WARP indiscriminately.\footnote{Since standard WARP requires ``$c$ satisfies WARP over $\mathcal{\mathcal{A}}$'',
which in turn implies ``$c$ satisfies WARP over $\mathcal{S}$''
for any $\mathcal{S\subseteq A}$, it is stronger than \axmref{ord_warp}.}

To illustrate further, consider the following choice correspondence
on $Y=\left\{ a,b,c,d\right\} $.
\begin{center}
\begin{tabular}{cc|cc|cc}
$A$ & $c\left(A\right)$ & $A$ & $c\left(A\right)$ & $A$ & $c\left(A\right)$\tabularnewline
\hline 
$\left\{ a,b,c,d\right\} $ & $b$ & $\left\{ b,c,d\right\} $ & $b$ & $\left\{ b,c\right\} $ & $b$\tabularnewline
$\left\{ a,b,c\right\} $ & $b$ & $\left\{ b,d\right\} $ & $b$ &  & \tabularnewline
$\left\{ a,b,d\right\} $ & $b$ & $\left\{ c,d\right\} $ & $c$ &  & \tabularnewline
$\left\{ a,c,d\right\} $ & $d$ &  &  &  & \tabularnewline
$\left\{ a,b\right\} $ & $b$ &  &  &  & \tabularnewline
$\left\{ a,c\right\} $ & $a$ &  &  &  & \tabularnewline
$\left\{ a,d\right\} $ & $d$ &  &  &  & \tabularnewline
\end{tabular}
\par\end{center}

Note that this choice correspondence fails to satisfy WARP globally
because $d$ is chosen from $\left\{ a,c,d\right\} $ but $c$ is
chosen from $\left\{ c,d\right\} $. To check whether it satisfies
\axmref{ord_warp}, we have to visit every choice set. Starting with
$A=\left\{ a,b,c,d\right\} $, note that there is no WARP violations
among subsets of $A$ that contain $a$, i.e., $a$ is a reference
alternative, so choice set $A$ passes the test. These subsets have
been conveniently placed in the left column. Moreover, note that when
we visit any of these subsets, $a$ continues to be a reference alternative,
so they too pass the test. For the remaining choice sets, we begin
with $A'=\left\{ b,c,d\right\} $ and note that is no WARP violations
among subsets of $A'$ that contain $d$, so $A'$ and these subsets
pass the test; they are conveniently positioned in the middle column.
The only choice set left is $\left\{ b,c\right\} $ where WARP is
trivial because the only non-singleton subset of $\left\{ b,c\right\} $
is itself. The axiom is thus satisfied. It amounts to a structured
partitioning of choice sets\textemdash the left, middle, and right
columns\textemdash so that within each part there is no WARP violation.\footnote{\axmref{ord_warp} is falsifiable whenever when $|Y|\geq3$. Using
notation $\left\{ a,\underline{b},c\right\} $ to denote ``$b$ is
chosen from the choice set $\left\{ a,b,c\right\} $'', the choice
correspondence $\left\{ a,\underline{b},c\right\} ,\left\{ \underline{a},b\right\} ,\left\{ b,\underline{c}\right\} ,\left\{ \underline{a},c\right\} $
have two instances of WARP violations, (i) between $\left\{ a,\underline{b},c\right\} $
and $\left\{ \underline{a},b\right\} $ and (ii) between \textbf{$\left\{ a,\underline{b},c\right\} $}
and $\left\{ b,\underline{c}\right\} $, so none of $a,b,c$ can be
the reference alternative of $A=\left\{ a,b,c\right\} $. Relatedly,
a cardinal measure of falsifiability is to count the minimum number
of observations required for falsification. For standard WARP, that
number is 2: for example, when WARP is violated between $\left\{ a,\underline{b},c\right\} $
and $\left\{ \underline{a},b\right\} $. For \axmref{ord_warp}, that
number is 3: for example $\left\{ a,\underline{b},c\right\} $, $\left\{ \underline{a},b\right\} $,
and $\left\{ \underline{a},c\right\} $, since the reference of $\left\{ a,b,c\right\} $
is in $\left\{ a,b\right\} $ and/or $\left\{ a,c\right\} $, but
WARP is violated both between $\left\{ a,\underline{b},c\right\} $,
$\left\{ \underline{a},b\right\} $ and between $\left\{ a,\underline{b},c\right\} $,
$\left\{ \underline{c},b\right\} $. Under this measure, reference
dependence makes \axmref{ord_warp} harder to reject relative to WARP
by one additional observation.}

The highlight of this approach is not the rationality assumption WARP
per se, but the way WARP as a behavioral postulate was generalized
in an attempt to call for its compliance locally. More generally,
it follows the template ``for every choice set $A$, the choice correspondence
$c$ satisfies $\mathcal{T}$ over $\left\{ B\in\mathcal{A}:x\in B\subseteq A\right\} $
for some $x\in\Psi\left(A\right)$'' where $\mathcal{T}$ can be
a behavioral postulate of interest and $\Psi$ can be an objective
range in which reference points lie. This general approach is referred
to as \emph{Reference Dependence} (RD), which is formally introduced
and analyzed in \oappxref{tau-and-proofs} and used in Sections \ref{sec:risk}-\ref{sec:social}.
Related studies like \citet{kibris2018theory} propose alternative
characterization designed for WARP and cannot be directly extended
in this way (see \fnref{sra}).

\section{\label{sec:risk}Risk Preference}

Consider a decision maker whose willingness to take risk is dynamic
and dependent on how much of it is avoidable. The safest alternative
in a choice set provides a natural measure for this context. Sometimes,
we have the option to fully avoid risk by keeping our assets in cash
or by buying an insurance policy, and so the safest option is quite
safe. But in other situations, such as a carefully designed lab experiment
in which all options involve risk, taking some risk becomes unavoidable.
The premise of my model is a decision maker whose risk aversion systematically
differs between different set-dependent contexts\textemdash greater
risk aversion when risk is increasingly avoidable.

\subsection{Preliminaries and axioms}

Consider a finite set of prizes $X\subset\mathbb{R}$, where $|X|>2$,
with the largest and smallest prizes denoted by $b$ and $w$\textbf{
}respectively.\footnote{If $|X|\leq2$, either the only choice set is a singleton set or choice
sets contain only lotteries related by first order stochastic dominance,
and \axmref{FOSD} full pins down choices.} Let $Y=\Delta\left(X\right)$ be the set of all probability measures
over $X$, called lotteries. Everything else follows \secref{ORD}.
Per convention, $\delta$ denotes a degenerate lottery and $\delta_{x}$
denotes the degenerate lottery that gives prize $x\in X$. For $p,q\in\Delta\left(X\right)$
and $\alpha\in\left[0,1\right]$, $p^{\alpha}q$ denotes the convex
combination $\alpha p\oplus\left(1-\alpha\right)q$. For $p\in\Delta\left(X\right)$,
$p\left(x\right)$ denotes the probability that lottery $p$ gives
prize $x$. I assume throughout that $c$ satisfies first order stochastic
dominance (FOSD):
\begin{ax}
\label{axm:FOSD}If $p$ first order stochastically dominates $q$
(where $p\ne q$) and $p\in A$, then $q\notin c\left(A\right)$.
\end{ax}
Next, \emph{Reference Dependence} (\secref{ORD}) is applied to both
WARP and the von Neumann-Morgenstern's Independence condition, beginning
with a definition that applies Independence selectively.
\begin{defn}
$c$ \emph{satisfies Independence over $\mathcal{S}\subseteq\mathcal{A}$}
if for all $A,B\in\mathcal{S}$ and $\alpha\in\left(0,1\right)$,
if $p\in c\left(A\right)$, $q\in A$, $q^{\alpha}s\in c\left(B\right)$,
and $p^{\alpha}s\in B$, then $p^{\alpha}s\in c\left(B\right)$ and
$q\in c\left(A\right)$.
\end{defn}
In standard expected utility, $c$ satisfies WARP and Independence
over $\mathcal{S}=\mathcal{\mathcal{A}}$. I depart from standard
expected utility by allowing for preferences to depend on the \emph{safest
available alternatives}\textemdash the reference\textemdash but demand
compliance with WARP and Independence whenever a collection of choice
sets share a reference. When is that? If $p$ ($\ne q$) is a mean-preserving
spread of\emph{ $q$} ($p\text{MPS}q$), it is clearly not the safest.
Additionally, a second order partially compensates for the incomplete
nature of MPS by also deeming lotteries with increased probabilities
of the most extreme prizes (but keeping the relative probability of
intermediate prizes the same) to be riskier. Formally, $p$ is an\emph{
}extreme spread of $q$ ($p\text{ES}q$) if $p=\beta q\oplus\left(1-\beta\right)\left(\text{\ensuremath{\alpha}}\delta_{b}\oplus\left(1-\alpha\right)\delta_{w}\right)$
for some $\beta\in[0,1)$ and $\alpha\in\left(q\left(b\right),1-q\left(w\right)\right)$.\footnote{The two risk orders are non-contradictory and typically non-nested.
Extreme spread is intuitively related to \citet{aumann2008economic}'s
risk index, where lotteries are deemed safer in the ``economics
sense''\textemdash under standard expected utility, the extreme spreads
of $q$ are lotteries in $\text{conv}\left(\left\{ q,\delta_{b},\delta_{w}\right\} \right)$
that are preferred to $q$ by a more-risk-loving decision maker if
a more-risk-averse decision maker does so. Non-contradictory: extreme
spreads of $q$ live in $\text{conv}\left(\left\{ q,\delta_{b},\delta_{w}\right\} \right)$,
which does not contain any mean preserving contraction of $q$. Non-nested:
extreme spreads need not preserve mean, mean preserving spreads need
not maintain relative probability of intermediate prizes; in the special
case where $|X|\leq3$, mean preserving spreads are nested in extreme
spreads.}
\begin{defn}
\label{def:psi_risk}Let $\Psi\left(A\right):=\left\{ p\in A:\text{for all \ensuremath{q\in A}, neither }p\text{MPS}q\text{ nor }p\text{ES}q\right\} $
be the set of\emph{ least risky lotteries} in $A$.
\end{defn}
\begin{ax}[Risk Reference Dependence]
\label{axm:risk_axiom1}For every $A\in\mathcal{\mathcal{A}}$, $c$
satisfies WARP and Independence over $\left\{ B\in\mathcal{\mathcal{A}}:p\in B\subseteq A\right\} $
for some $p\in\Psi\left(A\right)$.
\end{ax}
The next and last axiom captures changing risk aversion when more
options become available. It is standard to say that a preference
relation $\succsim_{1}$ is \emph{more-risk-averse} than another preference
relation $\succsim_{2}$ if, for any degenerate lottery $\delta$
and lottery $p$, $\delta\succsim_{2}p$ implies $\delta\succsim_{1}p$.
This definition is often studied alongside expected utility, but it
is, in fact, independent of it. \axmref{risk_monotoneRA} extends
this definition to lotteries that differ by a degenerate component:
where $p^{\alpha}s$ can be obtained from $\delta^{\alpha}s$ by reallocating
probabilities from one prize to one or more prizes. Then, it posits
that a decision maker cannot be more-risk-loving when a choice set
expands. The underlying intuition is that additional alternatives
should only be able to increase the extent to which risk is avoidable,
and if the avoidability of risk (weakly) increases risk aversion,
then the additions must not result in increased risk tolerance. We
say the pair of lotteries $\left(\delta^{*},p^{*}\right)$ is a common
mixture of the pair of lotteries $\left(\delta,p\right)$ if there
exist $\alpha\in\left[0,1\right]$ and $s\in\Delta\left(X\right)$
such that $\delta^{*}=\delta^{\alpha}s$ and $p^{*}=p^{\alpha}s$.
\begin{ax}
\label{axm:risk_monotoneRA}Suppose $B\subset A$ and $\left(\delta_{1},p_{1}\right),\left(\delta_{2},p_{2}\right)$
are common mixtures of $\left(\delta,p\right)$. If $\delta_{2}\in c\left(B\right)$
and $p_{2}\in B\backslash c\left(B\right)$, then $\delta_{1}\in A$
implies $p_{1}\notin c\left(A\right)$.
\end{ax}

\subsection{Model}
\begin{defn}
$c$ admits an Avoidable Risk Expected Utility (AREU) representation
if it admits an ORDU representation $\left(\left\{ U_{r}\right\} _{r\in Y},R\right)$
such that for some set of strictly increasing functions $\left\{ u_{r}:X\rightarrow\mathbb{R}\right\} _{r\in Y}$,
\begin{itemize}
\item $U_{r}\left(p\right)=\sum_{x}p\left(x\right)u_{r}\left(x\right)$,
\item $p\text{MPS}q$ and $p\text{ES}q$ each implies $qRp$,
\item $qRp$ implies $u_{q}=f\circ u_{p}$ for some concave function $f$:$\mathbb{R}\rightarrow\mathbb{R}$.
\end{itemize}
\end{defn}
\begin{thm}
\label{thm:AREU}$c$ satisfies Axioms \ref{axm:FOSD}-\ref{axm:risk_monotoneRA}
and Continuity if and only if it admits an AREU representation.
\end{thm}
When choice behavior admits an AREU representation, it is as if the
reference alternative $r\left(A\right)$ is first determined by $R$,
which ranks safer alternatives higher, and then the decision maker
maximizes expected utility using the associated context-dependent
(Bernoulli) utility function $u_{r\left(A\right)}$. Moreover, a safer
reference leads to a (weakly) more concave utility function. This
generalizes the standard model where a decision maker maximizes expected
utility using a single utility function throughout, but departure
from expected utility is limited to systematic changes in risk attitude.
It can be shown that (for a fixed $R$) each $u_{r}$ is unique up
to positive affine transformation, except possibly when $r=b^{\alpha}w$.\footnote{Uniqueness is demonstrated in \thmref{AREU}. When $r=b^{\alpha}w$,
it is possible that $u_{r}$ is only used to evaluate lotteries that
first order stochastically dominates / dominated by $r$, so that
any strictly increasing transformation of $u_{r}$ is acceptable.}

\paragraph*{Allais in WARP violations}

Perhaps because the Allais paradox is a direct failure of the structural
assumption Independence, many models that seek to explain this anomaly
weaken Independence but maintain basic rationality. AREU considers
an arguably different approach by linking the Allais paradox to a
completely different class of failures, WARP violations from non-binary
choice sets.

To see the intuition, consider the common ratio effect in binary comparisons:
the sure prize\emph{ }of $\$3000$ ($p_{1}$) is preferred to a lottery
that yields $\$4000$ with $80\%$ chance ($p_{2}$), but a lottery
that yields $\$4000$ with $20\%$ chance ($q_{2}$) is preferred
to a lottery that yields $\$3000$ with $25\%$ chance ($q_{1}$).
If treated as separate decisions, the former decision entails a (Bernoulli)
utility function that is more concave than the latter's under the
expected utility functional.\footnote{\label{fn:lotteries}Let $A=\left\{ p_{1},p_{2}\right\} $ and $B=\left\{ q_{1},q_{2}\right\} $.
Suppose $u_{A}$ (resp. $u_{B}$) explains the choice from $A$ (resp.
$B$) under expected utility. After normalization (for example $u_{A}\left(0\right)=u_{B}\left(0\right)=0$
and $u_{A}\left(4000\right)=u_{B}\left(4000\right)=1$), choice pattern
$\left(p_{1},q_{2}\right)$ arises if and only if $u_{A}\left(3000\right)>0.8$
and $u_{B}\left(3000\right)<0.8$, which in turn implies $u_{A}$
is a concave transformation of $u_{B}$.} But the expected utility theory rules out the use of different utility
functions for the same decision maker.\footnote{More precisely, expected utility allows for different utility functions
as long as they are related by positive affine transformations, but
these utility functions make identical predictions.} AREU builds on this observation. Given a reference order that deems
$r\left(\left\{ p_{1},p_{2}\right\} \right)$ safer than $r\left(\left\{ q_{1},q_{2}\right\} \right)$,
the utility function for the first choice set is more concave, which
in consequence allows for the observed pair of choices ($p_{1}$ and
$q_{2}$) but rules out the opposite pair ($p_{2}$ and $q_{1}$).\footnote{Continuing from \fnref{lotteries}, the opposite behavior requires
$u_{B}\left(3000\right)>u_{A}\left(3000\right)$ and is ruled out.
This observation resembles the \emph{Negative Certainty Independence}
postulate in \citet{dillenberger2010preferences,cerreia2015cautious}.} The same prediction applies to the common consequence effect and
the lotteries involved can be generalized.\footnote{Consider a degenerate lottery $\delta$ and a lottery $p$ such that
neither of them first order stochastically dominates another. Consider
the lotteries $\delta'=\delta^{\alpha}q$ and $p'=p^{\alpha}q$ where
$q$ is a lottery and $\alpha\in\left(0,1\right)$, and suppose $|X|=3$.
If $\delta\in c\left(\left\{ \delta,p\right\} \right)$ and $p'\in c\left(\left\{ \delta',p'\right\} \right)$,
then for all $u_{1},u_{2}:X\rightarrow\mathbb{R}$ such that $u_{1}$
explains the first choice and $u_{2}$ explains the second choice,
it is straightforward to show that $u_{1}=f\circ u_{2}$ for some
concave function $f:\mathbb{R}\rightarrow\mathbb{R}$. Moreover, these
choices can always be explained by an AREU representation such that
$r\left(\left\{ \delta,p\right\} \right)Rr\left(\left\{ \delta',p'\right\} \right)$.
Conversely, suppose the choices $c\left(\left\{ \delta,p\right\} \right)$
and $c\left(\left\{ \delta',p'\right\} \right)$ admit an AREU representation
such that $r\left(\left\{ \delta,p\right\} \right)Rr\left(\left\{ \delta',p'\right\} \right)$,
then $p\in c\left(\left\{ \delta',p'\right\} \right)$ whenever $p\in c\left(\left\{ \delta,p\right\} \right)$
(and equivalently $\delta\in c\left(\left\{ \delta,p\right\} \right)$
whenever $\delta'\in c\left(\left\{ \delta',p'\right\} \right)$).}

Because different utility functions are involved, AREU predicts a
novel manifestation of the common ratio effect\textemdash typically
formulated in binary comparisons\textemdash as WARP violations. Consider
the lotteries $p_{1}=\delta_{3000}$, $p_{2}=0.5\delta_{4000}\oplus0.5\delta_{0}$,
$q_{1}=0.2\delta_{4000}\oplus0.7\delta_{3000}\oplus0.1\delta_{0}$,
and $q_{2}=0.4\delta_{4000}\oplus0.3\delta_{3000}\oplus0.3\delta_{0}$,
related by common mixture.\footnote{$q_{1}=\frac{2}{5}p_{1}\oplus\frac{3}{5}s$ and $q_{2}=\frac{2}{5}p_{2}\oplus\frac{3}{5}s$
where $s=\frac{1}{3}\delta_{4000}\oplus\frac{1}{2}\delta_{3000}\oplus\frac{1}{6}\delta_{0}$.} A decision maker who chooses $p_{1}$ over $p_{2}$, $q_{2}$ over
$q_{1}$, and $q_{1}$ over $p_{1}$ in binary comparisons commits
the common ratio effect (between the first two choices), reconciled
in AREU by a reference order that ranks $p_{1}$ highest. Now, consider
the choice set $\left\{ p_{1},q_{1},q_{2}\right\} $, for which $p_{1}$
must be the reference. The decision maker treats this choice set as
having the same context as $\left\{ p_{1},p_{2}\right\} $ and use
the same utility function that ranks $p_{1}$ over $p_{2}$, which,
due to expected utility, requires her to choose $q_{1}$ from $\left\{ p_{1},q_{1},q_{2}\right\} $.
However, the decision maker chose $q_{2}$ from $\left\{ q_{1},q_{2}\right\} $,
so she has committed a WARP violation. This simple observation introduces
a direct link between structural violations and basic rationality
violations.

\paragraph*{Other evidence}

While the Allais paradox takes center stage among anomalies in the
risk domain, the evidence and intuition for increased risk aversion
in the presence of safer options are also found in a wide range of
studies. In a setting meant to test for the compromise effect, \citet{herne1999effects}
found that the presence of a safer option results in WARP violations
in the direction of greater risk aversion. \citet{Wakker1996} introduces
the tradeoff\emph{ method }to elicit risk aversion without using a
sure prize and found that the estimated utility functions are less
concave relative to standard methods that involve sure prizes. \citet{andreoni2011uncertainty}
found similar effects when the safest option is close enough to certainty.
Restricted to binary comparisons, \citet{bleichrodt2002context} studies
a model of context-dependent gambling effect where a decision maker
has two utility functions and uses the more concave one whenever the
binary comparison involves a riskless option.

\paragraph*{Linking structural properties to basic rationality}

It turns out that compliance with WARP or Independence would independently
bring us back to standard expected utility, stated in \propref{WARP}.
\begin{prop}
\label{prop:WARP}If $c$ admits an AREU representation, then the
following are equivalent:
\begin{enumerate}
\item $c$ satisfies WARP (over $\mathcal{A}$).
\item $c$ satisfies Independence (over $\mathcal{A}$).
\item $c$ admits an expected utility representation.
\item $c$ admits a utility representation.
\end{enumerate}
\end{prop}
This also means that if $c$ admits \emph{any} utility representation,
then it must also have an expected utility representation.\footnote{As is standard, we say $c$ admits a utility representation if there
exists a real valued function $U:Y\rightarrow\mathbb{R}$ such that
$c\left(A\right)=\arg\max_{y\in A}U\left(y\right)$ for all $A\in\mathcal{A}$.} This observation provides a formal separation between AREU and non-expected
utility models that uphold basic rationality and further suggests
that violation of Independence in this model is a matter of changing
preferences.

It can be shown that imposing transitivity achieves the same outcome.
Moreover, if transitivity is only satisfied locally, that is, applying
only to a region of lotteries, then the model gives rise to betweenness
behavior in that region and further implies fanning out if behavior
is risk averse and fanning in when it is risk loving. These in-depth
analyses are relegated to \citet{lim2023avoidable}.

\paragraph*{Model specification and identification}

In applications, keeping track of so many utility functions can be
challenging, an issue shared in \citet{cerreia2015cautious}, \citet{chakraborty2021present},
and \citet{ellis2022choice}.\footnote{Relatedly, models of ambiguity aversion also use a collection of subjective
priors \citep{gilboa1989maxmin}.} AREU provides a middle ground: Knowing that utility functions are
related by concave transformations, an analyst might reasonably assume
that a decision maker's utility functions come from a set of constant
absolute risk aversion (CARA) utility functions given by a subjective
range of Arrow-Pratt coefficients $\alpha\in\left[\underline{\alpha},\bar{\alpha}\right]$.
More generally, it is also possible for risk attitude to progress
from risk loving (convex utility functions) to risk averse (concave
utility functions). The range of risk attitudes is ultimately subjective
and could vary across individuals or demographics; one individual
may be moderately but consistently risk averse, with a very small
range of CARA coefficients, whereas another individual may be occasionally
risk loving but sometimes very risk averse.

Partial subjectivity in the reference order allows for more individual
differences but burdens identification. In the extreme case where
behavior is consistent with standard expected utility, it is impossible
to pin down $R$, although analysis can proceed with standard expected
utility. Fortunately, as long as two reference points index different
utility functions, identification of $R$ between them is guaranteed.
First, if two choice sets differ only by $p$, and choices are inconsistent
with expected utility maximization, then we identify that $p$ ranks
higher in $R$ than the other alternatives in the choice set. It turns
out that the converse is also true. As long as $p$ and $q$ index
different utility functions, if $pRq$, then we can find choice sets
$A,B$ such that $p,q\in A$ and $B=A\backslash\left\{ p\right\} $
where choices from $A$ and $B$ violate WARP, meaning we revealed
$pRq$.\footnote{The proof of \propref{WARP} contains this observation. Essentially,
it relies on a less obvious property implied the model that guarantees
existence of a full-dimensional subset of lotteries that rank below
$p$ and $q$ in $R$ but are better than $p$ and $q$ when they
act as the reference points.}

\section{\label{sec:time}Time Preference}

The canonical model for time preference is Discounted Utility, where
a decision maker evaluates each payment-time pair $\left(x,t\right)$
using exponential discounting, i.e., $\delta^{t}u\left(x\right)$.
But the Stationarity condition within this model is routinely challenged
by lab and field subjects who switch their choices between two payments
when the decision is made in advance, typically favoring the later
option for long-term decisions, an actively studied behavioral phenomenon
termed \emph{present bias} \citep{laibson1997golden,frederick2002time,benhabib2010present,halevy2015,chakraborty2021present,chambers2023decreasing}.
This section studies how present bias is related WARP-violating preference
changes. The original axioms in \citet{fishburn1982time} are imposed
only among choice sets that share a reference point, which in this
case is the soonest available payment, as it partially captures how
early in advance a decision maker is making the decision.

\subsection{Preliminaries and axioms}

Let $X=\left[a,b\right]\subset\mathbb{R}_{>0}$ be a non-degenerate
interval of payments and let $T=\left[0,\bar{t}\right]\subset\mathbb{\mathbb{R}}_{\geq0}$
be a non-degenerate interval of time points. Let $Y=X\times T$ be
the set of all timed payments, where each option $\left(x,t\right)\in X\times T$
is a payment of $x$ that arrives at time $t$. Everything else follows
\secref{ORD}. To simplify analysis, I assume the upper bound of payments
is large enough so that some payment at time $\bar{t}$ is better
than the worst payment at time $0$, specifically $\left(b,\bar{t}\right)\in c\left(\left\{ \left(a,0\right),\left(b,\bar{t}\right)\right\} \right)$.
The first axiom is standard; greater payments and sooner payments
are better.
\begin{ax}
\label{axm:time_monotonicity=000026impatience}~
\begin{enumerate}
\item If $x>y$, then $c\left(\left\{ \left(x,t\right),\left(y,t\right)\right\} \right)=\left\{ \left(x,t\right)\right\} $.
\item If $t<s$, then $c\left(\left\{ \left(x,t\right),\left(x,s\right)\right\} \right)=\left\{ \left(x,t\right)\right\} $.
\end{enumerate}
\end{ax}
The well-known Stationarity condition posits that a decision maker's
preference between two future payments is consistent regardless of
when the decision is made. Consider the following definition that
allows for selective application.
\begin{defn}
$c$ \emph{satisfies Stationarity over $\mathcal{S}\subseteq\mathcal{A}$}
if for all $A,B\in\mathcal{S}$ and $a>0$, if $\left(x,t\right)\in c\left(A\right)$,
$\left(y,q\right)\in A$, $\left(y,q+a\right)\in c\left(B\right)$,
and $\left(x,t+a\right)\in B$, then $\left(x,t+a\right)\in c\left(B\right)$.
\end{defn}
Whereas global compliance with Stationarity is captured by $\mathcal{S}=\mathcal{A}$,
the next axiom demands local compliance. Specifically, it requires
Stationarity to be satisfied between any two choice sets that share
an earliest payment.
\begin{defn}
\label{def:psi_time}Let $\Psi\left(A\right):=\left\{ \left(x,t\right)\in A:t\leq q\text{ for all }\left(y,q\right)\in A\right\} $
be the set of \emph{earliest payments} in $A$.
\end{defn}
\begin{ax}[Time Reference Dependence]
\label{axm:time_axiom1}If $\Psi\left(A\right)\cap\Psi\left(B\right)\ne\emptyset$,
then $c$ satisfies WARP and Stationary over $\left\{ A,B\right\} $.
\end{ax}
It turns out that \axmref{time_axiom1} is an application of \emph{Reference
Dependence} (\secref{ORD}), formalized by \lemref{timeistime}, which
assures us that the proposed approach is related to demanding compliance
between certain pairs of choice sets.
\begin{lem}
\label{lem:timeistime}$c$ satisfies \axmref{time_axiom1} if and
only if for every $A\in\mathcal{\mathcal{A}}$ and $\left(x,t\right)\in\Psi\left(A\right)$,
$c$ satisfies WARP and Stationarity over $\left\{ B\in\mathcal{\mathcal{A}}:\left(x,t\right)\in B\subseteq A\right\} $.
\end{lem}
The next postulate rules out increased patience when more options
become available. The intuition is that additional options can only
tempt the decision maker to become more impatient, so if an impatient
decision is already made from $B$, for example if $\left(x_{1},t_{1}\right)$
is (strictly) chosen over $\left(x_{2},t_{2}\right)$ where $t_{1}<t_{2}$,
then there is no superset $A\supset B$ such that the decision maker
becomes more patient by choosing $\left(x_{2},t_{2}+d\right)$ in
the presence of $\left(x_{1},t_{1}+d\right)$.
\begin{ax}
\label{axm:patience1}Suppose $B\subset A$, $t_{1}<t_{2}$, and $d\in\mathbb{R}$.
If $\left(x_{1},t_{1}\right)\in c\left(B\right)$ and $\left(x_{2},t_{2}\right)\in B\backslash c\left(B\right)$,
then $\left(x_{1},t_{1}+d\right)\in A$ implies $\left(x_{2},t_{2}+d\right)\notin c\left(A\right)$.
\end{ax}
However, this falls short of definitively capturing changes in patience.
Even in a completely standard world where every individual maximizes
exponentially discounted utility, behavioral differences in delay
aversion (among individuals) cannot be definitively decomposed into
differences in discounting and differences in consumption utility,
an issue discussed in \citet{ok2007delay}. Meaning an individual
who prefers the sooner alternative could have \emph{greater} patience
paired with lower marginal utility for money.

The last postulate addresses this issues by capturing fixed\emph{
}consumption utilities under varying discounting/patience: Suppose
a decision maker is indifferent between all options in the choice
set $\left\{ \left(x_{1},t_{1}\right),\left(x_{2},t_{2}\right),\left(x_{3},t_{3}\right)\right\} $,
where $x_{1}<x_{2}<x_{3}$ and $t_{1}<t_{2}<t_{3}$. Then in the
choice set $\left\{ \left(x_{1},\lambda t_{1}\right),\left(x_{2},\lambda t_{2}\right),\left(x_{3},\lambda t_{3}\right)\right\} $
where $0<\lambda<1$, since the delays between options have shortened,
a standard exponential discounting decision maker would pick $\left(x_{3},\lambda t_{3}\right)$
as the new choice. Yet, our decision maker will face competing forces.
On one hand, the possibility of sooner consumption makes her more
impatient; on the other hand, shorter delays between options make
later payments more attractive. Allowing her the freedom to resolve
these competing forces, the next postulate requires that if she ends
up choosing both $\left(x_{1},\lambda t_{1}\right)$ and $\left(x_{3},\lambda t_{3}\right)$\textemdash as
if the competing forces are balanced\textemdash then she must also
choose the intermediate option $\left(x_{2},\lambda t_{2}\right)$.
The same requirement applies when a common delay (or advancement)
$d$ is additionally imposed. Both \axmref{patience1} and \axmref{patience2}
are trivially satisfied in exponential discounting.
\begin{ax}
\label{axm:patience2}Consider $A=\left\{ \left(x_{1},t_{1}\right),\left(x_{2},t_{2}\right),\left(x_{3},t_{3}\right)\right\} $
such that $t_{1}<t_{2}<t_{3}$ and $A'=\left\{ \left(x_{1},\lambda t_{1}+d\right),\left(x_{2},\lambda t_{2}+d\right),\left(x_{3},\lambda t_{3}+d\right)\right\} $
such that $0<\lambda<1$ and $d\in\mathbb{R}$. If $c\left(A\right)=A$,
then either $c\left(A'\right)=\left(x_{1},\lambda t_{1}+d\right)$,
$c\left(A'\right)=\left(x_{3},\lambda t_{3}+d\right)$, or $c\left(A'\right)=A'$.
\end{ax}

\subsection{Model}

Since we consider the standard environment where sooner is always
better, discount factors are restricted to non-negative real numbers
strictly less than 1, with the exception of $r=\left(x,\bar{t}\right)$
for which $\delta_{r}=1$ is possible.
\begin{defn}
$c$ admits a Present-Biased Exponentially Discounted Utility (PEDU)
representation if it admits an ORDU representation $\left(\left\{ U_{r}\right\} _{r\in Y},R\right)$
such that for some strictly increasing function $u:X\rightarrow\mathbb{R}$
and set of discount factors $\left\{ \delta_{r}\right\} _{r\in Y}$,
\begin{itemize}
\item $U_{r}\left(x,t\right)=\delta_{r}^{t}u\left(x\right)$,
\item $t<t'$ implies $\left(x,t\right)R\left(x',t'\right)$ and $\delta_{\left(x,t\right)}\leq\delta_{\left(y,t'\right)}$,
\item $t=t'$ implies $\delta_{\left(x,t\right)}=\delta_{\left(y,t\right)}$.
\end{itemize}
\end{defn}
\begin{thm}
\label{thm:pbdu}$c$ satisfies Axioms \ref{axm:time_monotonicity=000026impatience}-\ref{axm:patience2}
and Continuity if and only if it admits a PEDU representation.
\end{thm}
In this model, it is as if the decision maker maximizes exponentially
discounted utility, but with discount factors that depend on the timing
of the earliest available payment. When it is possible to choose an
early payment, the decision maker uses a lower discount factor, resulting
in behavior that reflects reduced patience. The model thus delivers
present bias behavior using familiar technologies\textemdash since
the exponential discounting form is preserved in every instance of
decision-making, changes in patience are simply captured by set-dependent
discount factors. Intuitively, with the entire set of possible payments
progressively postponed, the decision maker begins to treat them more
akin to long-term concerns than before, resulting in increased patience.

It can be shown that $\delta_{r}$ is unique given $u$, except possibly
when $r=\left(x,\bar{t}\right)$.\footnote{Uniqueness is demonstrated in the proof of \thmref{pbdu}. When $r=\left(x,\bar{t}\right)$,
$\delta_{r}$ is only used to evaluate alternatives that also arrive
at time $\bar{t}$, so any $\delta_{r}$ paired with a strictly increasing
$u$ can explain those choices. It could still be unique if $\lim_{t\rightarrow\bar{t}}\delta_{\left(x,t\right)}=1$,
since a PEDU representation requires $\delta_{\left(x,t\right)}\leq\delta_{\left(x,\bar{t}\right)}\leq1$.} In applications, since the reference order and the discount factors
depend only on the timing of a payment, it is without loss to consider
discount factors that are based on time rather than on alternatives.
This is achieved by setting $\tilde{\delta}_{t}:=\delta_{\left(x,t\right)}$
for all $t\in T$ and then using the earliest available time of a
payment as reference point.

\paragraph*{Generalized single-switching}

Changes in preferences are tractable due to a generalized single-switching
property. In binary comparisons, a unique threshold captures the postponement
beyond which the later payment will be chosen and before which the
sooner payment will be chosen. In more general choice sets, this threshold
no longer guarantees a choice between the two timed payments but continues
to stipulate the point of postponement beyond which the sooner payment
cannot be chosen (because the later payment is available) and before
which the later payment cannot be chosen (because the sooner payment
is available). This generalized single-switching property thus extends
our understanding of present bias in binary comparisons to arbitrary
choice sets\textemdash even in the absence of basic rationality assumptions\textemdash and
it is closely tied to the unified framework in which references are
ordered and preference shifts systematically along this established
order.

\paragraph*{Present bias in WARP violations}

Although present bias is typically viewed as a structural violation,
PEDU predicts a novel manifestation of present bias as WARP violations.
Consider the present bias behavior where ``\$20 in 4 days'' is chosen
over ``\$18 in 3 days'', but ``\$18 today'' is chosen over ``\$20
tomorrow''. In PEDU, this behavior is explained using a lower discount
factor for the latter choice set. However, notice that under this
lower discount factor, ``\$18 in 3 days'' is preferred to ``\$20
in 4 days'', so the introduction of a third option that induces this
discount factor but is not itself chosen, for example ``\$15 today'',
will result in a reversal where ``\$18 in 3 days'' is chosen over
``\$20 in 4 days''. This is now a WARP violation that shares the
same underlying driver as present bias behavior, even though present
bias is typically studied in binary comparisons. In fact, consistent
with the spirit of present bias, WARP violations in PEDU are restricted
to decreased patience, and only when sooner payments are added.

\paragraph*{Linking structural properties to basic rationality}

To further ascertain the aforementioned connection, \propref{WARP2}
shows that relaxing just one of the two conditions would fully recover
standard exponential discounting. Consequently, if a PEDU decision
maker has \emph{any} utility representation, then she must also have
a standard exponential discounting utility representation. This adds
to the suggestion that anomalies captured by PEDU are rooted in systematic
changes in preferences.
\begin{prop}
\label{prop:WARP2}If $c$ admits a PEDU representation, then the
following are equivalent:
\begin{enumerate}
\item $c$ satisfies WARP (over $\mathcal{A}$).
\item $c$ satisfies Stationarity (over $\mathcal{A}$).
\item $c$ admits an exponential discounting utility representation.
\item $c$ admits a utility representation.
\end{enumerate}
\end{prop}

\paragraph*{Hyperbolic discounting}

\propref{WARP2} separates PEDU from hyperbolic discounting, quasi-hyperbolic
discounting, and related generalizations \citep{phelps1968second,loewenstein1992anomalies,laibson1997golden,frederick2002time,chambers2023decreasing,chakraborty2021present}
due to their adherence to basic rationality, but the empirically informed
intuition that discount factors can vary is shared. In contrast, PEDU
varies discount factors at the choice problem level whereas hyperbolic
discounting does so at the alternative level. Binary comparisons hold
similar behavioral implications: when two options are gradually advanced,
there may be a point where the choice is switched from the sooner
to the later.\footnote{\citet{chakraborty2021present} calls this \emph{Weak Present Bias}
and studies its implications.} But for larger choice sets, unlike PEDU, hyperbolic discounting predicts
that the preference ranking between any two options stays the same
regardless of the presence of a third alternative.\textcolor{blue}{}

\paragraph*{WARP violations in other time preference settings}

Beyond the conventional time preference setting, an active literature
on menu preference applies \citet{gul2001temptation}'s temptation
model to decision makers who prefer a smaller menu in order to prevent
their future selves from committing undesirable present bias behaviors
\citep{noor2011temptation,lipman2013temptation,ahn2019behavioural}.
In these models, past and future selves prefer to choose differently
from the same set of alternatives, which could manifest as a reversal
if played out, therefore PEDU and these models tackle dynamic inconsistency
using related intuitions about long-term and short-term attitudes.

\citet{freeman2016revealing}'s task completion study, which is related
to the above literature and closer to PEDU's setting, considers a
time-inconsistent decision maker who exhibits choice reversals when
additional opportunities for completions are introduced. In particular,
a sophisticated decision maker ends up completing the task earlier,
therefore choosing a sooner option when choice set expands is a common
theme between our work. However, the manifestation of this behavior
is different; a reversal in PEDU can only occur when an alternative
earlier than any other is added, yet in \citet{freeman2016revealing},
adding this kind of alternatives either results in the addition chosen
or the choice remains unchanged, therefore WARP will hold.

\paragraph*{Consumption streams}

Focusing on one time payment helps glean the intuition of this framework,
but the approach already suggests how an extension to consumption
streams can be conducted, where a decision maker maximizes $\sum_{t}\delta_{r\left(A\right)}^{t}u\left(x_{t}\right)$
(for discrete time). If $r\left(A\right)$ is the consumption stream
that offers the soonest payment, then the characterization amounts
to adding \citet{koopmans1960stationary}'s axioms alongside WARP
and Stationarity using \emph{Reference Dependence} (\secref{ORD}).
\oappxref{tau-and-proofs} clarifies what axioms can be accommodated,
and it includes common versions of separability.

\section{\label{sec:social}Social Preference}

Consider a decision maker whose willingness to share is greater when
the situation allows for greater equality. It departs from models
of other-regarding preferences that capture a fixed inequality aversion
\citep{fehr1999theory,bolton2000erc,charness2002understanding}.
To illustrate, suppose a decision maker is endowed with $\$10$ and
is asked to share it with another individual. However, instead of
choosing any split of this $\$10$, she was only given a few options.
When asked to choose between giving $\$2$ and giving $\$3$, giving
$\$2$ may seem like a fair decision. However, when the choice is
between giving $\$2$, $\$3$, or $\$5$, she may opt for giving $\$3$
instead. The choices $c\left(\left\{ \left(\$8,\$2\right),\left(\$7,\$3\right)\right\} \right)=\left\{ \left(\$8,\$2\right)\right\} $
and $c\left(\left\{ \left(\$8,\$2\right),\left(\$7,\$3\right),\left(\$5,\$5\right)\right\} \right)=\left\{ \left(\$7,\$3\right)\right\} $
violate WARP, and hence a fixed utility function, even if it captures
other-regarding preferences and inequality aversion, is incapable
of explaining this behavior.

\subsection{Preliminaries and axioms}

Let $Y=[w,+\infty)\times[w,+\infty)$, where $w\in\mathbb{R}_{>0}$,
be a set of income distributions. For each option $\left(x,y\right)\in Y$,
$x$ is the dollar amount received by the decision maker and $y$
is the dollar amount given to another individual. Everything else
follows \secref{ORD}. The first axiom assumes that an income distribution
is strictly preferred when it gives someone more and no one less.
\begin{ax}
\label{axm:social_monotonicity}If $x\geq x'$, $y\geq y',$ and $\left(x,y\right)\ne\left(x',y'\right)$,
then $c\left(\left\{ \left(x,y\right),\left(x',y'\right)\right\} \right)=\left\{ \left(x,y\right)\right\} $.
\end{ax}
\emph{Reference Dependence} (\secref{ORD}) adapts to this domain
and characterizes choices that conform with \emph{quasi-linear preferences}
when the underlying choice sets have the same level of \emph{attainable}
\emph{equality}. Since the impending model involves reference-dependent
preferences, using quasi-linear utilities as baseline (rather than
using more general models of other-regarding preferences) provides
meaningful restrictions.
\begin{defn}
\emph{$c$ satisfies Quasi-linearity over $\mathcal{S}\subseteq\mathcal{A}$}
if for all $A,B\in\mathcal{S}$ and $a\in\mathbb{R}\backslash\left\{ 0\right\} $,
if $\left(x,y\right)\in c\left(A\right)$, $\left(x',y'\right)\in A$,
$\left(x'+a,y'\right)\in c\left(B\right)$, and $\left(x+a,y\right)\in B$,
then $\left(x+a,y\right)\in c\left(B\right)$.
\end{defn}
The measure of attainable equality is based on the Gini coefficient,
\[
G\left(\left(x,y\right)\right)=\frac{\left|x-y\right|+\left|y-x\right|}{4\left(x+y\right)},
\]
which ranges from $0$ (most balanced) to $0.5$ (least balanced)
for our 2-agents setting. Analogous to other domains, compliance with
WARP and Quasi-linearity is called for when two choice sets share
a Gini-minimizing income distribution.
\begin{defn}
\label{def:psi_social}Let $\Psi\left(A\right):=\left\{ \left(x,y\right)\in A:G\left(\left(x,y\right)\right)\leq G\left(\left(x',y'\right)\right)\text{ for all }\left(x',y'\right)\in A\right\} $
be the set of \emph{most-balanced income distributions} in $A$.
\end{defn}
\begin{ax}[Equality Reference Dependence]
\label{axm:equality_reference_dependence}For any $A\in\mathcal{\mathcal{A}}$
and any most-balanced income distribution $\left(x,y\right)\in\Psi\left(A\right)$,
$c$ satisfies WARP and Quasi-linearity over $\left\{ B\in\mathcal{\mathcal{A}}:\left(x,y\right)\in B\subseteq A\right\} $.
\end{ax}
The next and last postulate regulates changes in preferences. Suppose
$y>y'$ and a decision maker chooses to share more $\left(x,y\right)$
than to share less $\left(x',y'\right)$. I postulate that making
more options available will not cause the decision maker to switch
to sharing less, since the added options can only increase attainable
equality.
\begin{ax}
\label{axm:social_increasingaltruism}Suppose $B\subset A$ and $y>y'$.
If $\left(x,y\right)\in c\left(B\right)$ and $\left(x',y'\right)\in B\backslash c\left(B\right)$,
then $\left(x',y'\right)\notin c\left(A\right)$.
\end{ax}

\subsection{Model}
\begin{defn}
$c$ admits a Fairness-based Social Preference Utility (FSPU) representation
if it admits an ORDU representation $\left(\left\{ U_{r}\right\} _{r\in Y},R\right)$
such that for some set of strictly increasing functions $\left\{ v_{r}:[w,+\infty)\rightarrow\mathbb{R}\right\} _{r\in Y}$,
\begin{itemize}
\item $U_{r}\left(x,y\right)=x+v_{r}\left(y\right)$,
\item $G\left(r\right)<G\left(r'\right)$ implies $rRr'$ and $v_{r}\left(y\right)-v_{r}\left(y'\right)\geq v_{r'}\left(y\right)-v_{r'}\left(y'\right)$
for all $y>y'$,
\item $G\left(r\right)=G\left(r'\right)$ implies $v_{r}\left(y\right)=v_{r'}\left(y\right)$.
\end{itemize}
\end{defn}
\begin{thm}
\label{thm:fspu}$c$ satisfies Axioms \ref{axm:social_monotonicity}-\ref{axm:social_increasingaltruism}
and Continuity if and only if it admits an FSPU representation.
\end{thm}
FSPU combines an objective measure of \emph{equality} with a subjective
interpretation of \emph{fairness}. Every decision maker bases her
choice on the Gini-minimizing option, $r\left(A\right)$, as it captures
the amount of attainable equality in a choice set. When attainable
equality is higher ($G\left(r\left(A\right)\right)$ is lower), utility
difference between sharing more and sharing less increases, reflecting
increased willingness to share. The amount of increase depends on
the decision maker's subjective sense of fairness. A very large increase
causes WARP violations, where the decision maker switches from an
option that shares less to an option that shares more even though
both options are always present. Like the other domains, preference
parameters $\left\{ v_{r}\right\} _{r\in Y}$ are unique.\footnote{Uniqueness is demonstrated in the proof of \thmref{fspu}.}

For applications, it is without loss to further simplify FSPU by using
Gini coefficient\textemdash rather than alternatives\textemdash to
index context-dependent utility from sharing. To do so, for all $\bar{G}\in[0,0.5)$,
set $\tilde{v}_{\bar{G}}:=v_{\left(x,y\right)}$ where $\bar{G}=G\left(\left(x,y\right)\right)$,
and then use the lowest attainable \emph{Gini coefficients} as reference
points.

\paragraph*{Menu-dependent altruism}

As in the motivating example, the model explains context-dependent
willingness to share when distributing a fixed pie with different
splitting options. Suppose a decision maker is allocating $\$M$ between
herself and another individual, and each choice set is characterized
by a set of splitting fractions $D\subset\left[0,1\right]$. That
is, she can allocate $\alpha\cdot\$M$ to herself and $\left(1-\alpha\right)\cdot\$M$
to other party if and only if $\alpha\in D$. Consider $D=\left\{ 0.6,0.7\right\} $
and $D'=\left\{ 0.5,0.6,0.7\right\} $. Since attainable equality
is greater in $D'$ (it contains an equal split), a decision maker
who chooses $0.7$ from $D$ may exhibit increased willingness to
share that results in choosing $0.6$ from $D'$, even if this violates
WARP. But the model rules out the opposite behavior: A decision maker
who chooses $0.6$ from $D$ cannot choose $0.7$ from $D'$, since
it would imply decreased willingness to share. Also, a reversal cannot
happen between $D=\left\{ 0.6,0.7\right\} $ and $D''=\left\{ 0.6,0.7,0.8\right\} $
since they have the same level of attainable equality.

\paragraph*{Equality over generosity}

Willingness to share is maximized when a perfectly balanced income
distribution is available. In particular, the model captures increased
altruism not due to the opportunity to \emph{give more} per se, but
due to the opportunity to \emph{be equal}. To illustrate the difference,
consider the same example but with $D=\left\{ 0.5,0.3,0.2\right\} $
and $D'=\left\{ 0.3,0.2\right\} $. Even though $D$ contains alternatives
that achieve greater equality, the decision maker's ability to give
is the same across the two choice sets. Yet, since the feasible allocations
are always unfavorable to her, higher attainable equality results
from her ability to take more. In this example, the decision maker
may be interpreted as being less altruistic when the world is unfair
to her, but becomes more altruistic when more greater equality becomes
possible.

\paragraph*{Fairness over efficiency}

Consider one last application where FSPU allows for willingness to
forgo a greater total surplus in favor of sharing. Suppose the decision
maker must choose between $\left(\$30,\$20\right)$ and $\left(\$60,\$0\right)$.
The second option is appealing in that the total amount of money is
greater, whereas the first option sacrifices both total surplus and
payment to oneself in order to provide a share to the other individual.
Suppose $\left(\$60,\$0\right)$ is chosen. In FSPU, adding $\left(\$25,\$25\right)$
as an option can cause the decision maker to switch from $\left(\$60,\$0\right)$
to $\left(\$30,\$20\right)$ due to increased altruism. While this
behavior seems reasonable, it is inconsistent with any model that
complies with WARP.

\paragraph*{Empirical evidence}

The vast literature on distributional preferences provides suggestive
evidence for FSPU behavior. Moreover, unlike the case of risk and
time domains, they do focus on basic rationality violations. In dictator
games, \citet{list2007interpretation,bardsley2008dictator,korenok2014taking}
find that changes to a dictator's choice set affect her willingness
to give and result in WARP-violating choices. \citet{dana2006you}
investigate the underlying mechanism by making the dictator game an
option and \citet{dana2007exploiting} do so by manipulating the visibility
of the choice set. They find the audience effect, where fair behavior
is the result of subjects' desire to be perceived (by themselves and
others) as fair. \citet{rabin1993incorporating} studies an intention-based
explanation in game theoretic settings where kindness is reciprocated.
Although existing studies motivate FSPU, the model does not distinguish
between willingness to share that depends intrinsically on outcomes
and that resulting from intentions.\footnote{More on outcome-based vs intention-based inequality aversion can be
found in \citet{ainslie1992picoeconomics}, \citet{nelson2002equity},
\citet{fehr2006economics}, \citet{sutter2007outcomes}, and \citet{kagel2016handbook}.}

In a more recent study, \citet{cox2016moral} conduct experiments
that explicitly test for basic rationality violations in dictator
games and, consistent with FSPU, find that shrinking a choice set
results in WARP violations in the direction of keeping more for oneself.
They propose a modification to basic rationality by introducing a
testable prediction based on a definition of moral reference points,
which depend on the framing of the problem (e.g., ``Give'' and ``Take'')
and features of the feasible distributions. When moral reference points
are fixed, rationality postulates are satisfied; otherwise, violations
favor the party who benefits from the new moral reference point. Their
work provides empirical support for FSPU, which in turn offers a theory
that complements their findings.

\paragraph*{Observable contexts and menu preference}

The intuitions contained in FSPU resonates with other studies that,
unlike FSPU, exploit a richer setting. In settings that include multiple
actors, \citet{cox2008revealed} study how the generosity of a first
mover affects the altruism of a second mover. \citet{cheung2023reciprocity}
focuses on a second mover who, more generally, makes different decisions
from the same choice set based on how the underlying choice set was
chosen by a first mover. Relatedly, \citet{vanBruggen2023giving}
consider a decision maker whose social preference depends on exogeneous
contexts like ``selfish'' and ``generous''. In a menu preference
setting, \citet{dillenberger2012ashamed} study a decision maker who
has shame concern and prefers a smaller menu that excludes normatively
better allocations that entail lower self-payoffs, since not choosing
those options can induce shame.

\paragraph*{Linking structural properties to basic rationality}

Like before, \propref{WARP3} shows that WARP violation and failure
of standard postulate (Quasi-linearity) are linked. In this setting,
it also suggests that wealth effects may be in part contributed by
reference dependent preferences.\footnote{Quasi-linear utility in wealth is often interpreted as the absence
of wealth effects. In this domain, it means an individual's willingness
to give does not depend on how much she would have left\textemdash her
wealth\textemdash because if giving $t$ is better than giving $t'$
with a base wealth $w$, i.e., $\left(w-t\right)+v\left(t\right)>\left(w-t'\right)+v\left(t'\right)$,
then the same holds true at a different wealth level $w'$, i.e.,
$\left(w'-t\right)+v\left(t\right)>\left(w'-t'\right)+v\left(t'\right)$.}
\begin{prop}
\label{prop:WARP3}If $c$ admits a FSPU representation, then the
following are equivalent:
\begin{enumerate}
\item $c$ satisfies WARP (over $\mathcal{A}$).
\item $c$ satisfies Quasi-linearity (over $\mathcal{A}$).
\item $c$ admits a quasi-linear utility representation.
\item $c$ admits a utility representation.
\end{enumerate}
\end{prop}

\section{\label{sec:Conclusion}Conclusion}

This paper presents a single, unifying, framework for reference-based
context-dependent preferences. The key innovation, \emph{Reference
Dependence} (RD), provides a way to jointly and systematically weaken
multiple postulates even if they are conceptually distinct. The method
is then applied to the risk, time, and social domains where \emph{basic
rationality postulates} and \emph{structural postulates} are jointly
relaxed, upholding the core principles of normative postulates by
demanding their local compliance. In each setting, behavior can be
understood as the result of canonical models when reference points
are fixed, and deviations from these models are accounted for by systematic
changes in reference-dependent preference parameters. Reference points
in this framework are determined by the maximization of a reference
order, which can be viewed as an instrument that captures the relevant
context of a choice problem.

Building upon decades of domain-specific research on seemingly independent
structural anomalies, including but not limited to the Allais paradox
and present bias behavior, this paper studies a possible link that
could relate them to WARP violations. This, in turn, informs more
fundamentally on the relationship between rationality postulates and
structural postulates. The exercise adds to our understanding of why
normative postulates fail, offers new ways to introduce assumptions,
and suggests new avenues for empirical research.

\begin{singlespace}
\noindent \bibliographystyle{econometrica}
\phantomsection\addcontentsline{toc}{section}{\refname}\bibliography{database}

\end{singlespace}

\appendix

\section{\label{appx:proofs}Appendix: Proofs}

\thmref{ORD}, \thmref{AREU}, \thmref{pbdu}, and \thmref{fspu}
require a technical result, \lemref{ordC}, that generalizes a large
class of behavioral postulates called \emph{finite theories} in a
reference dependent manner. The result is stated now but formally
introduced and proved in \oappxref{tau-and-proofs}. The definition
of a \emph{finite theory} is also given in \oappxref{tau-and-proofs},
it includes WARP, Independence, Stationarity, and Quasi-linearity.

A correspondence $\Psi:\mathcal{B}\rightarrow\mathcal{A}$ where $\Psi\left(A\right)\subseteq A$
is called an \emph{$\alpha-$correspondence} if for all $A,B\in\mathcal{\mathcal{B}}$,
if $a\in\Psi\left(A\right)$ and $a\in B\subset A$, then $a\in\Psi\left(B\right)$.
Given a linear order $\left(R,Y\right)$, let $r\left(A\right)$ denote
the unique element $x\in A$ such that $xRy$ for all $y\in A$. A
linear order $\left(R,Y\right)$ is called \emph{$\Psi$-consistent}
if for all $A\in\mathcal{B}$, $r\left(A\right)\in\Psi\left(A\right)$. 
\begin{lem}
\label{lem:ordC}Consider a choice correspondence $c$, a finite theory
$\mathcal{T}$, and an $\alpha-$correspondence $\Psi$. The following
are equivalent:
\begin{enumerate}
\item (Reference Dependence) For every choice set $A\in\mathcal{B}$, $c$
satisfies $\mathcal{T}$ over $\left\{ B\in\mathcal{B}:x\in B\subseteq A\right\} $
for some $x\in\Psi\left(A\right)$.
\item There exists a $\Psi$-consistent linear order $\left(R,Y\right)$
such that for all $x\in Y$, $c$ satisfies $\mathcal{T}$ over $\left\{ B\in\mathcal{B}:r\left(B\right)=x\right\} $.
\end{enumerate}
\end{lem}

\subsection{Proof of Theorem \ref{thm:ORD}}
\begin{lem}
\label{lem:ORDfinite}Suppose $Y$ is finite. A choice correspondence
$c$ satisfies \axmref{ord_warp} if and only if it admits an ORDU
representation.
\end{lem}
\begin{proof}
``\textbf{If}'' is straightforward. I prove ``\textbf{only if}''.
Denote by $\Gamma\left(A\right)$ the set of alternatives $x$ in
$A$ such that $c$ satisfies WARP over $\mathcal{S}=\left\{ B\subseteq A:x\in B\right\} $,
guaranteed to be non-empty by \axmref{ord_warp}. Create a list in
the following way: List elements of $\Gamma\left(Y\right)$ with an
arbitrary order. Since $Y\backslash\Gamma\left(Y\right)$ is again
finite, continue listing elements of $\Gamma\left(Y\backslash\Gamma\left(Y\right)\right)$
with an arbitrary order; continue until every $x\in Y$ is listed.
Finally, let $i_{x}$ denote the position of $x$ in the list. For
any $x,y\in Y$, construct $xRy$ if $i_{x}\geq i_{y}$.

For each $x\in Y$, it maximizes $R$ among alternatives in $R^{\downarrow}\left(x\right):=\left\{ y:xRy\right\} $,
hence by construction $c$ satisfies WARP over $\mathbb{A}_{R^{\downarrow}\left(x\right)}^{x}=\left\{ A\in\mathcal{A}:r\left(A\right)=x\right\} $.
Now construct $\left(\succsim_{x},Y\right)$. Set $y\succsim_{x}y$
for all $y\in Y$. For each $y\in R^{\downarrow}\left(x\right)$,
since $\left\{ x,y\right\} \in\mathbb{A}_{R^{\downarrow}\left(x\right)}^{x}$,
we set $y\succsim_{x}x$ or $x\succsim_{x}y$ or both according to
$c\left(\left\{ x,y\right\} \right)$. For each $y_{1},y_{2}\in R^{\downarrow}\left(x\right)$
such that $y_{1}\succsim_{x}x$ and $y_{2}\succsim_{x}x$, since $\left\{ x,y_{1},y_{2}\right\} \in\mathbb{A}_{R^{\downarrow}\left(x\right)}^{x}$,
we set $y_{1}\succsim_{x}y_{2}$ or $y_{2}\succsim_{x}y_{1}$ or both
according to $c\left(\left\{ x,y_{1},y_{2}\right\} \right)$, this
is guaranteed by the fact that $c$ satisfies WARP over $\mathbb{A}_{R^{\downarrow}\left(x\right)}^{x}$.
Now, $\succsim_{x}$ is complete on the set $\mathbb{P}^{x}:=\left\{ y:y\succsim_{x}x\right\} \equiv\left\{ y\in R^{\downarrow}\left(x\right):y\in c\left(\left\{ x,y\right\} \right)\right\} $,
which we call the prediction set of $x$. Now consider $Y\backslash\mathbb{P}^{x}=\left\{ y:yRx\text{ or }x\succ_{x}y\right\} $.
Set $y_{1}\sim_{x}y_{2}$ for all $y_{1},y_{2}\in Y\backslash\mathbb{P}^{x}$
and set $y_{1}\succ_{x}y_{2}$ for all $y_{1}\in\mathbb{P}^{x}$,
$y_{2}\in Y\backslash\mathbb{P}^{x}$. The constructed $\left(\succsim_{x},Y\right)$
is now complete. For transitivity, suppose $y_{1}\succsim_{x}y_{2}$
and $y_{2}\succsim_{x}y_{3}$, and that $y_{1},y_{2},y_{3}\in\mathbb{P}^{x}$
(if any of them is in $Y\backslash\mathbb{P}^{x}$ then the argument
is straightforward by $\sim_{x}$), hence $y_{1}\in c\left(\left\{ x,y_{1},y_{2}\right\} \right)$
and $y_{2}\in c\left(\left\{ x,y_{2},y_{3}\right\} \right)$. Furthermore,
since $y_{1},y_{2},y_{3}\in\mathbb{P}^{x}$, we have $\left\{ x,y_{1},y_{2},y_{3}\right\} \in\mathbb{A}_{R^{\downarrow}\left(x\right)}^{x}$,
and $c$ satisfies WARP over $\mathbb{A}_{R^{\downarrow}\left(x\right)}^{x}$
implies $y_{1}\in c\left(\left\{ x,y_{1},y_{2},y_{3}\right\} \right)$,
and hence $y_{1}\in c\left(\left\{ x,y_{1},y_{3}\right\} \right)$,
which implies $y_{1}\succsim_{x}y_{3}$. So $\left(\succsim_{x},Y\right)$
is transitive.

Finally, we show that $\left(R,Y\right)$ and $\left\{ \left(\succsim_{x},Y\right)\right\} _{x\in Y}$
explain $c$. For any $A\in\mathcal{A}$, since $A$ is finite and
$R$ is a linear order, there is a unique $R-$maximizer $x\in A$,
hence $A\in\mathbb{A}_{R^{\downarrow}\left(x\right)}^{x}$. Suppose
for contradiction $y_{1}\in c\left(A\right)$ but $y_{1}\notin\left\{ y\in A:y\succsim_{x}z\,\forall z\in A\right\} $,
so $y_{2}\succ_{x}y_{1}$ for some $y_{2}\in A$. Then $y_{1}\notin c\left(\left\{ x,y_{1},y_{2}\right\} \right)$.
Since $\left\{ x,y_{1},y_{2}\right\} $ is a subset of $A$, and both
are in $\mathbb{A}_{R^{\downarrow}\left(x\right)}^{x}$, this is a
violation of the statement $c$ satisfies WARP on $\mathbb{A}_{R^{\downarrow}\left(x\right)}^{x}$,
hence a contradiction. Suppose for contradiction $y_{2}\in\left\{ y\in A:y\succsim_{x}z\,\forall z\in A\right\} $
but $y_{2}\notin c\left(A\right)$. Consider any $y_{1}\in c\left(A\right)$,
since $y_{2}\succsim_{x}y_{1}$, $y_{2}\in c\left(\left\{ x,y_{1},y_{2}\right\} \right)$.
Since $\left\{ x,y_{1},y_{2}\right\} $ is a subset of $A$, and both
are in $\mathbb{A}_{R^{\downarrow}\left(x\right)}^{x}$, this is a
violation of the statement $c$ satisfies WARP on $\mathbb{A}_{R^{\downarrow}\left(x\right)}^{x}$.
Hence $c\left(A\right)=\left\{ y\in A:y\succsim_{x}z\,\forall z\in A\right\} $.
It remains to show that each $\succsim_{x}$ can be represented by
a utility function, but this is standard since $Y$ is finite and
$\succsim_{x}$ is complete and transitive.
\end{proof}
Now we prove the general case where $Y$ is not finite. ``\textbf{If}''
is straightforward. I prove ``\textbf{only if}''. Using \lemref{ordC},
let $\mathcal{T}$ be WARP and let $\Psi$ be the identify function,
then there exists a linear order $\left(R,Y\right)$ such that $c$
satisfies WARP over $\mathbb{A}_{R^{\downarrow}\left(x\right)}^{x}=\left\{ A\in\mathcal{A}:r\left(A\right)=x\right\} $
for every $x\in Y$. It is obviously $\Psi$-consistent. Proceed to
build $\left\{ \left(\succsim_{x},Y\right)\right\} _{x\in Y}$ using
the method outlined in the proof of \lemref{ORDfinite}, which gives
us a complete and transitive $\succsim_{x}$ for each $x$ such that
$c\left(A\right)=\left\{ y\in A:y\succsim_{r\left(A\right)}z\,\forall z\in A\right\} $.

It remains to show that each $\succsim_{x}$ can be represented by
a utility function. Based on our construction, $\succsim_{x}$ is
complete and transitive on $Y$. Moreover, it is continuous ($y_{n}\rightarrow y$,
$z_{n}\rightarrow z$, and $y_{n}\succsim_{x}z_{n}$ for each $n$
implies $y\succsim_{x}z$) when restricted to the prediction set $\mathbb{P}^{x}$,
otherwise a contradiction of Continuity would be detected in the choices
from a sequence of choice problems of form $\left\{ x,y_{n},z_{n}\right\} $
that converges to $\left\{ x,y,z\right\} $ ($\mathbb{P}^{x}$ guarantees
that $x$ will not be the only one chosen in any of these sets, so
that a contradiction of, say, $z\succ_{x}y$, will be substantiated
in choice: $z\in c\left(\left\{ x,y,z\right\} \right)$) . Therefore,
along with the fact $\mathbb{P}^{x}$ is a subset of the separable
metric space $Y$, $\succsim_{x}$ admits a (continuous) utility function
$U:\mathbb{P}^{x}\rightarrow\left[0,1\right]$ that represents $\succsim_{x}$
when restricted to the alternatives in $\mathbb{P}^{x}$. Now define
$U\left(z\right)=-1$ for all $z\in Y\backslash\mathbb{P}^{x}$. Now
$U$ also represents $y\succ_{x}z$ for all $y\in\mathbb{P}^{x}$
and $z\in Y\backslash\mathbb{P}^{x}$ and $z\sim_{x}z'$ for all $z,z'\in Y\backslash\mathbb{P}^{x}$.
And we are done. Finally, since our system of $\left(R,Y\right)$
and $\left\{ \left(\succsim_{x},Y\right)\right\} _{x\in Y}$ explains
$c$, which satisfies Continuity, so $c$ has a closed graph.

\subsection{Proof Outline of Theorems \ref{thm:AREU}, \ref{thm:pbdu}, \ref{thm:fspu}}

The proofs for these theorems are repetitive and cannot be streamlined
due to domain-specific details, I outline key ideas here and relegate
complete proofs to \oappxref{tau-and-proofs}.

\subsubsection*{Step 1: Reference order $R$}

In their respective domains, \defref{psi_risk}, \defref{psi_time},
and \defref{psi_social} prescribe $\Psi$'s that are $\alpha$-correspondences.
Moreover, WARP, Independence, Stationarity, and Quasi-linearity are
finite theories. Therefore, \emph{Risk Reference Dependence }(\axmref{risk_axiom1}),
\emph{Time Reference Dependence} (\axmref{time_axiom1}), and \emph{Equality
Reference Dependence} (\axmref{equality_reference_dependence}) each
qualifies as a special case of the ``meta'' axiom \emph{Reference
Dependence}. By invoking \lemref{ordC}, we obtain a linear order
$\left(R,Y\right)$ that is $\Psi$-consistent such that for all $r\in\Delta\left(X\right)$
(resp. $r\in X\times T$ and $r\in[w,+\infty)\times[w,+\infty)$),
$c$ satisfies WARP and Independence (resp. Stationarity, Quasi-linearity)
over $\mathbb{A}_{R^{\downarrow}\left(r\right)}^{r}$.

\subsubsection*{Step 2: Fixed reference, standard representation}

Next is to show that for each alternative $r\in Y$, the subcorrespondence
$\left(c,\mathbb{A}_{R^{\downarrow}\left(r\right)}^{r}\right)$ admits
a standard representation of its respective domain (i.e., expected
utility, exponential discounting, quasi-linear utility). This is not
obvious; for example in the risk domain, $c$ satisfies WARP and Independence
(and Continuity) over $\mathbb{A}_{R^{\downarrow}\left(r\right)}^{r}$,
which is a strict subset of all choice problems, so standard postulates
could be insufficient.\footnote{For example, if $c\left(\left\{ p,q\right\} \right)=\left\{ p\right\} $
and $c\left(\left\{ p',q'\right\} \right)=\left\{ q'\right\} $ where
$p=\frac{1}{2}x_{1}\oplus\frac{1}{2}x_{2}$, $q=\frac{3}{4}x_{1}\oplus\frac{1}{4}x_{3}$,
$p'=\frac{1}{2}x_{2}\oplus\frac{1}{2}x_{3}$ and $q'=\frac{1}{4}x_{1}\oplus\frac{3}{4}x_{3}$,
then $c$ satisfies WARP and Independence over $\left\{ \left\{ p,q\right\} ,\left\{ p',q'\right\} \right\} $
but does not admit an expected utility representation (because, even
though the lines $pq$ and $p'q'$ are parallel, $p,q$ are not related
to $p',q'$ by a common mixture).} This issue is resolved by exploiting the structure provided by a
$\Psi$-consistent linear order. In each domain, it guarantees that
(for each $r\in Y$) the strict prediction set $\mathbb{P}_{+}^{r}:=\left\{ p\in R^{\downarrow}\left(r\right):c\left(\left\{ p,r\right\} \right)=\left\{ p\right\} \right\} $
is rich in a sense that behavior inconsistent with the structural
properties can always be substantiated with observations from within
$\left(c,\mathbb{A}_{R^{\downarrow}\left(r\right)}^{r}\right)$. For
example in the risk domain, we first show that an expected utility
representation, with $u_{r}$, can be obtained for subcorrespondence
$\left(c,\mathbb{A}_{\mathbb{P}}^{r}\right)$ where $\mathbb{P}$
is a subset of $\mathbb{P}_{+}^{r}$ and is a linear transformation
of a $|X|-1$ dimensional simplex set; the existence of $\mathbb{P}$
is given by the $\Psi$-consistent linear order, which determines
which alternatives are in $R^{\downarrow}\left(r\right)$ and in turn
determines which choice sets are in $\mathbb{A}_{R^{\downarrow}\left(r\right)}^{r}$.
Then, for $p,q$ in $R^{\downarrow}\left(r\right)$ but possibly outside
$\mathbb{P}$, if $\arg\max_{z\in\left\{ p,q\right\} }\mathbb{E}_{z}u_{r}\left(x\right)=\left\{ p\right\} $,
it can be shown that there exist common mixtures $p'=p^{\alpha}s$,
$q'=q^{\alpha}s$ in $\mathbb{P}$ such that $c\left(\left\{ r,p',q'\right\} \right)=\left\{ p'\right\} $,
and Independence requires $c\left(\left\{ r,p,q\right\} \right)=\left\{ p\right\} $
(assuming $c\left(\left\{ r,p,q\right\} \right)\ne\left\{ r\right\} $).
Analogous methods, all derived using features of $\Psi$-consistent
linear orders, guarantee the sufficiency of standard postulates in
the time and social preference domains.

\subsubsection*{Step 3: Reference-dependent preferences}

\axmref{risk_monotoneRA}, \axmref{patience1}, and \axmref{social_increasingaltruism}
each provides a ``direction'' for preference change, along the reference
order, that must has been satisfied in the constructed representations.
\axmref{risk_monotoneRA}, \axmref{patience1}, and \axmref{social_increasingaltruism}
impose restrictions on behavior when a choice set expands, which necessarily
imply that a reference point, if it changes, ranks higher in $R$.
If the constructed representations violate the direction of preference
change from reference $r$ to $r'$ where $rRr'$, then it can be
shown that there exist choice problems $A\in\mathbb{A}_{R^{\downarrow}\left(r\right)}^{r}$
and $B\in\mathbb{A}_{R^{\downarrow}\left(r'\right)}^{r'}$ such that
$B\subset A$ where \axmref{risk_monotoneRA} / \axmref{patience1}
/ \axmref{social_increasingaltruism} is violated when we compare
$c\left(A\right)$ and $c\left(B\right)$. Like in Step 2, the existence
of axiom-violating choice behavior in the underlying subcorrespondences
$\left(c,\mathbb{A}_{R^{\downarrow}\left(r\right)}^{r}\right)$ and
$\left(c,\mathbb{A}_{R^{\downarrow}\left(r'\right)}^{r'}\right)$
is guaranteed by $\Psi$-consistent linear orders. For the time domain,
\axmref{patience2} additionally guarantees a persistent consumption
utility, so that reference effect can be summarized by changes in
discount factors (in general, both discount factor and consumption
utility can change). 

\subsection{Proof of Lemma \ref{lem:timeistime}}

\textbf{Only if}: Fix $A\in\mathcal{\mathcal{A}}$ and $\left(x,t\right)\in\Psi\left(A\right)$.
Consider $\mathcal{S}=\left\{ B\in\mathcal{\mathcal{A}}:\left(x,t\right)\in B\subseteq A\right\} $.
For any $B_{1},B_{2}\in\mathcal{S}$, since $\left(x,t\right)\in\Psi\left(B_{1}\right)\cap\Psi\left(B_{2}\right)$,
$c$ satisfies WARP and Stationarity over $\left\{ B_{1},B_{2}\right\} $.
Therefore, $c$ satisfies WARP and Stationarity over $\mathcal{S}$
(because WARP and Stationarity are restrictions on pairs of choices).
\textbf{If}: Take any $B_{1},B_{2}$ such that $\Psi\left(B_{1}\right)\cap\Psi\left(B_{2}\right)\ne\emptyset$.
Take $\left(x,t\right)\in\Psi\left(B_{1}\right)\cap\Psi\left(B_{2}\right)$.
Consider $A=B_{1}\cup B_{2}$. Since $B_{1}$ and $B_{2}$ are both
finite, $A$ is finite, and therefore $A\in\mathcal{A}$. Since $\left(x,t\right)\in\Psi\left(B_{1}\right)\cap\Psi\left(B_{2}\right)$,
$\left(x,t\right)\in\Psi\left(A\right)$, and so $c$ satisfies WARP
and Stationarity over $\left\{ B\subseteq A:\left(x,t\right)\in B\right\} $,
which contains $B_{1}$ and $B_{2}$ by construction and we are done.

\newpage{}

~{\huge{}\vskip 10em}{\huge\par}
\begin{center}
{\huge{}}%
\noindent\begin{minipage}[c]{1\columnwidth}%
\begin{center}
\textbf{\Huge{}Supplemental Materials}{\Huge\par}
\par\end{center}
\begin{center}
\textbf{\Huge{}(Online)}{\Huge\par}
\par\end{center}%
\end{minipage}{\huge\par}
\par\end{center}

\newpage{}

\section{\label{oappx:tau-and-proofs}Online Appendix: Omitted Proofs and
Results}

Let $Y$ be an arbitrary set of alternatives and let $\mathcal{A}$
be the set of all finite and nonempty subsets of $Y$, called choice
problems or choice sets. Let $\mathcal{C}$ be the set of all \emph{general
choice correspondences} $c:\mathcal{B}\rightarrow\mathcal{A}$ such
that $\mathcal{B}\subseteq\mathcal{A}$ and $c\left(B\right)\subseteq B$
for all $B\in\mathcal{B}$. For a general choice correspondence with
domain $\mathcal{A}$, we simply call it a\emph{ choice correspondence}.
Call $\hat{c}:\mathcal{S}\rightarrow\mathcal{A}$ a \emph{subcorrespondence}
of $c:\mathcal{B}\rightarrow\mathcal{A}$ if $\mathcal{S}\subseteq\mathcal{B}$
and $\hat{c}\left(B\right)=c\left(B\right)$ whenever defined. If,
furthermore, $\mathcal{S}$ is finite, then call $\hat{c}$ a \emph{finite
subcorrespondence} of $c$.

A behavioral postulate imposed on general choice correspondences can
be captured using a subset $\mathcal{T}$ of $\mathcal{C}$, where
some general choice correspondences are admitted and others excluded.
In line with how behavioral postulates are typically introduced, I
focus on postulates that are easier to satisfy when fewer observations
are considered, and call them \emph{theories}.
\begin{defn}
\label{def:property}~
\begin{enumerate}
\item $\mathcal{T}\subseteq\mathcal{C}$ is a \emph{theory} if for all $c\in\mathcal{T}$,
every subcorrespondence of $c$ is in $\mathcal{T}$.
\item $\mathcal{T}\subseteq\mathcal{C}$ is a \emph{finite theory} if it
is a theory and for all $c\in\mathcal{\mathcal{C}\backslash\mathcal{T}}$,
there exists a finite subcorrespondence of $c$ that is not in $\mathcal{T}$.
\end{enumerate}
\end{defn}
Postulates that place restrictions on finitely many choice sets at
a time are finite theories, such as the common definitions of WARP,
monotonicity, transitivity, convexity, betweenness, separability,
independence, stationarity, and many others. These are the cases where
non-compliance can always be concluded using finitely many observations.
An empirically falsifiable property need not be a finite theory, but
a finite theory is empirically falsifiable unless it is trivial (i.e.,
$\mathcal{T}=\mathcal{C}$).\footnote{It is commonly understood that an empirically falsifiable property
is one that can be falsified with finitely many observations (i.e.,
there exists $c\in\mathcal{C}\backslash\mathcal{T}$ such that $|\text{dom}\left(c\right)|<\infty$).
Consider the combination of WARP and some version of continuity, it
is a theory, and it is empirically falsifiable since WARP needs just
two observations to falsify. Yet in the absence of WARP violations,
a choice correspondence can still violate continuity, which is a non-compliance
that cannot be substantiated with finitely many observations.} Non-examples include various versions of continuity and infinite
acyclicity since they require an infinite number of observations to
substantiate a violation. When $Y$ is finite, every theory is trivially
a finite theory.

Imposing multiple postulates, $\mathcal{T}_{1}$ and $\mathcal{T}_{2}$,
is equivalent to taking the intersection $\mathcal{T}_{1}\cap\mathcal{T}_{2}\subseteq\mathcal{C}$.
Because taking intersection of theories (resp. finite theories) yields
a theory (resp. finite theory), this characterization can simultaneously
account for multiple postulates (or, a model).\footnote{\textbf{Theory}: Consider any $c\in\mathcal{T}_{1}\cap\mathcal{T}_{2}$.
For any $\hat{c}\in\mathcal{C}$ where $\hat{c}\subset c$, since
$\mathcal{T}_{1},\mathcal{T}_{2}$ are theories, we have $\hat{c}\in\mathcal{T}_{1},\mathcal{T}_{2}$,
and hence $\hat{c}\in\mathcal{T}_{1}\cap\mathcal{T}_{2}$, so $\mathcal{T}_{1}\cap\mathcal{T}_{2}$
is a theory. \textbf{Finite theory}: Suppose $\mathcal{\mathcal{T}}_{1}$
and $\mathcal{\mathcal{T}}_{2}$ are finite theories, which are theories,
and so $\mathcal{T}_{1}\cap\mathcal{T}_{2}$ is a theory. Consider
any $c\in\mathcal{C}\backslash\left(\mathcal{T}_{1}\cap\mathcal{T}_{2}\right)$.
Without loss of generality say $c\notin\mathcal{T}_{1}$, so by definition
of finite theory we can find a finite subcorrespondence $\hat{c}$
of $c$ where $\hat{c}\notin\mathcal{T}_{1}$, which means $\hat{c}\notin\mathcal{T}_{1}\cap\mathcal{T}_{2}$.}

\subsection{Reference-dependent $\mathcal{T}$}

In general, it is possible that $c:\mathcal{B}\rightarrow\mathcal{A}$
is not in $\mathcal{T}$ but its subcorrespondence $\hat{c}:\mathcal{S}\rightarrow\mathcal{A}$
is in $\mathcal{T}$, for which I say ``$c$ satisfies $\mathcal{T}$
over $\mathcal{S}$''. \lemref{ordC} provides the foundation for
all four models in this paper. It introduces a reference-dependent
generalization of a generic behavioral postulate, $\mathcal{T}$,
and shows that it is equivalent to a representation in which observations
are partitioned using a reference order $R$ such that $\mathcal{T}$
holds within each part.\footnote{\lemref{ordC} falls short of delivering the target utility representation
(of $\mathcal{T})$ due to the well-known limitation of an incomplete
dataset\textemdash when only a subset of choices are observed, canonical
postulates may not be sufficient for canonical utility representation.}

When $\Psi$ is the identity function, the first condition in \lemref{ordC}
is satisfied when, for each choice problem $A$, some alternative
$x\in A$ serves as an anchor that guarantees compliance with finite
theory $\mathcal{T}$ in subsets of $A$. In anticipation, this anchor
is a potential reference alternative for $A$, so the condition can
be understood as ``there is a reference in every $A$''. When $\Psi$
is not the identity function, we further demand that a potential reference
alternative can be found in a predetermined subset of the choice problem,
$\Psi\left(A\right)\subseteq A$, making reference formation less
subjective. The case of fully objective reference is captured when
$\Psi\left(A\right)$ is a singleton for all $A$, since it fully
pins down the reference.

Note that since every choice set $A\in\mathcal{B}$ is finite, \emph{Reference
Dependence} is both falsifiable (whenever $\mathcal{T}$ is) and can
be written without an explicit existential quantifier. However, the
current formulation may be most suitable for describing a universal
template of \emph{reference-dependent} \emph{generalization}. Applications
of this formulation without an existential quantifier are considered
in \secref{time} (time preference) and \secref{social} (social preference).

\subsection{Proof of Lemma \ref{lem:ordC}}

The proof for (2) implies (1) is straightforward: For every (finite)
set $A\in\mathcal{A}$, the maximizer of the linear order $R$ is
an ``$x$'' in (1). We focus on the proof for (1) implies (2). The
proof for (2) implies (1) begins with an observation using Zermelo\textquoteright s
well-ordering theorem and transfinite recursion, and then uses it
build a reference order given an arbitrary finite theory $\mathcal{T}$
(\defref{property}).
\begin{lem}
\label{lem:transfiniterecursion}Let $Z$ be a set and let $\mathbb{Z}$
be the set of all finite and nonempty subsets of $Z$. Let $\mathcal{R}$
be a self-map on $\mathbb{Z}$ such that $\mathcal{R}\left(S\right)\subseteq S$.
Suppose for all $T,S\in\mathbb{Z}$ and $x\in Z$ such that $x\in T\subseteq S$,
if $x\in\mathcal{R}\left(S\right)$, then $x\in\mathcal{R}\left(T\right)$
(property $\alpha$). Then, there exists a self-map $\mathcal{R}^{*}$
on $\mathbb{Z}$ such that
\begin{enumerate}
\item[(i)] For all $S\in\mathbb{Z}$, $\mathcal{R}^{*}\left(S\right)\subseteq\mathcal{R}\left(S\right)$.
\item[(ii)] For all $T,S\in\mathbb{Z}$ and $x\in Z$ such that $x\in T\subseteq S$,
if $x\in\mathcal{R}^{*}\left(S\right)$, then $x\in\mathcal{R}^{*}\left(T\right)$
(property $\alpha$), and
\item[(iii)] For all $S\in\mathbb{Z}$, $|\mathcal{R}^{*}\left(S\right)|=1$
\end{enumerate}
\end{lem}
\begin{proof}
We prove this by construction. Assume and invoke Zermelo's theorem
(also known as the well-ordering theorem) to well-order the set of
all doubletons in the domain of $\mathcal{R}$. Now we start the transfinite
recursion using this order.

\textbf{In the zero case}, we have $\mathcal{R}_{0}=\mathcal{R}$.
This correspondence satisfies $\alpha$ and is nonempty-valued ($\mathcal{R}_{0}\left(S\right)\ne\emptyset$
for all $S\in\mathbb{Z}$). 

\textbf{For the successor ordinal} $\sigma+1$, having supposed $\mathcal{R}_{\sigma}$
satisfies $\alpha$ and is nonempty-valued, we take the corresponding
doubleton $B_{\sigma+1}$ and take $x\in B_{\sigma+1}$ such that
$\forall S\supset B_{\sigma+1}$, $\mathcal{R}\left(S\right)\backslash\left\{ x\right\} \ne\emptyset$.
Suppose such an $x$ does not exist, then for both $x,y\in B_{\sigma+1}$,
there are $S_{x}\supset B_{\sigma+1}$ and $S_{y}\supset B_{\sigma+1}$
such that $\mathcal{R_{\sigma}}\left(S_{x}\right)=\left\{ x\right\} $
and $\mathcal{R_{\sigma}}\left(S_{y}\right)=\left\{ y\right\} $ since
$\mathcal{R}_{\sigma}$ is nonempty-valued. Consider $S_{x}\cup S_{y}\in\mathbb{Z}$.
Since $\mathcal{R}_{\sigma}$ is nonempty-valued, $\mathcal{R}_{\sigma}\left(S_{x}\cup S_{y}\right)\ne\emptyset$.
But since $\mathcal{R}_{\sigma}$ satisfies $\alpha$, it must be
that $\mathcal{R}_{\sigma}\left(S_{x}\cup S_{y}\right)\subseteq\mathcal{R}_{\sigma}\left(S_{x}\right)\cup\mathcal{R}_{\sigma}\left(S_{y}\right)$,
hence $\mathcal{R}_{\sigma}\left(S_{x}\cup S_{y}\right)\subseteq\left\{ x,y\right\} $.
Suppose without loss $x\in\mathcal{R}_{\sigma}\left(S_{x}\cup S_{y}\right)$,
then due to $\alpha$ again and that $x\in B_{\sigma+1}\subset S_{y}$,
it must be that $x\in\mathcal{R}_{\sigma}\left(S_{y}\right)$, which
contradicts $\mathcal{R_{\sigma}}\left(S_{y}\right)=\left\{ y\right\} $.
(That is, we showed that with nonempty-valuedness and $\alpha$, no
two elements can each have a unique appearance in the $\mathcal{R}_{\left(\cdot\right)}$-image
of a set containing those two elements.) Hence, $\exists x\in B_{\sigma+1}$
such that $\forall S\supset B_{\sigma+1}$, $\mathcal{R}\left(S\right)\backslash\left\{ x\right\} \ne\emptyset$.
Define $\mathcal{R}_{\sigma+1}$ from $\mathcal{R}_{\sigma}$ in the
following way: $\forall S\supset B_{\sigma+1}$, $\mathcal{R}_{\sigma+1}\left(S\right):=\mathcal{R}_{\sigma}\left(S\right)\backslash\left\{ x\right\} $.
Note: (i) Since $x$ is deleted from $\mathcal{R}_{\sigma}\left(T\right)$
only if it is also deleted (if it is in it at all) from $\mathcal{R}_{\sigma}\left(S\right)$
$\forall S\supset T$, we are preserving $\alpha$, and (ii) since
$x$ is never the unique element in $\mathcal{R}_{\sigma}\left(S\right)$
$\forall S\supset B_{\sigma+1}$, we preserve nonempty-valuedness.

\textbf{For a limit ordinal} $\lambda$, define $\mathcal{R}_{\lambda}=\cap_{\sigma<\lambda}\mathcal{R}_{\sigma}$.
Note that since $\mathcal{R}_{\sigma'}\subset\mathcal{R}_{\sigma''}$
$\forall\sigma'>\sigma''$, $\cap_{\sigma\leq\bar{\sigma}}=\mathcal{R}_{\bar{\sigma}}$.
Furthermore, for any $\sigma<\lambda$, $\mathcal{R}_{\sigma}$ is
constructed such that $\alpha$ and nonempty-valuedness are preserved.
Hence $\mathcal{R}_{\lambda}$ satisfies $\alpha$ and is nonempty-valued.

Note that this process terminates when all the doubletons have been
visited, for we would otherwise have constructed an injection from
the class of all ordinals to the set of all doubletons in $\mathbb{Z}$,
which is impossible.

Finally, we check that $|\mathcal{R}_{\lambda}\left(S\right)|=1$
for all $S\in\mathbb{Z}$. Suppose not, hence $\exists S\in\mathbb{Z}$
such that $\left\{ x,y\right\} \subseteq\mathcal{R}_{\lambda}\left(S\right)$.
Then by $\alpha$ we have $\left\{ x,y\right\} =\mathcal{R}_{\lambda}\left(\left\{ x,y\right\} \right)$,
which is not possible as the recursion process has visited $\left\{ x,y\right\} $
and deleted something from $\mathcal{R}\left(\left\{ x,y\right\} \right)$.
Now set $\mathcal{R}_{\lambda}=\mathcal{R}^{*}$ and we are done.
\end{proof}
For notational convenience, subcorrespondence $\hat{c}:\mathcal{S}\rightarrow\mathcal{A}$
of $c:\mathcal{B}\rightarrow\mathcal{A}$ is referred to as $\left(c,\mathcal{S}\right)$,
as in ``$c$ restricted to $\mathcal{S}$''. Given $\mathcal{B}\subseteq\mathcal{A}$,
for any $S\subseteq Y$ and $x\in S$, define 
\[
\mathbb{A}_{S}^{x}:=\left\{ A\in\mathcal{B}:x\in A\subseteq S\right\} .
\]
Given $\mathcal{T}\subseteq\mathcal{C}$ and a general choice correspondence
$c:\mathcal{B}\rightarrow\mathcal{A}$, let $\Gamma\left(S\right):=\left\{ x\in S:\left(c,\mathbb{A}_{S}^{x}\right)\in\mathcal{T}\right\} $
denote the set of \emph{reference alternatives} \emph{of $S$} (note
that $S$ need not be in $\mathcal{B}$). The following observations
are obtained when $\mathcal{T}$ is a finite theory.
\begin{lem}
\label{lem:finiteproperty}Let $c:\mathcal{B}\rightarrow\mathcal{A}$
be a general choice correspondence and $\mathcal{T}$ a finite theory.
Consider $A,B,D\subseteq Y$.
\begin{enumerate}
\item If $x\in\Gamma\left(A\right)$ and $B\subset A$, then $x\in\Gamma\left(B\right)$.
\item If $x\in\Gamma\left(A\right)$ for all finite $A\subseteq D$, then
$x\in\Gamma\left(D\right)$.
\end{enumerate}
\end{lem}
\begin{proof}
Since $B\subseteq A$ implies $\mathbb{A}_{B}^{x}\subseteq\mathbb{A}_{A}^{x}$
and since $\left(c,\mathbb{A}_{A}^{x}\right)\in\mathcal{\mathcal{T}}$,
$\left(c,\mathbb{A}_{B}^{x}\right)\in\mathcal{\mathcal{T}}$ is a
direct consequence of the definition of a theory. For (2), suppose
for contradiction $x\notin\Gamma\left(D\right)$. Because $\mathcal{T}$
is a finite theory, we can find a finite set of choice problems $\mathcal{S}=\left\{ A_{1},...,A_{n}\right\} \subseteq\mathbb{A}_{D}^{x}$
such that $\left(c,\mathcal{S}\right)\notin\mathcal{T}$. Since the
set $A:=\cup_{i=1}^{n}A_{i}\subseteq D$ is finite, $x\in\Gamma\left(A\right)$.
Note that $\mathcal{S}\subseteq\mathbb{A}_{A}^{x}$, so the definition
of a theory gives $\left(c,\mathcal{S}\right)\in\mathcal{T}$, a contradiction.
\end{proof}
Now I prove (1) implies (2) in \lemref{ordC}. Let $\mathcal{R}':\mathcal{A}\rightarrow\mathcal{A}\cup\left\{ \emptyset\right\} $
be a set-valued function that picks out reference alternatives, formally
$\mathcal{R}'\left(A\right):=\left\{ x\in A:\left(c,\mathbb{A}_{A}^{x}\right)\in\mathcal{T}\right\} $.
Since $\mathcal{T}$ is a finite theory, by point 1 of \lemref{finiteproperty},
$\mathcal{R}'$ satisfies property $\alpha$ (defined in \lemref{transfiniterecursion}).
Furthermore, (1) in \lemref{ordC} guarantees that $\mathcal{R}'\left(A\right)\cap\Psi\left(A\right)$
is nonempty for all $A\in\mathcal{A}$. Finally, define $\mathcal{R}:\mathcal{A}\rightarrow\mathcal{A}$
by $\mathcal{R}\left(A\right):=\mathcal{R}'\left(A\right)\cap\Psi\left(A\right)$.
Since both $\mathcal{R}'\left(A\right)$ and $\Psi\left(A\right)$
satisfy property $\alpha$, $\mathcal{R}\left(A\right)$ satisfies
property $\alpha$.

Putting the $\mathcal{R}$ we just built through \lemref{transfiniterecursion},
we get a function $\mathcal{R}^{*}$ that picks one thing from every
set and satisfies property $\alpha$. With this, we build the order
$\left(R,Y\right)$ by setting $xRy$ if $\left\{ x\right\} =\mathcal{R}^{*}\left(\left\{ x,y\right\} \right)$
and $xRx$ for all $x\in Y$. It is well-known that this results in
a linear order $\left(R,Y\right)$ such that $\mathcal{R}^{*}\left(A\right)=\left\{ x\in A:xRy\,\forall y\in A\right\} $
for all $A\in\mathcal{A}$. Since $\mathcal{R}^{*}\left(A\right)\subseteq\mathcal{R}\left(A\right)\subseteq\Psi\left(A\right)$
for all $A\in\mathcal{A}$, this means $\left(R,Y\right)$ is also
$\Psi$-consistent.

Finally, consider the set of alternatives that are ``reference dominated''
by $x$ according to $R$ (including $x$ itself), denoted by
\[
R^{\downarrow}\left(x\right):=\left\{ y\in Y:xRy\right\} .
\]
For any finite subset $A\subseteq R^{\downarrow}\left(x\right)$ such
that $x\in A$, we have $x$$\in\mathcal{R}^{*}\left(A\right)\subseteq\mathcal{R}\left(A\right)\subseteq\mathcal{R}'\left(A\right)$,
which by definition implies $x$ is a reference alternative of $A$.
Using point 2 of \lemref{finiteproperty}, we conclude that $x$ is
reference alternative for $R^{\downarrow}\left(x\right)$, which need
not be finite.

To summarize, we have created a partition of $\mathcal{A}$ where
the parts are characterized by $\left\{ \mathbb{A}_{R^{\downarrow}\left(x\right)}^{x}\right\} _{x\in Y}$.
To see this, take any $A\in\mathcal{A}$, since $R$ is a linear order,
there is a unique $z\in A$ such that $zRy$ for all $y\in A$, and
so $A\in\mathbb{A}_{R^{\downarrow}\left(z\right)}^{z}$ and $A\notin\mathbb{A}_{R^{\downarrow}\left(y\right)}^{y}$
for any $y\ne z$. Furthermore for each part $\mathbb{A}_{R^{\downarrow}\left(x\right)}^{x}$,
$\left(c,\mathbb{A}_{R^{\downarrow}\left(x\right)}^{x}\right)$ is
in $\mathcal{T}$. Since $\left\{ B\in\mathcal{A}:r\left(B\right)=z\right\} $
is simply $\mathbb{A}_{R^{\downarrow}\left(z\right)}^{z}$, the proof
is complete.

\subsection{Proof of Theorem \ref{thm:AREU}}

``\textbf{If}'' is straightforward. I prove ``\textbf{only if}''.
We interpret $\Delta\left(X\right)$ as a $|X|-1$ dimensional simplex,
and full-dimensional means $|X|-1$ dimensional. Also, where $\text{conv}\left(\left\{ \delta_{b},\delta_{w}\right\} \right)$
denotes the set of lotteries that only put non-zero probabilities
on prizes $b$ and $w$, we partition $\Delta\left(X\right)$ into
three parts: $I=\Delta\left(X\right)\backslash\text{conv}\left(\left\{ \delta_{b},\delta_{w}\right\} \right)$,
$E_{1}=\left\{ r\in\text{conv}\left(\left\{ \delta_{b},\delta_{w}\right\} \right):c\left(\left\{ r,p\right\} \right)=\left\{ p\right\} \text{ for some }p\in R^{\downarrow}\left(r\right)\cap I\right\} $,
and $E_{2}=\text{conv}\left(\left\{ \delta_{b},\delta_{w}\right\} \right)\backslash E_{1}$.\textbf{
Stage 1} builds the reference order $R$.\textbf{ Stage 2} provides
basic results about the prediction set of each reference lottery.
\textbf{Stage 3} builds a (Bernoulli) utility function for each $r\in I\cup E_{1}$
and \textbf{Stage 4} shows that they are related by concave transformations.
\textbf{Stage 5} deals with $r\in E_{2}$.

For any $r\in\Delta\left(X\right)$, let $\mathbb{P}_{+}^{r}:=\left\{ p\in R^{\downarrow}\left(r\right)\backslash\left\{ r\right\} :c\left(\left\{ p,r\right\} \right)=\left\{ p\right\} \right\} $.
For any $r,p\in\Delta\left(X\right)$, let $\mathbb{P}_{+p}^{r}:=\left\{ q\in R^{\downarrow}\left(r\right)\backslash\left\{ r,p\right\} :c\left(\left\{ r,p,q\right\} \right)=\left\{ q\right\} \right\} $.
We call these \emph{prediction sets}. Note that if $rRp$, then the
fact that $c$ satisfies WARP over $\mathbb{A}_{R^{\downarrow}\left(r\right)}^{r}$
implies $\mathbb{P}_{+p}^{r}\subseteq\mathbb{P}_{+}^{r}$. For any
$\mathbb{P}\subseteq\Delta\left(X\right)$ and lotteries $p,q\in\Delta\left(X\right)$,
we call $\left(p',q'\right)$ a\emph{ $\mathbb{P}$-common mixture}
of $\left(p,q\right)$ if for some $s\in\Delta\left(X\right)$ and
$\alpha\in\left[0,1\right]$, we have $p'=p^{\alpha}s$, $q'=q^{\alpha}s$,
and $p',q'\in\mathbb{P}$.

\subsubsection*{Stage 1: Reference order $R$}

A binary relation $R$ is said to be \emph{risk-consistent} if $qRp$
whenever $p\text{MPS}q$ or $p\text{ES}q$. Note that $\Psi$ is an
$\alpha-$correspondence. By \lemref{ordC}, \axmref{risk_axiom1}
gives a linear order $\left(R,\Delta\left(X\right)\right)$ where
$c$ satisfies WARP and Independence over $\mathbb{A}_{R^{\downarrow}\left(r\right)}^{r}$
for any $r\in\Delta\left(X\right)$. Since $R$ is $\Psi$-consistent
(i.e., $\max\left(A,R\right)\in\Psi\left(A\right)$) and $\Psi\left(\left\{ p,q\right\} \right)=\left\{ q\right\} $
if $p\text{MPS}q$ or $p\text{ES}q$, so $R$ is risk-consistent.

\subsubsection*{Stage 2: Technical Preparations}

The next results guarantee that the revealed preference relation constructed
using subcorrespondence $\left(c,\mathbb{A}_{R^{\downarrow}\left(r\right)}^{r}\right)$,
where $r$ is given, is complete and transitive on a full-dimensional
convex subset of $\Delta\left(X\right)$. This is due in large part
to $R$ being risk-consistent, and because of it, choices that further
satisfy Independence will have an expected utility representation.
\begin{lem}
\label{lem:FDC1}For any $r\in I$ and any open ball $B_{r}$ that
contains $r$, $B_{r}\cap R^{\downarrow}\left(r\right)$ contains
a full-dimensional convex subset of $\Delta\left(X\right)$.
\end{lem}
\begin{proof}
Take any $r\in I$. By definition, $r\left(x\right)\ne0$ for some
$x\in X\backslash\left\{ b,w\right\} $. Consider the set $\mathbb{C}\left(r\right):=\left\{ r\right\} \cup ES\left(\left\{ r\right\} \right)\cup MPS\left(\left\{ r\right\} \cup ES\left(\left\{ r\right\} \right)\right)$.
It consists of $r$, all extreme spreads of $r$, and all of their
mean-preserving spreads.

To see $\mathbb{C}\left(r\right)$ is convex: First note that since
$ES\left(\left\{ r\right\} \right)$ is a convex set and $r$ is on
the boundary of $ES\left(\left\{ r\right\} \right)$, so $\left\{ r\right\} \cup ES\left(\left\{ r\right\} \right)$
is convex. Take any two lotteries $p_{1},p_{2}\in\mathbb{C}\left(r\right)$
and consider their convex combination $\left(p_{1}\right)^{\alpha}\left(p_{2}\right)$
for some $\alpha\in\left(0,1\right)$. Since $p_{1},p_{2}\in\mathbb{C}\left(r\right)$,
there exist $e_{1},e_{2}\in\left\{ r\right\} \cup ES\left(\left\{ r\right\} \right)$
such that either $p_{1}=e_{1}$ or $p_{1}\text{MPS}e_{1}$ and either
$p_{2}=e_{2}$ or $p_{2}\text{MPS}e_{2}$. If $p_{i}=e_{i}$ for both
$i=1,2$, then $\left(p_{1}\right)^{\alpha}\left(p_{2}\right)=\left(e_{1}\right)^{\alpha}\left(e_{2}\right)$
and by convexity of $\left\{ r\right\} \cup ES\left(\left\{ r\right\} \right)$
we are done. Suppose $p_{i}\ne e_{i}$ for some $i=1,2$, then since
the mean-preserving spread relation is preserved under convex combinations,
we have $\left(p_{1}\right)^{\alpha}\left(p_{2}\right)\text{MPS}\left(e_{1}\right)^{\alpha}\left(e_{2}\right)$.
Then, since $\left(e_{1}\right)^{\alpha}\left(e_{2}\right)\in\left\{ r\right\} \cup ES\left(\left\{ r\right\} \right)$
by the convexity of $\left\{ r\right\} \cup ES\left(\left\{ r\right\} \right)$,
we have $\left(p_{1}\right)^{\alpha}\left(p_{2}\right)\in MPS\left(\left\{ r\right\} \cup ES\left(\left\{ r\right\} \right)\right)\subseteq\mathbb{C}\left(r\right)$.

To see $\mathbb{C}\left(r\right)$ is full-dimensional: For any $p\in I$,
$MPS\left(\left\{ p\right\} \right)$ is $|X-2|$ dimensional, and
it is a subset of the $|X-2|$ dimensional space defined by lotteries
that have the same mean as $p$. But $ES\left(\left\{ p\right\} \right)$
contains lotteries that do not have the same mean as $p$, and therefore
$ES\left(\left\{ p\right\} \right)\cup MPS\left(\left\{ p\right\} \right)$
is full-dimensional. This means $\mathbb{C}\left(r\right)$ is full
dimensional as well since it contains $ES\left(\left\{ p\right\} \right)\cup MPS\left(\left\{ p\right\} \right)$
for some $p\in I$.

To see $\mathbb{C}\left(r\right)\subseteq R^{\downarrow}\left(r\right)$:
If $p\in ES\left(\left\{ r\right\} \right)$, $rRp$ since $R$ is
risk-consistent. If $q\in MPS\left(\left\{ r\right\} \cup ES\left(\left\{ r\right\} \right)\right)$,
$qRp$ for some $p\in\left\{ r\right\} \cup ES\left(\left\{ r\right\} \right)$
since $R$ is risk-consistent, and by transitivity of $R$ we have
$qRr$. Since $B_{r}$ is also a full-dimensional and convex set,
$B_{r}\cap\mathbb{C}\left(r\right)$ is a full-dimensional convex
subset of $B_{r}\cap R^{\downarrow}\left(r\right)$.
\end{proof}
\begin{lem}
\label{lem:FDC2}For any $r\in I$, $\mathbb{P}_{+}^{r}$ contains
a full-dimensional convex subset of $\Delta\left(X\right)$.
\end{lem}
\begin{proof}
Fix $r\in I$. Note that $\mathbb{P}_{+}^{r}$ contains an extreme
spread $e$ of $r$ (else, there is a sequence of alternatives $e_{k}=\left(\delta_{w}\right)^{\alpha_{k}}\left(\delta_{b}\right)$
such that $\alpha_{k}$ converges from above to $r\left(w\right)$
such that $r\in c\left(\left\{ r,e_{k}\right\} \right)$ for all $k$,
which by Continuity means $r\in c\left(\left\{ r,\left(\delta_{w}\right)^{r\left(w\right)}\left(\delta_{b}\right)\right\} \right)$,
a violation of FOSD (\axmref{FOSD})). Consider $q=r^{0.5}e\in I$.
Since $q\text{ES}r$, $q\in R^{\downarrow}\left(r\right)$. Since
$c$ satisfies Independence over $\mathbb{A}_{R^{\downarrow}\left(r\right)}^{r}$
and $c\left(\left\{ r,e\right\} \right)=\left\{ e\right\} $, we establish
$q\in\mathbb{P}_{+}^{r}\cap I$. By Continuity, there exists an open
ball $B_{q}$ around $q$ such that $c\left(\left\{ r,q'\right\} \right)=\left\{ q'\right\} $
for all $q'\in B_{q}$. By \lemref{FDC1}, $B_{q}\cap R^{\downarrow}\left(q\right)$
contains a full-dimensional convex subset of $\Delta\left(X\right)$.
Moreover, $B_{q}\cap R^{\downarrow}\left(q\right)\subseteq B_{q}\cap R^{\downarrow}\left(r\right)\subseteq\mathbb{P}_{+}^{r}$,
hence $\mathbb{P}_{+}^{r}$ contains a full-dimensional convex subset
of $\Delta\left(X\right)$.
\end{proof}
\begin{lem}
\label{lem:FDC3}Fix any $r\in\Delta\left(X\right)$. If $p\in\mathbb{P}_{+}^{r}\cap I$,
then $\mathbb{P}_{+p}^{r}$ contains a full-dimensional convex subset
of $\Delta\left(X\right)$. If $\mathbb{P}_{+}^{r}\cap I\ne\emptyset$,
then $\mathbb{P}_{+}^{r}$ contains a full-dimensional convex subset
of $\Delta\left(X\right)$.
\end{lem}
\begin{proof}
Fix $r\in\Delta\left(X\right)$ and $p\in\mathbb{P}_{+}^{r}\cap I$.
Since $p\in I$, the set of extreme spreads of $p$, $ES\left(p\right)$,
is nonempty. Also, $ES\left(p\right)\subseteq R^{\downarrow}\left(p\right)\subseteq R^{\downarrow}\left(r\right)$.
Since $c$ satisfies WARP over $\mathbb{A}_{R^{\downarrow}\left(r\right)}^{r}$
and $p\in\mathbb{P}_{+}^{r}$, $r\notin c\left(\left\{ r,p,s\right\} \right)$
for all $s\in R^{\downarrow}\left(r\right)$, so we can use the same
technique in the proof of \lemref{FDC2} to establish that $\mathbb{P}_{+p}^{r}$
contains an extreme spread $e$ of $p$ (else, $p\in c\left(\left\{ r,p,\left(\delta_{w}\right)^{p\left(w\right)}\left(\delta_{b}\right)\right\} \right)$
by Continuity, which violates FOSD (\axmref{FOSD})). Consider $q=p^{0.5}e\in I$.
Because $q\text{ES}p$ and $p\in R^{\downarrow}\left(r\right)$, so
$q\in R^{\downarrow}\left(r\right)$. Since $c$ satisfies Independence
over $\mathbb{A}_{R^{\downarrow}\left(r\right)}^{r}$ and $c\left(\left\{ r,p,e\right\} \right)=\left\{ e\right\} $,
we establish $q\in\mathbb{P}_{+p}^{r}\cap I$. By Continuity, there
exists an open ball $B_{q}$ around $q$ such that $c\left(\left\{ r,p,q'\right\} \right)=\left\{ q'\right\} $
for all $q'\in B_{p}$. By \lemref{FDC1}, $B_{q}\cap R^{\downarrow}\left(q\right)$
contains a full-dimensional convex subset of $\Delta\left(X\right)$.
Moreover, $B_{q}\cap R^{\downarrow}\left(q\right)\subseteq B_{q}\cap R^{\downarrow}\left(r\right)\subseteq$$\mathbb{P}_{+p}^{r}$,
hence $\mathbb{P}_{+p}^{r}$ contains a full-dimensional convex subset
of $\Delta\left(X\right)$. The second statement is given by the first
statement and the observation that $c$ satisfies WARP over $\mathbb{A}_{R^{\downarrow}\left(r\right)}^{r}$
implies $\mathbb{P}_{+p}^{r}\subseteq\mathbb{P}_{+}^{r}$.
\end{proof}

\subsubsection*{Stage 3: Expected utility when $r\in I\cup E_{1}$}

\lemref{FDC2} and \lemref{FDC3} establish that when $r\in I\cup E_{1}$,
$\mathbb{P}_{+}^{r}$ contains a full-dimensional convex subset of
$\Delta\left(X\right)$. The next result shows that for every $r\in I\cup E_{1}$,
the subcorrespondence $\left(c,\mathbb{A}_{R^{\downarrow}\left(r\right)}^{r}\right)$
admits an expected utility representation.
\begin{lem}
\label{lem:EUconstruction}For any $r\in\Delta\left(X\right)$, if
$\mathbb{P}_{+}^{r}$ contains a full-dimensional convex subset of
$\Delta\left(X\right)$, then there exists a strictly increasing utility
function $u_{r}:X\rightarrow\mathbb{R}$, unique up to a positive
affine transformation, such that $c\left(A\right)=\arg\max_{p\in A}\mathbb{E}_{p}u_{r}\left(x\right)$
for all $A\in\mathbb{A}_{R^{\downarrow}\left(r\right)}^{r}$.
\end{lem}
\begin{proof}
Since $\mathbb{P}_{+}^{r}$ contains a full-dimensional convex subset
of $\Delta\left(X\right)$, consider a subset $\mathbb{P}\subseteq\mathbb{P}_{+}^{r}$
that is a linear transformation of a $|X|-1$ dimensional simplex
(hence also full-dimensional and convex). First, notice that for all
$p,q\in\mathbb{P}$, we have $\left\{ r,p,q\right\} \in\mathbb{A}_{R^{\downarrow}\left(r\right)}^{r}$
and $r\notin c\left(\left\{ r,p,q\right\} \right)$. Recall that $c$
satisfies WARP and Independence over $\mathbb{A}_{R^{\downarrow}\left(r\right)}^{r}$.
By letting $p\succsim_{r}q$ if $p\in c\left(\left\{ r,p,q\right\} \right)$,
we obtain a binary relation $\left(\succsim_{r},\mathbb{P}\right)$
that is complete, transitive, continuous, and satisfies the standard
von Neumann-Morgenstern Independence, and it is well-known that there
exists a utility function $u_{r}:X\rightarrow\mathbb{R}$, unique
up to a positive affine transformation, such that $c\left(A\right)=\arg\max_{p\in A}\mathbb{E}_{p}u_{r}\left(x\right)$
for all $A\in\mathbb{A}_{\mathbb{P}}^{r}$. Since $\left(\succsim_{r},\mathbb{P}\right)$
satisfies FOSD (\axmref{FOSD}), $u_{r}$ is strictly increasing.
We normalize this function to $u_{r}:X\rightarrow\left[0,1\right]$
where $u_{r}\left(w\right)=0$ and $u_{b}\left(b\right)=1$.

We now show that this utility function can explain $\left(c,\mathbb{A}_{R^{\downarrow}\left(r\right)}^{r}\right)$.
First, note that for any two lotteries $p,q\in\Delta\left(X\right)$,
there exist two (possibly different) lotteries $p',q'\in\mathbb{P}$
such that $\left(p',q'\right)$ is a $\mathbb{P}$-common mixture
of $\left(p,q\right)$. This can be done by taking an arbitrary $s\in\text{Int }\mathbb{P}$
and $\alpha$ large enough so that both $p'$ and $q'$ enter $\mathbb{P}$
(this is why we need $\mathbb{P}$ to be full-dimensional and convex).
Now consider any $p\in R^{\downarrow}\left(r\right)$ and let $\left(r',p'\right)$
be a $\mathbb{P}$-common mixture of $\left(r,p\right)$. Since $c$
satisfies Independence over $\mathbb{A}_{R^{\downarrow}\left(r\right)}^{r}$,
for $i=r,p$, $i'\in c\left(\left\{ r,r',p'\right\} \right)$ if and
only if $i\in c\left(\left\{ r,p\right\} \right)$. Now take any $p,q\in R^{\downarrow}\left(r\right)$
such that $p\in c\left(\left\{ r,p\right\} \right)$ and $q\in c\left(\left\{ r,q\right\} \right)$,
then again by Independence over $\mathbb{A}_{R^{\downarrow}\left(r\right)}^{r}$,
$p'\in c\left(\left\{ r,p',q'\right\} \right)$ if and only if $p\in c\left(\left\{ r,p,q\right\} \right)$,
where $\left(p',q'\right)$ is a $\mathbb{P}$-common mixture of $\left(p,q\right)$.
We have thus shown that $c\left(\left\{ r,p\right\} \right)=\arg\max_{s\in\left\{ r,p\right\} }\mathbb{E}_{s}u_{r}\left(x\right)$
for all $\left\{ r,p\right\} \in\mathbb{A}_{R^{\downarrow}\left(r\right)}^{r}$
and $c\left(\left\{ r,p,q\right\} \right)=\arg\max_{s\in\left\{ r,p,q\right\} }\mathbb{E}_{s}u_{r}\left(x\right)$
for all $\left\{ r,p,q\right\} \in\mathbb{A}_{R^{\downarrow}\left(r\right)}^{r}$
with $p\in c\left(\left\{ r,p\right\} \right)$ and $q\in c\left(\left\{ r,q\right\} \right)$.
Since $c$ satisfies WARP over $\mathbb{A}_{R^{\downarrow}\left(r\right)}^{r}$,
showing $c\left(A\right)=\arg\max_{p\in A}\mathbb{E}_{p}u_{r}\left(x\right)$
for all $A\in\mathbb{A}_{R^{\downarrow}\left(r\right)}^{r}$ is straightforward
from here.
\end{proof}

\subsubsection*{Stage 4: Concave transformations when $r_{1},r_{2}\in I\cup E_{1}$}
\begin{lem}
\label{lem:Concave}For any $r_{1},r_{2}\in I$, if $r_{1}Rr_{2}$,
then $u_{r_{1}}=f\circ u_{r_{2}}$ for some concave and strictly increasing
function $f:\left[0,1\right]\rightarrow\left[0,1\right]$.
\end{lem}
\begin{proof}
This proof uses \axmref{risk_monotoneRA}. Take any $r_{1},r_{2}\in I$
such that $r_{1}Rr_{2}$. Consider the function $\bar{f}$ whose domain
is the set of numbers $\left\{ u_{r_{2}}\left(x\right):x\in X\right\} $
such that $u_{r_{1}}\left(x\right)=\bar{f}u_{r_{2}}\left(x\right)$.
Since $u_{r_{1}}$ and $u_{r_{2}}$ are strictly increasing, $\bar{f}$
is strictly increasing in its domain.

We show that if $x_{1}<x_{2}<x_{3}$, then $\bar{f}\left(\alpha u_{r_{2}}\left(x_{1}\right)+\left(1-\alpha\right)u_{r_{2}}\left(x_{3}\right)\right)\geq$
$\alpha\bar{f}\left(u_{r_{2}}\left(x_{1}\right)\right)+\left(1-\alpha\right)\bar{f}\left(u_{r_{2}}\left(x_{3}\right)\right)$
where $\alpha$ solves $\alpha u_{r_{2}}\left(x_{1}\right)+\left(1-\alpha\right)u_{r_{2}}\left(x_{3}\right)=u_{r_{2}}\left(x_{2}\right)$.
Suppose not, then there exists $\beta$, strictly greater than $\alpha$,
such that $\bar{f}\left(\alpha u_{r_{2}}\left(x_{1}\right)+\left(1-\alpha\right)u_{r_{2}}\left(x_{3}\right)\right)<$
$\beta\bar{f}\left(u_{r_{2}}\left(x_{1}\right)\right)+\left(1-\beta\right)\bar{f}\left(u_{r_{2}}\left(x_{3}\right)\right)<$
$\alpha\bar{f}\left(u_{r_{2}}\left(x_{1}\right)\right)+\left(1-\alpha\right)\bar{f}\left(u_{r_{2}}\left(x_{3}\right)\right)$.
Consider the lotteries $\delta=\delta_{x_{2}}$ and $p=\left(\delta_{x_{1}}\right)^{\beta}\left(\delta_{x_{3}}\right)$.
The above equations give $\mathbb{E}_{\delta}u_{r_{1}}\left(x\right)<\mathbb{E}_{p}u_{r_{1}}\left(x\right)$
and $\mathbb{E}_{\delta}u_{r_{2}}\left(x\right)>\mathbb{E}_{p}u_{r_{2}}\left(x\right)$.
Let $\left(\delta_{1},p_{1}\right)$ be a $\mathbb{P}$-common mixture
of $\left(\delta,p\right)$ where $\mathbb{P}$ is a full-dimensional
convex subset of $\mathbb{P}_{+r_{2}}^{r_{1}}$ if $c\left(\left\{ r_{1},r_{2}\right\} \right)=\left\{ r_{2}\right\} $
and of $\mathbb{P}_{+}^{r_{1}}$ otherwise (\lemref{FDC3} guarantees
the existence of $\mathbb{P}$). Let $\left(\delta_{2},p_{2}\right)$
be a $\mathbb{P}$-common mixture of $\left(\delta,p\right)$ where
$\mathbb{P}$ is a full-dimensional convex subset of $\mathbb{P}_{+}^{r_{2}}$.
Since $\mathbb{E}u_{r_{1}}$ explains $\left(c,\mathbb{A}_{R^{\downarrow}\left(r_{1}\right)}^{r_{1}}\right)$
and $\mathbb{E}u_{r_{2}}$ explains $\left(c,\mathbb{A}_{R^{\downarrow}\left(r_{2}\right)}^{r_{2}}\right)$,
we have $c\left(\left\{ r_{1},\delta_{1},p_{1}\right\} \right)=\left\{ p_{1}\right\} $
and $c\left(\left\{ r_{2},\delta_{2},p_{2}\right\} \right)=\left\{ \delta_{2}\right\} $.
Now consider $A=\left\{ r_{1},r_{2},\delta_{1},\delta_{2},p_{1},p_{2}\right\} $,
which is in $\mathbb{A}_{R^{\downarrow}\left(r_{1}\right)}^{r_{1}}$,
and so $c\left(A\right)=\arg\max_{q\in A}\,\mathbb{E}_{q}u_{r_{1}}\left(x\right)$.
Because we have established $\mathbb{E}_{r_{2}}u_{r_{1}}\left(x\right)<\mathbb{E}_{p_{1}}u_{r_{1}}\left(x\right)$,
$\mathbb{E}_{r_{1}}u_{r_{1}}\left(x\right)<\mathbb{E}_{p_{1}}u_{r_{1}}\left(x\right)$,
and $\mathbb{E}_{\delta_{i}}u_{r_{1}}\left(x\right)<\mathbb{E}_{p_{i}}u_{r_{1}}\left(x\right)$
for $i=1,2$ (the first two inequality are due to the way $p_{1}$
was picked), so we know $c\left(A\right)\subseteq\left\{ p_{1},p_{2}\right\} $
. But $c\left(A\right)\subseteq\left\{ p_{1},p_{2}\right\} $ and
$c\left(\left\{ r_{2},\delta_{2},p_{2}\right\} \right)=\left\{ \delta_{2}\right\} $
jointly violate \axmref{risk_monotoneRA}. 

To complete the proof, extend $\bar{f}$ to a concave function $f:\left[0,1\right]\rightarrow\left[0,1\right]$
(for example by connecting points with lines).
\end{proof}
\begin{lem}
\label{lem:rinE}For any $r\in E_{1}\cup E_{2}$ and $p\in R^{\downarrow}\left(r\right)\backslash\left\{ r\right\} $,
either $p$ first-order stochastically dominates $r$ or $r$ first-order
stochastically dominates $p$.
\end{lem}
\begin{proof}
Take $r\in E_{1}\cup E_{2}$ and $p\in R^{\downarrow}\left(r\right)$,
$p\ne r$. Let $\alpha=r\left(b\right)$, then $r\left(w\right)=1-\alpha$.
If $p\left(b\right)<\alpha$ and $p\left(w\right)<\left(1-\alpha\right)$,
then $r$ is an extreme spread of $p$ and $pRr$, so $p\notin R^{\downarrow}\left(r\right)$.
Furthermore, it is not possible that $p\left(b\right)\geq\alpha$
and $p\left(w\right)\geq\left(1-\alpha\right)$ since $p\ne r$. Hence
either $p\left(b\right)\geq\alpha$ and $p\left(w\right)\leq\left(1-\alpha\right)$
with at least one strict inequality, so $p$ first-order stochastically
dominates $r$, or $p\left(b\right)\leq\alpha$ and $p\left(w\right)\geq\left(1-\alpha\right)$
with at least one strict inequality, so $r$ first-order stochastically
dominates $p$.
\end{proof}
\begin{lem}
\label{lem:I=000026E1}For any $r_{1},r_{2}\in I\cup E_{1}$, if $r_{1}Rr_{2}$,
then $u_{r_{1}}=f\circ u_{r_{2}}$ for some concave and increasing
function $f:\left[0,1\right]\rightarrow\left[0,1\right]$.
\end{lem}
\begin{proof}
We use the proof in \lemref{Concave} with the following modifications.
When $r_{2}\in E_{1}$, let $\left(\delta_{1},p_{1}\right)$ be a
$\mathbb{P}$-common mixture of $\left(\delta,p\right)$, where $\mathbb{P}$
is a full-dimensional convex subset of $\mathbb{P}_{+}^{r_{1}}$.
(Before, we let $\mathbb{P}$ be a full-dimensional convex subset
of $\mathbb{P}_{+r_{2}}^{r_{1}}$ when $c\left(\left\{ r_{1},r_{2}\right\} \right)=\left\{ r_{2}\right\} $,
but now such a subset may not exist since $r_{2}\notin I$). Since
$\delta_{2},p_{2}\in\mathbb{P}_{+}^{r_{2}}$ and \lemref{rinE} guarantees
$\delta_{2}$ and $p_{2}$ each first-order stochastically dominates
$r_{2}$, we replace the argument ``$\mathbb{E}_{r_{2}}u_{r_{1}}\left(x\right)<\mathbb{E}_{p_{1}}u_{r_{1}}\left(x\right)$''
with ``$\mathbb{E}_{r_{2}}u_{r_{1}}\left(x\right)<\mathbb{E}_{p_{2}}u_{r_{1}}\left(x\right)$''.
Everything else goes through according to the proof in \lemref{Concave},
giving the desired result.
\end{proof}

\subsubsection*{Stage 5: Expected utility when $r\in E_{2}$ and concave transformations
by construction}

We are left with $r\in E_{2}$, the alternatives in $\text{conv}\left(\left\{ \delta_{b},\delta_{w}\right\} \right)$
that are weakly preferred to everything they reference dominate. The
construction of $u_{r}$ can be partly arbitrary, where the main goal
is to make sure they are related by concave transformations to other
utility functions.

By definition of $E_{2}$, $\mathbb{P}_{+}^{r}\cap I=\emptyset$,
so by \lemref{rinE} and FOSD (\axmref{FOSD}), $r$ first-order stochastically
dominates $p$ for all $p\in R^{\downarrow}\left(r\right)\cap I$.
For any $p\in R^{\downarrow}\left(r\right)\cap\text{conv}\left(\left\{ \delta_{b},\delta_{w}\right\} \right)$,
FOSD requires the choice $c\text{\ensuremath{\left(\left\{  r,p\right\}  \right)}}$
to obey first order stochastic dominance. Together, any strictly increasing
utility function $u_{r}:X\rightarrow\left[0,1\right]$ will accomplish
$c\left(A\right)=\arg\max_{p\in A}\mathbb{E}_{p}u_{r}\left(x\right)$
for all $A\in\mathbb{A}_{R^{\downarrow}\left(r\right)}^{r}$. 

We now construct $u_{r}$ so that it is related to other utility functions
by concave transformations. For any strictly increasing utility function
$u_{p}$, consider the object $\rho^{p}=\left(\rho_{2}^{p},...,\rho_{|X|-1}^{p}\right)\in\left(0,1\right)^{|X|-2}$
such that for all $i\in\left\{ 2,...,|X|-1\right\} $,
\begin{equation}
\rho_{i}^{p}=\frac{u_{p}\left(x_{i}\right)-u_{p}\left(x_{i-1}\right)}{u_{p}\left(x_{i+1}\right)-u_{p}\left(x_{i-1}\right)}\label{eq:rhos}
\end{equation}
(so $\rho_{i}^{p}$ satisfies $u_{p}\left(x_{i}\right)=\rho_{i}^{p}u_{p}\left(x_{i+1}\right)+\left(1-\rho_{i}^{p}\right)u_{p}\left(x_{i-1}\right)$).
There is a one-to-one relationship between $u_{p}$ and $\rho^{p}$.
Also, it is an algebraic exercise to show that $u_{p}=f\circ u_{q}$
for some concave and strictly increasing $f:\left[0,1\right]\rightarrow\left[0,1\right]$
if and only if $\rho_{i}^{p}\geq\rho_{i}^{q}$ for all $i$.

Fix $r\in E_{2}$. Let $\rho^{r}=\left(\inf_{p\in K_{r}}\left(\rho_{2}^{p}\right),...,\inf_{p\in K_{r}}\left(\rho_{|X|-1}^{p}\right)\right)$,
where $K_{r}:=\left(I\cup E_{1}\right)\cap\left\{ p:pRr\right\} $,
and subsequently construct $u_{r}$ using \eqref{rhos}, which is
possible as long as $K_{r}$ is nonempty. Note that when $r\notin\left\{ \delta_{b},\delta_{w}\right\} $,
$r$ must be the mean preserving spread of something in $I$, so $I\cap\left\{ p:pRr\right\} $
is nonempty, and so $K_{r}$ is nonempty. In the exception where $r\in\left\{ \delta_{b},\delta_{w}\right\} $
and $K_{r}$ is empty, this implies $rRp$ for all $p\in\Delta\left(X\right)\backslash\left\{ \delta_{b},\delta_{w}\right\} $.
Then, we let 
\[
\rho_{i}^{r}=\frac{1}{2}\left(1\right)+\frac{1}{2}\sup_{p\in\Delta\left(X\right)\backslash\left\{ \delta_{b},\delta_{w}\right\} }\rho_{i}^{p}
\]
for all $i$ and construct $u_{r}$ using \eqref{rhos}. For any $p\in\Delta\left(X\right)\backslash\left\{ r\right\} $,
this construction results in $\rho_{i}^{r}\geq\rho_{i}^{p}$ for all
$i$, with equality for $p$ that also falls into this exception (there
are at most two of them, $\delta_{b}$ and $\delta_{w}$).

We now show that for any $r_{1},r_{2}\in\Delta\left(X\right)$ where
$r_{1}Rr_{2}$, we have $\rho_{i}^{r_{1}}\geq\rho_{i}^{r_{2}}$ for
all $i$. This is already shown for any $r_{1},r_{2}\in I\cup E_{1}$
by \lemref{I=000026E1}. It is also already shown for the special
cases in the preceding paragraph, by careful construction. Hence,
we restrict attention to the remaining cases. Suppose $\rho_{i}^{r_{1}}<\rho_{i}^{r_{2}}$
for some $i$. Then $\inf_{p\in K_{r_{1}}}\left(\rho_{i}^{p}\right)<\rho_{i}^{r_{2}}$,
so $\rho_{i}^{p}<\rho_{i}^{r_{2}}$ for some $p\in K_{r_{1}}$. However,
since $R$ is transitive, $p\in K_{r_{1}}$ implies $pRr_{2}$; and
since $p\in I\cup E_{1}$, this contradicts \lemref{I=000026E1}.
Say $r_{1}\in I\cup E_{1}$, $r_{2}\in E_{2}$, but $\rho_{i}^{r_{1}}<\rho_{i}^{r_{2}}$
for some $i$. Then $\rho_{i}^{r_{1}}<\inf_{p\in K_{r_{2}}}\left(\rho_{i}^{p}\right)$,
so $\rho_{i}^{r_{1}}<\rho_{i}^{p}$ for all $p\in K_{r_{2}}$. But
$r_{1}\in K_{r_{2}}$, a contradiction. Finally, for $r_{1},r_{2}\in E_{2}$,
either $K_{r_{1}}=K_{r_{2}}$ or $K_{r_{1}}\subsetneq K_{r_{2}}$.
If it is the former, it is immediate that $\rho^{r_{1}}=\rho^{r_{2}}$.
If it is the later, then $\rho_{i}^{r_{1}}=\inf_{p\in K_{r_{1}}}\left(\rho_{i}^{p}\right)\geq\inf_{p\in K_{r_{2}}}\left(\rho_{i}^{p}\right)=\rho_{i}^{r_{2}}$
for all $i$, as desired.

Thus, we have now shown that for any $r_{1},r_{2}\in\Delta\left(X\right)$
such that $r_{1}Rr_{2}$, $\rho_{i}^{r_{1}}\geq\rho_{i}^{r_{2}}$
for all $i$, or equivalently $u_{r_{1}}=f\circ u_{r_{2}}$ for some
concave and strictly increasing $f:\left[0,1\right]\rightarrow\left[0,1\right]$.

\subsection{Proof of Theorem \ref{thm:pbdu}}

``\textbf{If}'' is straightforward, where compliance with \axmref{patience2}
is shown in a footnote. I prove ``\textbf{only if}''. In \textbf{Stage
1}, we show that with \axmref{time_monotonicity=000026impatience}
and \axmref{time_axiom1}, for any time $\tau\in T$, the set of all
choice problems such that the earliest payment arrives at time $\tau$
can be explained by a nonempty set of Discounted Utility specifications,
where a typical element of this set is a utility function and a discount
factor. In \textbf{Stage 2}, we show that at least one (consumption)
utility function $u$ can be supported for all $\tau\in T$, and for
each $\tau\in T$ we set as $\hat{\delta}_{\tau}$ the corresponding
discount factor associated with $u$ for $\tau$; this is the more
involved portion of the proof and it uses \axmref{patience2}. In\textbf{
Stage 3}, with \axmref{patience1}, we show the desired relationship
between $\hat{\delta}_{\tau}$ and $\hat{\delta}_{\tau'}$ for any
two $\tau,\tau'$$.$ Note that the representation constructed has
discount factors indexed by time, not alternatives, so in \textbf{Stage
4 }we convert them back to alternatives.

\subsubsection*{Stage 1: DU representation for each $\tau\in T$}

By \lemref{timeistime} and \lemref{ordC}, for any $x\in X$ and
$\tau\in T$, $c$ satisfies WARP and Stationarity over $S_{\left(x,\tau\right)}:=\left\{ A\in\mathcal{A}:\left(x,\tau\right)\in\Psi\left(A\right)\right\} $
(the collection of choice sets such that the earliest timed payment
is $\left(x,\tau\right)$). In fact, WARP and Stationarity hold even
when we consolidate the collection of choice problems where the earliest
payment arrives at the same time (although the payments themselves
may be different), which we now show. Let $S_{\left(\cdot,\tau\right)}:=\cup_{x\in X}S_{\left(x,\tau\right)}$.
\begin{lem}
\label{lem:DUfixedtime}For any $\tau\in T$, $c$ satisfies WARP
and Stationarity over $S_{\left(\cdot,\tau\right)}$.
\end{lem}
\begin{proof}
Take any two choice sets $A,B\in S_{\left(\cdot,\tau\right)}$. Suppose
it is not true that $c$ satisfies WARP or Stationarity over $\left\{ A,B\right\} $.
Therefore, it must be that $\Psi\left(A\right)\cap\Psi\left(B\right)=\emptyset$.
Now let's take the worse payment at $\tau$ for each set: $\left(x^{*},\tau\right)\in A$
such that $x^{*}\leq x$ for all $\left(x,\tau\right)\in A$ and $\left(y^{*},\tau\right)\in B$
such that $y^{*}\leq y$ for all $\left(y,\tau\right)\in B$. Suppose
without loss of generality $x^{*}<y^{*}$ (due to $\Psi\left(A\right)\cap\Psi\left(B\right)=\emptyset$).
By \axmref{time_monotonicity=000026impatience}, adding $\left(x^{*},\tau\right)$
to $B$ would not alter the choice, i.e., $c\left(B\cup\left\{ \left(x^{*},\tau\right)\right\} \right)=c\left(B\right)$.
Let $B^{*}:=B\cup\left\{ \left(x^{*},\tau\right)\right\} $; note
that $A$ and $B^{*}$ are both in $S_{\left(x^{*},\tau\right)}$,
and therefore $c$ satisfies WARP or Stationarity over $\left\{ A,B^{*}\right\} $.
If it is Stationarity that is violated between $A$ and $B$, then
it is also violated between $A$ and $B^{*}$, a contradiction. If
it is WARP that is violated between $A$ and $B$, it remains to show,
due to $\left(x^{*},\tau\right)\in A$, if $A\subseteq B$ then $A\subseteq B^{*}$
and there is a contradiction, whereas if $A\supseteq B$ then $A\supseteq B^{*}$
and there is a contradiction.
\end{proof}
We just established that $c$ satisfies WARP and Stationarity over
$S_{\left(\cdot,\tau\right)}$. This will give us, from the choices
in $\left(c,S_{\left(\cdot,\tau\right)}\right)$, a revealed preference
relation on $\left\{ \left(x,t\right)\in X\times T:t\geq\tau\right\} $
that is complete, transitive, continuous, and satisfies stationarity,
and then it is well-known (\citet{fishburn1982time}) that along with
\axmref{time_monotonicity=000026impatience} we obtain (many) Discounted
Utility (DU) representations, for instance by translating the time-index
by $-\tau$ so that time $\tau$ is, in that instance, time $0$.

\subsubsection*{Stage 2: $u_{\tau}$ can coincide with $u_{0}$ for each $\tau\in T$}

With existence guaranteed, arbitrarily pick a DU representation with
parameters $\left(\hat{\delta}_{0},u_{0}\right)$ that explains $\left(c,S_{\left(\cdot,0\right)}\right)$.
Define $U_{0}:X\times\mathbb{\mathbb{R}}_{\geq0}\rightarrow\mathbb{R}$
by $U_{0}\left(x,t\right):=\hat{\delta}_{0}^{t}u_{0}\left(x\right)$.
For every $\tau\in\left(0,\bar{t}\right)$, arbitrarily pick a DU
representation $\left(\tilde{\delta}_{\tau},\tilde{u}_{\tau}\right)$
that explains $\left(c,S_{\left(\cdot,\tau\right)}\right)$ and define
$U_{\tau}:X\times\mathbb{\mathbb{R}}_{\geq0}\rightarrow\mathbb{R}$
by $U_{\tau}\left(x,t\right):=\tilde{\delta}_{\tau}^{t}\tilde{u}_{\tau}\left(x\right)$.
We proceed to show that for every $\tau\in\left(0,\bar{t}\right)$,
there exists a DU representation $\left(\hat{\delta}_{\tau},u_{\tau}\right)$
that explains $\left(c,S_{\left(\cdot,\tau\right)}\right)$ where
$u_{\tau}=u_{0}$. Fix a $\tau$. This boils down to identifying a
certain relationship between $U_{0}$ and $U_{\tau}$ due to the fact
that they are DU representations and \axmref{patience2}\textemdash indifferences
are preserved under a common delay multiplier $\lambda$.
\begin{fact}
\label{fact:time-translation}For any $\tau\in[0,\bar{t})$, $t\geq0$,
and $q\geq0$, $U_{\tau}\left(x,0\right)=U_{\tau}\left(y,t\right)$
if and only if $U_{\tau}\left(x,q\right)=U_{\tau}\left(y,q+t\right)$.
\end{fact}
\begin{lem}
\label{lem:proportional}For any $x\in\left(a,b\right)$ (resp. $x=a$
and $x=b$), there exists an open interval $B=\left(x^{-},x^{+}\right)\subseteq\left(a,b\right)$
(resp. proper interval $B=[a,x_{a}^{+})$ where $x_{a}^{+}<b$ and
proper interval $B=(x_{b}^{-},b]$ where $x_{b}^{-}>a$) that contains
$x$ such that for some unique $\lambda\in\mathbb{R}$, $U_{0}\left(z_{1},\tilde{t}_{1}\right)=U_{0}\left(z_{2},\tilde{t}_{2}\right)$
if and only if $U_{\tau}\left(z_{1},\hat{t}_{1}\right)=U_{\tau}\left(z_{2},\hat{t}_{1}+\lambda\left(\tilde{t}_{2}-\tilde{t}_{1}\right)\right)$
for all $z_{1},z_{2}\in B$.
\end{lem}
\begin{proof}
Fix any $x\in\left(a,b\right)$. Consider $i\in\left\{ 0,\tau\right\} $.
Since $U_{i}\left(\cdot,\cdot\right)$ is continuous and decreasing
in it's second argument, there exists $q\in\left(i,\bar{t}\right)$
such that $c\left(\left\{ \left(a,i\right),\left(x,q\right)\right\} \right)=\left\{ \left(x,q\right)\right\} $.
Since there exists an open interval in $\left(i,\bar{t}\right)$ that
contains $q$, by continuity of $U_{i}\left(\cdot,\cdot\right)$,
there exists an open interval $O_{i}$ in $X$ that contains $x$
such that $x'\in O_{i}$ implies $c\left(\left\{ \left(a,i\right),\left(x,q\right),\left(x',q'\right)\right\} \right)=\left\{ \left(x,q\right),\left(x',q'\right)\right\} $
for some $q'\in\left(i,\bar{t}\right)$. \emph{Observation}: for every
$x_{1},x_{2}\in O_{i}$ such that $x_{1}<x_{2}$, since we have $c\left(\left\{ \left(a,i\right),\left(x,q\right),\left(x_{1},t_{1}\right)\right\} \right)=\left\{ \left(x,q\right),\left(x_{1},t_{1}\right)\right\} $
for some $t_{1}$, $c\left(\left\{ \left(a,i\right),\left(x,q\right),\left(x_{2},t_{2}\right)\right\} \right)=\left\{ \left(x,q\right),\left(x_{2},t_{2}\right)\right\} $
for some $t_{2}$, and \lemref{DUfixedtime}, we have $c\left(\left\{ \left(x_{1},i\right),\left(x_{2},i+t_{2}-t_{1}\right)\right\} \right)=\left\{ \left(x_{1},i\right),\left(x_{2},i+t_{2}-t_{1}\right)\right\} $
in $\left(c,S_{\left(\cdot,i\right)}\right)$. 

Now consider an open interval $\left(x^{-},x^{+}\right)\subseteq O_{\tau}\cap O_{0}$
that contains $x$. Consider any $x_{1},x_{2},z\in\left(x^{-},x^{+}\right)$
where $x_{1}<z<x_{2}$. We show an intermediate result that (i) $U_{0}\left(x_{1},0\right)=U_{0}\left(z,\alpha_{z}t\right)=U_{0}\left(x_{2},t\right)$
if and only if (ii) $U_{\tau}\left(x_{1},0\right)=U_{\tau}\left(z,\alpha_{z}t'\right)=U_{\tau}\left(x_{2},t'\right)$.
Say (i) holds (for some $\alpha_{z}$). Due to the \emph{observation},
$x_{1},x_{2}\in O_{0}$, and \lemref{DUfixedtime}, we have $c\left(A\right)=A$
where $A=\left\{ \left(x_{1},0\right),\left(z,\alpha_{z}t\right),\left(x_{2},t\right)\right\} $.
Due to the \emph{observation} and $x_{1},x_{2}\in O_{\tau}$, we have
$c\left(\left\{ \left(x_{1},\tau\right),\left(x_{2},\tau+t'\right)\right\} \right)$
$=\left\{ \left(x_{1},\tau\right),\left(x_{2},\tau+t'\right)\right\} $
for some $t'$. Consider the choice set $B=\left\{ \left(x_{1},\tau\right),\left(z,\tau+\alpha_{z}t'\right),\left(x_{2},\tau+t'\right)\right\} $,
and note that $B$ is related to $A$ by transforming the time of
each timed payment in $A$ from $\hat{t}$ to $\lambda^{*}\hat{t}+d^{*}$,
where $\lambda^{*}=\frac{t'}{t}$ and $d^{*}=\tau$. Then, invoking
\axmref{patience2} gives $c\left(B\right)=B$, which gives (ii) as
desired. The converse, (ii) implies (i), can be shown analogously.
Due to \factref{time-translation}, we also note that $U_{0}\left(x_{1},0\right)=U_{0}\left(z,\alpha_{z}t\right)=U_{0}\left(x_{2},t\right)$
if and only if $U_{\tau}\left(x_{1},0\right)=U_{\tau}\left(z,\alpha_{z}t'\right)=U_{\tau}\left(x_{2},t'\right)$.

Consider any $z_{1},z_{2},z_{3},z_{4}\in\left(x^{-},x^{+}\right)$.
There exist $x_{1},x_{2}\in\left(x^{-},x^{+}\right)$ such that $z_{i}\in\left(x_{1},x_{2}\right)$
for all $i$. The intermediate result gives, for all $i,j\in\left\{ 1,2,3,4\right\} $,
$U_{0}\left(x_{1},0\right)=U_{0}\left(z_{i},\alpha_{i}t\right)=U_{0}\left(x_{2},t\right)$
if and only if $U_{\tau}\left(x_{1},0\right)=U_{\tau}\left(z_{i},\alpha_{i}t'\right)=U_{\tau}\left(x_{2},t'\right)$,
so $U_{0}\left(z_{i},\alpha_{i}t\right)=U_{0}\left(z_{j},\alpha_{j}t\right)$
if and only if $U_{\tau}\left(z_{i},\alpha_{i}t'\right)=U_{\tau}\left(z_{j},\alpha_{j}t'\right)$,
so by \factref{time-translation}, $U_{0}\left(z_{i},0\right)=U_{0}\left(z_{j},\left(\alpha_{j}-\alpha_{i}\right)t\right)$
if and only if $U_{\tau}\left(z_{i},0\right)=U_{\tau}\left(z_{j},\left(\alpha_{j}-\alpha_{i}\right)t'\right)$,
which means $U_{0}\left(z_{i},0\right)=U_{0}\left(z_{j},\tilde{t}\right)$
if and only if $U_{\tau}\left(z_{i},0\right)=U_{\tau}\left(z_{j},\lambda\tilde{t}\right)$
where $\lambda=\frac{t'}{t}$. Note that $\lambda$ is independent
of $i,j$, hence the same $\lambda$ applies to relate $z_{1},z_{2}$
and to relate $z_{3},z_{4}$. Invoking \factref{time-translation}
once more completes the proof for the existence of $\lambda$. Since
$\lambda=\frac{t'}{t}$, where $t,t'$ are the unique solutions to
$U_{0}\left(x_{1},0\right)=U_{0}\left(x_{2},t\right)$ and $U_{\tau}\left(x_{1},0\right)=U_{\tau}\left(x_{2},t'\right)$,
therefore $\lambda$ is unique (for the given $x\in\left(a,b\right)$).

For $x=a$ and $x=b$, the proof is similar other than we replace
open intervals $\left(x^{+},x^{-}\right)$ with half-open intervals
$[a,x_{a}^{+})$ and $(x_{b}^{-},b]$.
\end{proof}
\begin{lem}
\label{lem:lambda}There exists $\lambda\in\mathbb{R}$ such that
for all $x^{*}\in X$, $U_{0}\left(a,0\right)=U_{0}\left(x^{*},t^{*}\right)$
if and only if $U_{\tau}\left(a,0\right)=U_{\tau}\left(x^{*},\lambda t^{*}\right)$.
Moreover, $\lambda$ is unique.
\end{lem}
\begin{proof}
Let $\text{\ensuremath{\mathbb{C}:=}\ensuremath{\left\{  [a,x_{a}^{+}),(x_{b}^{-},b]\right\} } \ensuremath{\cup}}\left\{ \left(x_{x}^{+},x_{x}^{-}\right):x\in\left(a,b\right)\right\} $
be the collection intervals guaranteed by \lemref{proportional}.
Note that $\mathbb{C}$ is an open cover of the closed and bounded
interval $\left[a,b\right]$, so a finite subcover $\bar{\mathbb{C}}$
is guaranteed by the Heine\textendash Borel theorem. Consider a finite
sequence of intervals in $\bar{\mathbb{C}}$, $\left(B_{k}\right)_{k=1}^{K}$,
such that the first interval is $B_{1}=[a,x_{a}^{+})$, last interval
is $B_{K}=(x_{b}^{-},b]$, and for all $k\in\left\{ 1,K-1\right\} $,
$B_{k}\cap B_{k+1}\ne\emptyset$. This is guaranteed by the fact that
$\bar{\mathbb{C}}$ is a cover of $\left[a,b\right]$ and the intervals
in $\bar{\mathbb{C}}$ are open except for $[a,x_{a}^{+})$ and $(x_{b}^{-},b]$.
Then, for every two consecutive intervals $B_{k},B_{k+1}$, the unique
$\lambda$'s guaranteed by \lemref{proportional}, one for $B_{k}$
and another for $B_{k+1}$, must coincide due to the nondegenerate
intersection $B_{k}\cap B_{k+1}$. Iterating through this finite sequence
of intersecting consecutive intervals guarantees, for every $x^{*}\ne a$,
an increasing sequence of payments $\left(x_{k}\right)_{k=1}^{M}$
such that $x_{1}=a$, $x_{M}=x^{*}$, and for some $\lambda$, $U_{0}\left(x_{k},0\right)=U_{0}\left(x_{k+1},t\right)$
if and only if $U_{\tau}\left(x_{k},0\right)=U_{\tau}\left(x_{k+1},\lambda t\right)$
for all $k\in\left\{ 1,...,M-1\right\} $. The rest is straightforward
using \factref{time-translation} (for example if $M=3$, we have
$U_{0}\left(a,0\right)=U_{0}\left(x_{1},t_{1}\right)=U_{0}\left(x^{*},t^{*}\right)$
if and only if $U_{\tau}\left(a,0\right)=U_{\tau}\left(x_{1},\lambda t_{1}\right)=U_{\tau}\left(x^{*},\lambda t_{1}+\lambda\left(t^{*}-t_{1}\right)\right)$,
which completes the proof since $\lambda t_{1}+\lambda\left(t^{*}-t_{1}\right)=\lambda t^{*}$). 
\end{proof}
To recover $\lambda$, take any $x_{1},x_{2}\in X$ such that $x_{1}<x_{2}$.
For some $t$ and $t'$, $U_{0}\left(x_{1},0\right)=U_{0}\left(x_{2},t\right)$
and $U_{\tau}\left(x_{1},0\right)=U_{\tau}\left(x_{2},t'\right)$.
Then since we must have $\lambda t=t'$, we have $\lambda=\frac{t'}{t}$.
With \lemref{lambda}, we conclude that $\left(\hat{\delta}_{\tau},u_{\tau}\right)$
where $u_{\tau}=u_{0}$ and $\hat{\delta}_{\tau}=\hat{\delta}_{0}^{-\lambda}$
is a DU representation for $\left(c,S_{\left(\cdot,\tau\right)}\right)$.

The analysis thus far was for $\tau\in\left(0,\bar{t}\right)$. When
$\tau=\bar{t}$, since every choice problem in $S_{\left(\cdot,\bar{t}\right)}$
contains only timed payments that pay at time $\bar{t}$, a DU representation
is trivially established with any positive $\hat{\delta}_{\bar{t}}$
and any strictly increasing $u_{\bar{t}}$. Therefore, we set $u_{\bar{t}}=u_{0}$
and $\hat{\delta}_{\bar{t}}=\sup_{\tau\in[0,\bar{t})}\hat{\delta}_{\tau}$
(this is why we cannot guarantee $\hat{\delta}_{\bar{t}}<1$, even
if \axmref{time_monotonicity=000026impatience} gives us $\hat{\delta}_{\tau}\in\left(0,1\right)$
for all $\tau$). From now on, we remove subscript $\tau$ from $u_{\tau}$
and simply write $u$.

\subsubsection*{Stage 3: $\hat{\delta}_{\tau}\protect\geq\hat{\delta}_{\tau'}$ for
all $\tau>\tau'$}

If $\tau=\bar{t}$, this is trivial from the construction of $\hat{\delta}_{\bar{t}}$.
Consider any $\tau,\tau'\in[0,\bar{t})$. Continuity of $U_{\tau}\left(x,q\right)=\hat{\delta}_{\tau}^{q}u\left(x\right)$
and $U_{\tau'}\left(x,q\right)=\hat{\delta}_{\tau'}^{q}u\left(x\right)$
guarantee the existence of $y>a$ such that $c\left(\left\{ \left(a,\tau\right),\left(y,t\right)\right\} \right)=\left\{ \left(a,\tau\right),\left(y,t\right)\right\} $
and $c\left(\left\{ \left(a,\tau'\right),\left(y,t'\right)\right\} \right)=\left\{ \left(a,\tau'\right),\left(y,t'\right)\right\} $
for some $t,t'\in T$, with which we obtain $\hat{\delta}_{\tau}^{\tau}u\left(a\right)=\hat{\delta}_{\tau}^{t}u\left(y\right)$
and $\hat{\delta}_{\tau'}^{\tau'}u\left(a\right)=\hat{\delta}_{\tau'}^{t'}u\left(y\right)$.
Note that by \axmref{time_monotonicity=000026impatience}, $\hat{\delta}_{\tau},\hat{\delta}_{\tau'}<1$,
so $t-\tau>0$ and $t'-\tau'>0$. Suppose for contradiction $\hat{\delta}_{\tau'}>\hat{\delta}_{\tau}$,
then $t'-\tau'>t-\tau$, or equivalently $t'>\tau'+t-\tau$. Note
also that $\tau'-\tau<0$ implies $\tau'+t-\tau<t$. So $t,t'\in T$
implies $\left(y,\tau'+t-\tau\right)\in X\times T$. Putting together
what we established, we have $\tau'<\tau<t$, $\tau'<\tau'+t-\tau$,
$\hat{\delta}_{\tau'}^{\tau}u\left(a\right)<\hat{\delta}_{\tau'}^{\tau'}u\left(a\right)=\hat{\delta}_{\tau'}^{t'}u\left(y\right)<\hat{\delta}_{\tau'}^{\tau'+t-\tau}u\left(y\right)$,
and $\hat{\delta}_{\tau'}^{t}u\left(y\right)<\hat{\delta}_{\tau'}^{\tau'+t-\tau}u\left(y\right)$,
which implies $c\left(\left\{ \left(a,\tau'\right),\left(y,\tau'+t-\tau\right),\left(a,\tau\right),\left(y,t\right)\right\} \right)=\left\{ \left(y,\tau'+t-\tau\right)\right\} $.
By Continuity of $U_{\tau}\left(x,q\right)$ and $U_{\tau'}\left(x,q\right)$,
if we consider $y-\epsilon$ for some $\epsilon>0$ sufficiently small,
we have $c\left(\left\{ \left(a,\tau'\right),\left(y-\epsilon,\tau'+t-\tau\right),\left(a,\tau\right),\left(y-\epsilon,t\right)\right\} \right)=\left\{ \left(y-\epsilon,\tau'+t-\tau\right)\right\} $
and $c\left(\left\{ \left(a,\tau\right),\left(y-\epsilon,t\right)\right\} \right)$$=\left\{ \left(a,\tau\right)\right\} $,
which jointly contradict \axmref{patience1} because $\tau'=\tau+d$
and $\tau'+t-\tau=t+d$ where $d=\tau'-\tau>0$.

\subsubsection*{Stage 4: $R$ and $\delta_{\left(x,t\right)}$}

Create a complete, transitive, and antisymmetric $R$ on $Y$ such
that $t<t'$ implies $\left(x,t\right)R\left(x',t'\right)$, which
involves an arbitrary completion between when $\left(x,t\right)$
and $\left(x',t'\right)$ when $t=t'$, and set, for every $\left(x,t\right)\in Y$,
$\delta_{\left(x,t\right)}:=\hat{\delta}_{t}$.

\subsection{Proof of Theorem \ref{thm:fspu}}

``\textbf{If}'' is straightforward. I prove ``\textbf{only if}''.
\textbf{Stage 1} and \textbf{Stage 2} show that with \axmref{equality_reference_dependence}
and \axmref{social_monotonicity}, for each Gini coefficient $g$,
the set of all choice problems where the most balanced alternative
has Gini coefficient $g$ can be explained by the maximization of
$\tilde{x}+\hat{v}_{g}\left(\tilde{y}\right)$ for some unique $\hat{v}_{g}:[w,+\infty)\rightarrow\mathbb{R}$.
\textbf{Stage 3} shows that $g<g'$ implies $\hat{v}_{g}\left(y\right)-\hat{v}_{g}\left(y'\right)\geq\hat{v}_{g'}\left(y\right)-\hat{v}_{g'}\left(y'\right)$
for all $y>y'$. \textbf{Stage 4} builds the reference order $R$
using Gini coefficient and arbitrary completion.

\subsubsection*{Stage 1: $x+v_{\left(x,y\right)}\left(y\right)$ for each alternative
$\left(x,y\right)\in Y$}

Fix $\left(x,y\right)\in Y$. Like before, let $R^{\downarrow}\left(\left(x,y\right)\right):=\left\{ \left(x',y'\right)\in Y:G\left(\left(x,y\right)\right)\leq G\left(\left(x',y'\right)\right)\right\} $,
$\mathbb{P}^{\left(x,y\right)}:=\left\{ \left(x',y'\right)\in R^{\downarrow}\left(\left(x,y\right)\right):\left(x',y'\right)\in c\left(\left\{ \left(x',y'\right),\left(x,y\right)\right\} \right)\right\} $,
and $\mathbb{A}:=\mathbb{A}_{R^{\downarrow}\left(\left(x,y\right)\right)}^{\left(x,y\right)}=\left\{ A\in\mathcal{A}:\left(x,y\right)\in\arg\min_{z\in A}G\left(z\right)\right\} $.

By \axmref{equality_reference_dependence}, $c$ satisfies WARP over
$\mathbb{A}$. By \thmref{ORD}, there exists a utility function $U:Y\rightarrow\mathbb{R}$
that explains $\left(c,\mathbb{A}\right)$.

\begin{figure}
\begin{centering}
\includegraphics[scale=0.3]{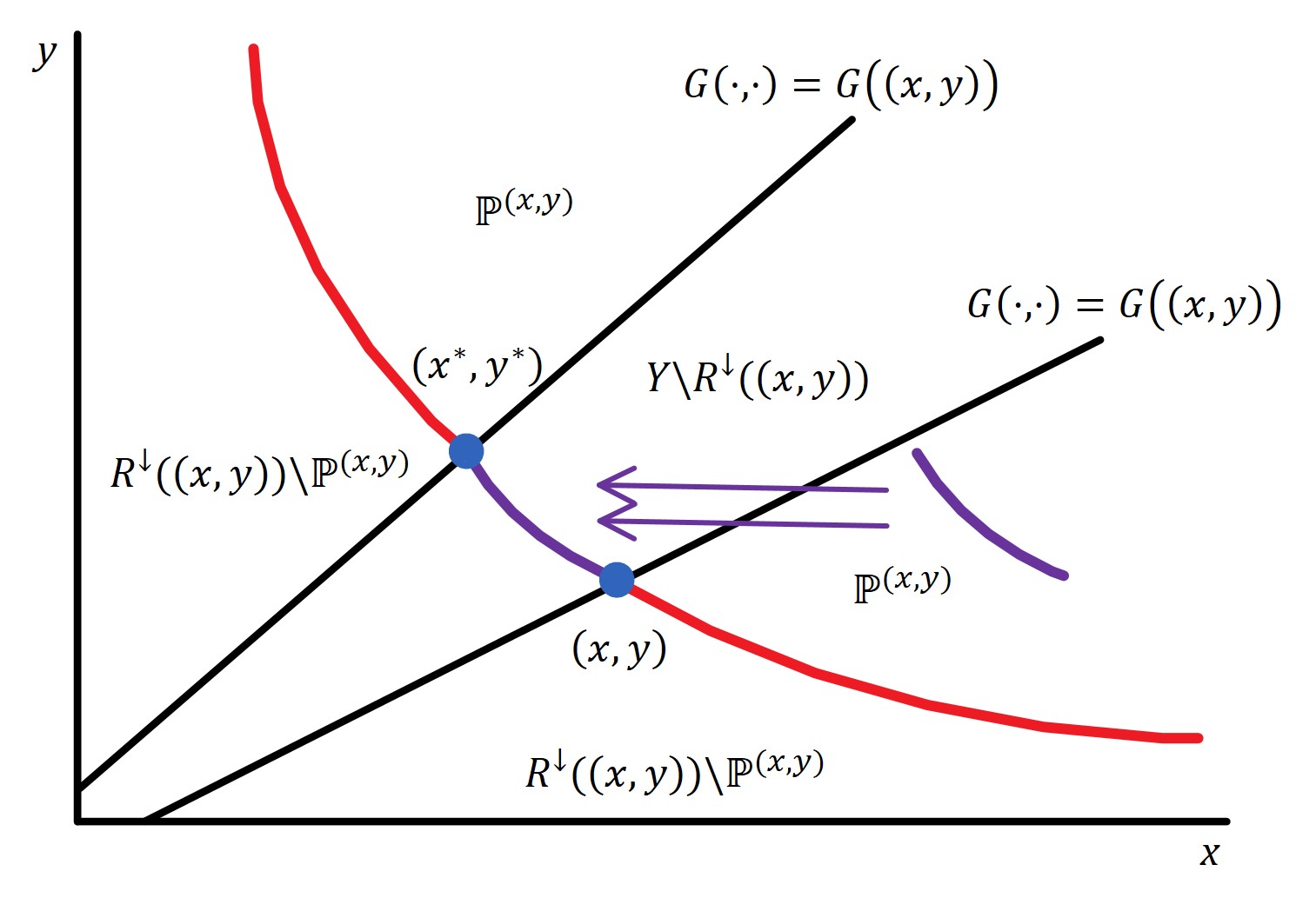}
\par\end{centering}
\caption{\label{fig:social_proof}This figure illustrates the construction
of $v_{\left(x,y\right)}$ for a fixed $\left(x,y\right)\in Y$. The
space $Y$ is divided into three regions: (1) Between the two diagonal
lines are alternatives in $Y\backslash R^{\downarrow}\left(\left(x,y\right)\right)$,
they have lower Gini coefficients than $\left(x,y\right)$, and therefore
they appear in a choice problem where $\left(x,y\right)$ is the reference.
The alternatives in $R^{\downarrow}\left(\left(x,y\right)\right)$
are then split into two groups: (2) those that are chosen when $\left(x,y\right)$
is the reference, $\mathbb{P}^{\left(x,y\right)}$, and (3) those
that are not, $R^{\downarrow}\left(\left(x,y\right)\right)\backslash\mathbb{P}^{\left(x,y\right)}$.
These two groups are separated by the indifference curve passing through
$\left(x,y\right)$, the red curve, which partially constructs $v_{\left(x,y\right)}$
(partial because Gini coefficient truncates the space). The rest of
$v_{\left(x,y\right)}$ can be constructed by using an indifference
curve that connects $\left(x+a,y\right)$ and $\left(x^{*}+a,y^{*}\right)$,
the purple curve, where $G\left(\left(x,y\right)\right)=G\left(\left(x^{*},y^{*}\right)\right)$,
$c\left(\left\{ \left(x,y\right),\left(x^{*},y^{*}\right)\right\} \right)=\left\{ \left(x,y\right),\left(x^{*},y^{*}\right)\right\} $,
and $\left(x+a,y\right),\left(x^{*}+a,y^{*}\right)\in\mathbb{P}^{\left(x,y\right)}$.}
\end{figure}

Note that for all $\left(x',y'\right)\in R^{\downarrow}\left(\left(x,y\right)\right)$,
$U\left(x',y'\right)\geq U\left(x,y\right)$ if and only if $\left(x',y'\right)\in\mathbb{P}^{\left(x,y\right)}$.
Since $c$ satisfies Quasi-linearity over $\mathbb{A}$ (\axmref{equality_reference_dependence}),
$U$ restricted to the domain $\mathbb{P}^{\left(x,y\right)}$ (which
contains $\left(x,y\right)$ itself) must be a strictly increasing
transformation of $\tilde{x}+v_{\left(x,y\right)}\left(\tilde{y}\right)$
for some unique $v_{\left(x,y\right)}:[w,+\infty)\rightarrow\mathbb{R}$.
\figref{social_proof} provides an illustration of how $v_{\left(x,y\right)}$
is constructed, and $\tilde{x}+v_{\left(x,y\right)}\left(\tilde{y}\right)$
is our target, quasi-linear, utility function. It is straightforward
that for all $A\in\mathbb{A}$ such that $A\subseteq\mathbb{P}^{\left(x,y\right)}$,
the maximization of $\tilde{x}+v_{\left(x,y\right)}\left(\tilde{y}\right)$
gives $c\left(A\right)$. Next, we show that this consistency applies
to other $A\in\mathbb{A}$. For any $\left(x',y'\right)\in R^{\downarrow}\left(\left(x,y\right)\right)\backslash\mathbb{P}^{\left(x,y\right)}$,
there is no $A\in\mathbb{A}$ such that $\left(x',y'\right)\in c\left(A\right)$,
so we just need to guarantee $x'+v_{\left(x,y\right)}\left(y'\right)<x+v_{\left(x,y\right)}\left(y\right)$.
Suppose for contradiction this inequality fails. Since for some $a$
we have $\left\{ \left(x+a,y\right),\left(x'+a,y'\right)\right\} \subseteq\mathbb{P}^{\left(x,y\right)}$,
and therefore $\left(x'+a,y'\right)\in c\left\{ \left(x,y\right),\left(x+a,y\right),\left(x'+a,y'\right)\right\} $,
the fact that $\left\{ \left(x,y\right),\left(x+a,y\right),\left(x'+a,y'\right)\right\} $
and $\left\{ \left(x,y\right),\left(x',y'\right)\right\} $ are both
in $\mathbb{A}$ but $\left(x',y'\right)\notin c\left(\left\{ \left(x,y\right),\left(x',y'\right)\right\} \right)$
(because $\left(x',y'\right)\notin\mathbb{P}^{\left(x,y\right)}$)
contradicts $c$ satisfies Quasi-linearity over $\mathbb{A}$. It
remains to consider the consistency of $\tilde{x}+v_{\left(x,y\right)}\left(\tilde{y}\right)$
for alternative $\left(x',y'\right)\notin R^{\downarrow}\left(\left(x,y\right)\right)$,
but this is immediate since there is no $A\in\mathbb{A}$ such that
$\left(x',y'\right)\in\mathbb{A}$. So $\tilde{x}+v_{\left(x,y\right)}\left(\tilde{y}\right)$
explains $\left(c,\mathbb{A}\right)$.

\subsubsection*{Stage 2: $x+\hat{v}_{g}\left(y\right)$ for each Gini coefficient
$g$}

Fix $g\in[0,0.5)$, we now show that $v_{\left(x,y\right)}$ must
coincide for all $\left(x,y\right)$ where $G\left(\left(x,y\right)\right)=g$.
Consider the collection of choice sets $\text{\ensuremath{\mathcal{S}}}:=\left\{ A\in\mathcal{A}:\min_{z\in A}G\left(z\right)=g\right\} $.
It turns out that $c$ satisfies WARP and Quasi-linearity over $\mathcal{S}$.
To see this, take any two choice problems $A_{1},A_{2}$ in $\mathcal{S}$.
For each $i=1,2$, there must be an alternative $\left(x_{i},y_{i}\right)\in A_{i}$
such that $G\left(\left(x_{i},y_{i}\right)\right)=g$ and $G\left(\left(x',y'\right)\right)\geq g$
for all other $\left(x',y'\right)$ in $A_{i}$. Consider an income
distribution $\left(x^{*},y^{*}\right)$ such that $x^{*}\leq\min$$\left\{ x_{1},x_{2}\right\} $
and $y^{*}\leq\min\left\{ y_{1},y_{2}\right\} $ and $G\left(\left(x^{*},y^{*}\right)\right)=g$.
Due to $\left(x_{i},y_{i}\right)\in\Psi\left(A_{i}\cup\left\{ \left(x^{*},y^{*}\right)\right\} \right)$,
\axmref{equality_reference_dependence}, and \axmref{social_monotonicity},
we have $c\left(A_{i}\right)=c\left(A_{i}\cup\left\{ \left(x^{*},y^{*}\right)\right\} \right)$
for $i=1,2$. But $\left(x^{*},y^{*}\right)\in\Psi\left(A_{1}\cup A_{2}\cup\left\{ \left(x^{*},y^{*}\right)\right\} \right)$,
so by \axmref{equality_reference_dependence} again, between $c\left(A_{1}\cup\left\{ \left(x^{*},y^{*}\right)\right\} \right)$
and $c\left(A_{2}\cup\left\{ \left(x^{*},y^{*}\right)\right\} \right)$,
which as established are equal to $c\left(A_{1}\right)$ and $c\left(A_{2}\right)$
respectively, WARP and Quasi-linearity must hold. Since $c$ satisfies
WARP and Quasi-linearity over $\mathcal{S}$, there is a unique $\hat{v}_{g}:[w,+\infty)\rightarrow\mathbb{R}$
such that the utility function $\tilde{x}+\hat{v}_{g}\left(\tilde{y}\right)$
explains $\left(c,\mathcal{S}\right)$. But every $v_{\left(x,y\right)}$
constructed in Stage 1 is also unique, and $\mathbb{A}_{R^{\downarrow}\left(\left(x,y\right)\right)}^{\left(x,y\right)}\subseteq\mathcal{S}$
if $G\left(\left(x,y\right)\right)=g$, so $v_{\left(x,y\right)}$
must coincide for all $\left(x,y\right)$ such that $G\left(\left(x,y\right)\right)=g$.

\subsubsection*{Stage 3: $g<g'$ implies $\hat{v}_{g}\left(y\right)-\hat{v}_{g}\left(y'\right)\protect\geq\hat{v}_{g'}\left(y\right)-\hat{v}_{g'}\left(y'\right)$
for all $y>y'$}

Finally we show that the constructed $\hat{v}_{g}\left(y\right)'s$
are systematically related. Consider any $g,g'\in[0,0.5)$ such that
$g<g'$ (reminder: lower $g$ implies greater attainable equality)
and any $y,y'\in\mathbb{R}_{\geq0}$ such that $y>y'$. Define $\bar{v}_{g}:=\hat{v}_{g}\left(y\right)-\hat{v}_{g}\left(y'\right)$
and $\bar{v}_{g'}:=\hat{v}_{g'}\left(y\right)-\hat{v}_{g'}\left(y'\right)$.
We want to show $\bar{v}_{g}\geq\bar{v}_{g'}$. Suppose not, our goal
is to find a contradiction of \axmref{social_increasingaltruism}
in choice behavior.

Let $z$ be a number such that $\bar{v}_{g}<z<\bar{v}_{g'}$. Consider
$\left(x_{g'},w\right),\left(x_{g},w\right)\in Y$ such that $G\left(\left(x_{g'},w\right)\right)=g'$
and $G\left(\left(x_{g},w\right)\right)=g$, which exist because $G\left(\left(\tilde{x},w\right)\right)$
is continuous and increasing in $\tilde{x}$ from $G\left(\left(w,w\right)\right)=0$
to $\lim_{x\rightarrow+\infty}G\left(\left(x,w\right)\right)=0.5$
and $g,g'\in[0,0.5)$. Consider $x:=z+\Delta$, $x':=2z+\Delta$ for
some $\Delta>0$ such that $g'\leq\min\left(\left\{ G\left(\left(x,y\right)\right),G\left(\left(x',y'\right)\right)\right\} \right)$
and $x'>x>\max\left(\left\{ x_{g'},x_{g}\right\} \right)$, where
$\Delta$ exists because for any fixed $\bar{y}$, $G\left(\left(\tilde{x},\bar{y}\right)\right)$
is asymptotically increasing in $\tilde{x}$ and $\lim_{x\rightarrow+\infty}G\left(\left(x,\bar{y}\right)\right)=0.5$,
and $g'\in[0,0.5)$. Essentially, we have introduced reference points
$\left(x_{g'},w\right),\left(x_{g},w\right)$ that will not be chosen
(due in part to \axmref{social_monotonicity}), forcing the choice
to be between $\left(x,y\right)$ and $\left(x',y'\right)$.

We now use the constructed alternatives, $\left(x,y\right),\left(x',y'\right),\left(x_{g'},w\right),\left(x_{g},w\right)$,
to demonstrate a violation of \axmref{social_increasingaltruism}.
For the choice problem $\left\{ \left(x,y\right),\left(x',y'\right),\left(x_{g'},w\right)\right\} $,
$\left(x_{g'},w\right)$ is the reference (so $\hat{v}_{g'}$ is used)
and cannot be chosen. Since $\bar{v}_{g'}>z$, or equivalently $z+\hat{v}_{g'}\left(y\right)>2z+\hat{v}_{g'}\left(y'\right)$,
we have $x+\hat{v}_{g'}\left(y\right)>x'+\hat{v}_{g'}\left(y'\right)$,
and therefore 
\begin{equation}
c\left(\left\{ \left(x,y\right),\left(x',y'\right),\left(x_{g'},w\right)\right\} \right)=\left\{ \left(x,y\right)\right\} .\label{eq:fspu_increasing_v_4}
\end{equation}
By analogous arguments, $z>\bar{v}_{g}$ gives $c\left(\left\{ \left(x,y\right),\left(x',y'\right),\left(x_{g},w\right)\right\} \right)=\left\{ \left(x',y'\right)\right\} $
($\hat{v}_{g}$ is used), which also gives
\begin{equation}
c\left(\left\{ \left(x,y\right),\left(x',y'\right),\left(x_{g'},w\right),\left(x_{g},w\right)\right\} \right)=\left\{ \left(x',y'\right)\right\} .\label{eq:fspu_increasing_v_5}
\end{equation}
due to \axmref{equality_reference_dependence} and $G\left(\left(x_{g},w\right)\right)=g\leq g'=G\left(\left(x_{g'},w\right)\right)$.
Since $y>y'$, \eqref{fspu_increasing_v_4} and \eqref{fspu_increasing_v_5}
jointly contradict \axmref{social_increasingaltruism}.

\subsubsection*{Stage 4: $R$ on $Y$}

Create a complete, transitive, and antisymmetric $R$ on $Y$ such
that $G\left(\left(x,y\right)\right)<G\left(\left(x',y'\right)\right)$
implies $\left(x,y\right)R\left(x',y'\right)$, which involves an
arbitrary completion when $G\left(\left(x,y\right)\right)=G\left(\left(x',y'\right)\right)$.

\subsection{Proof of Propositions \ref{prop:WARP}, \ref{prop:WARP2}, \ref{prop:WARP3}}

I focus on showing that WARP (1) and structural postulate (2) are
independently sufficient for the standard model (3). The remaining
statements, that WARP and structural postulates are necessary for
standard models ((1) if (3) and (2) if (3)), and that WARP is sufficient
and necessary for a (general) utility representation ((1) if and only
if (4)), are well-known and omitted.

\subsubsection*{Proof of \propref{WARP}: (1) / (2) implies (3)}

Suppose a choice correspondence $c$ admits an AREU representation
with specification $\left(R,\left\{ u_{r}\right\} _{r}\right)$. Suppose
$c$ satisfies WARP or Independence (or both). We first show that
$u_{r}=u_{s}$ for all $r,s\in\Delta\left(X\right)\backslash\text{conv}\left(\left\{ \delta_{b},\delta_{w}\right\} \right)$.
Suppose without loss of generality $rRs$. Suppose for contradiction
$u_{r}\ne u_{s}$, then the fact that $u_{r}$ is a concave transformation
of $u_{s}$ and that both functions are normalized to $\left[0,1\right]$
implies $u_{r}\left(x\right)>u_{s}\left(x\right)$ for all $x\in X\backslash\left\{ b,w\right\} $.
Consider the set $\tau_{s}:=\text{conv}\left(\left\{ s,\delta_{b},\delta_{w}\right\} \right)$.
The interior of this set, $\text{Int}\tau_{s}$, consists of lotteries
that are extreme spreads of $s$, hence $\text{Int}\tau_{s}\subseteq R^{\downarrow}\left(s\right)\subseteq R^{\downarrow}\left(r\right)$.
By \axmref{FOSD}, $c\left(\left\{ \delta_{b},r\right\} \right)=c\left(\left\{ \delta_{b},s\right\} \right)=\delta_{b}$.
Then by Continuity, there exist open balls around $\delta_{b}$, $B_{r}$
and $B_{s}$, such that they contain lotteries that are chosen over
$r$ and $s$ respectively. Now consider an open subset $S$ of $B_{r}\cap B_{s}\cap\text{Int}\tau_{s}$.
Since $u_{r}\left(x\right)>u_{s}\left(x\right)$ for all $x\in X\backslash\left\{ b,w\right\} $,
we can find lotteries $p,q\in S$ such that $\mathbb{E}_{p}u_{r}\left(x\right)>\mathbb{E}_{q}u_{r}\left(x\right)$
and $\mathbb{E}_{p}u_{s}\left(x\right)<\mathbb{E}_{q}u_{s}\left(x\right)$.
This means
\begin{align}
c\left(\left\{ r,s,p,q\right\} \right) & =\left\{ p\right\} \text{ and}\label{eq:warp1_1}\\
c\left(\left\{ s,p,q\right\} \right) & =\left\{ q\right\} .\label{eq:warp1_2}
\end{align}
Consider $t\in S$, $p'=\frac{1}{2}p\oplus\frac{1}{2}t$, and $q'=\frac{1}{2}q\oplus\frac{1}{2}t$,
then $p',q'\in S$, and therefore
\begin{equation}
c\left(\left\{ s,p',q'\right\} \right)=\left\{ q'\right\} .\label{eq:warp1_3}
\end{equation}
Finally we conclude that \eqref{warp1_1} and \eqref{warp1_2} jointly
violate WARP, whereas \eqref{warp1_1} and \eqref{warp1_3} jointly
violate Independence.

Next we turn to $r\in\text{conv}\left(\left\{ \delta_{b},\delta_{w}\right\} \right)$
and show that $u_{r}$ is either identical, or has the freedom to
be identical, to $u_{s}$ where $s\in\Delta\left(X\right)\backslash\text{conv}\left(\left\{ \delta_{b},\delta_{w}\right\} \right)$.
If $r=\delta_{b}$ or $r=\delta_{w}$ or $R^{\downarrow}\left(r\right)\subseteq\text{conv}\left(\left\{ \delta_{b},\delta_{w}\right\} \right)$,
then any strictly increasing $u_{r}$ can explain $c$ over $\mathbb{A}_{R^{\downarrow}\left(r\right)}^{r}:=\left\{ A\in\mathcal{A}:A\subseteq R^{\downarrow}\left(r\right)\text{ and }r\in A\right\} $,
so we can just pick one that is identical to $u_{s}$ for every $s\in\Delta\left(X\right)\backslash\text{conv}\left(\left\{ \delta_{b},\delta_{w}\right\} \right)$.
If $r$ doesn't satisfy any of those conditions, then there exists
$s_{2}\in\Delta\left(X\right)\backslash\text{conv}\left(\left\{ \delta_{b},\delta_{w}\right\} \right)$
such that $r$ is an extreme spread of $s_{2}$ and there exists $s_{1}\in R^{\downarrow}\left(r\right)\backslash\text{conv}\left(\left\{ \delta_{b},\delta_{w}\right\} \right)$.
This implies $u_{s_{2}}$ is a concave transformation of $u_{r}$
(because $s_{2}Rr$) and $u_{r}$ is a concave transformation of $u_{s_{1}}$
(because $rRs_{1}$), but we already showed that $u_{s_{1}}=u_{s_{2}}$
(since $s_{1},s_{2}\in\Delta\left(X\right)\backslash\text{conv}\left(\left\{ \delta_{b},\delta_{w}\right\} \right))$,
so $u_{r}$ is identical to $u_{s}$ for all $s\in\Delta\left(X\right)\backslash\text{conv}\left(\left\{ \delta_{b},\delta_{w}\right\} \right)$.
We conclude that if either WARP or Independence (or both) holds, then
$c$ admits an expected utility representation.

\subsubsection*{Proof of \propref{WARP2}: (1) / (2) implies (3)}

Suppose a choice correspondence $c$ admits a PEDU representation
with specification $\left(\left\{ \delta_{r}\right\} _{r},u\right)$.
We show that if $\delta_{r}\ne\delta_{r'}$ for some $r,r'\in[0,\bar{t})$,
then $c$ violates both WARP and Stationarity. ($\delta_{\bar{t}}$
only plays a role for choice problems $A\in\mathcal{A}$ where every
alternative has $t=\bar{t}$, so we set it as $\delta_{\bar{t}}=\delta_{0}$.)

Suppose for contradiction $\delta_{r}\ne\delta_{r'}$ for some $r,r'\in[0,\bar{t})$.
Say without loss of generality $r>r'\geq0$, then $\delta_{r}>\delta_{r'}\geq\delta_{0}$.
Recall that $X=\left[a,b\right]$. Consider alternatives $\left(b-\Delta_{x},0\right),\left(b,0+\Delta_{t}\right)\in X\times T$
such that (i) $\Delta_{x}\in\left(0,b-a\right)$, (ii) $\Delta_{t}\in\left(0,\bar{t}-r\right)$,
(iii) $\delta_{r}^{0}u\left(b-\Delta_{x}\right)<\delta_{r}^{0+\Delta_{t}}u\left(b\right)$,
and (iv) $\delta_{0}^{0}u\left(b-\Delta_{x}\right)>\delta_{0}^{0+\Delta_{t}}u\left(b\right)$,
which is possible due in part to the assumption that $\left(b,\bar{t}\right)\in c\left(\left\{ \left(a,0\right),\left(b,\bar{t}\right)\right\} \right)$.
Note that (i) and (ii) guarantee $\left(b-\Delta_{x},0\right),\left(b,0+\Delta_{t}\right),\left(b-\Delta_{x},r\right),\left(b,r+\Delta_{t}\right)\in X\times T$.
Then, (iii) gives 
\begin{equation}
c\left(\left\{ \left(b-\Delta_{x},r\right),\left(b,r+\Delta_{t}\right)\right\} \right)=\left\{ \left(b,r+\Delta_{t}\right)\right\} ,\label{eq:warp2_5}
\end{equation}
and (iv) gives 
\begin{align}
c\left(\left\{ \left(b-\Delta_{x},0\right),\left(b,0+\Delta_{t}\right)\right\} \right) & =\left\{ \left(b-\Delta_{x},0\right)\right\} \text{ and}\label{eq:warp2_6}\\
c\left(\left\{ \left(a,0\right),\left(b-\Delta_{x},r\right),\left(b,r+\Delta_{t}\right)\right\} \right) & =\left\{ \left(b-\Delta_{x},r\right)\right\} ,\label{eq:warp2_7}
\end{align}
where \eqref{warp2_7} is due in part to the assumption that $\left(b,\bar{t}\right)\in c\left(\left\{ \left(a,0\right),\left(b,\bar{t}\right)\right\} \right)$
as it excludes $\left(a,0\right)$ from being uniquely chosen. Now
note that \eqref{warp2_5} and \eqref{warp2_6} jointly violate Stationarity,
whereas \eqref{warp2_5} and \eqref{warp2_7} jointly violate WARP.
We conclude that if either WARP or Stationarity (or both) holds, then
$c$ admits an exponential discounting utility representation.

\paragraph*{Proof of \propref{WARP3}: (1) / (2) implies (3)}

Suppose a choice correspondence $c$ admits an FSPU representation
with specification $\left\{ v_{r}\right\} _{r}$. We show that if
$v_{r}\left(y\right)-v_{r}\left(y'\right)\ne v_{r'}\left(y\right)-v_{r'}\left(y'\right)$
for some $r,r'$ and $y>y'$, then $c$ violates both WARP and Quasi-linearity.

Suppose for contradiction $v_{r}\left(y\right)-v_{r}\left(y'\right)\ne v_{r'}\left(y\right)-v_{r'}\left(y'\right)$
for some $r,r'\in[0,0.5)$ and $y>y'$. Without loss of generality,
say $r>r'\geq0$. Then $v_{r}\left(y\right)-v_{r}\left(y'\right)<v_{r'}\left(y\right)-v_{r'}\left(y'\right)\leq v_{0}\left(y\right)-v_{0}\left(y'\right)$,
and therefore there exist $\tilde{x},\tilde{x}'\in[w,+\infty)$ such
that $\tilde{x}'+v_{r}\left(y'\right)>\tilde{x}+v_{r}\left(y\right)$
and $\tilde{x}'+v_{0}\left(y'\right)<\tilde{x}+v_{0}\left(y\right)$.
Consider $\left(x^{*},y^{*}\right)\in Y$ such that $y^{*}=w$ and
$G\left(\left(x^{*},y^{*}\right)\right)=r$, which is possible since
$G\left(\left(\cdot,w\right)\right)$ is continuous and increasing
in it's first argument from $G\left(\left(w,w\right)\right)=0$ to
$\lim_{x\rightarrow+\infty}G\left(\left(x,w\right)\right)=0.5$. Since
for any fixed $\bar{y}$, $G\left(\left(\cdot,\bar{y}\right)\right)$
is asymptotically increasing in it's first argument, there exists
$\Delta>0$ such that $\min\left(\left\{ G\left(\left(\tilde{x}+\Delta,y\right)\right),G\left(\left(\tilde{x}'+\Delta,y'\right)\right)\right\} \right)\geq r$
and $\min\left(\left\{ \tilde{x}+\Delta,\tilde{x}'+\Delta\right\} \right)>x^{*}$.
Let $x:=\tilde{x}+\Delta$ and $x':=\tilde{x}'+\Delta$. We have now
established that (i) $\min\left(\left\{ x,x'\right\} \right)>x^{*}\geq w,\min\left(\left\{ y,y'\right\} \right)\geq y^{*}=w$,
(ii) $\min\left(\left\{ G\left(\left(x,y\right)\right),G\left(\left(x',y'\right)\right)\right\} \right)\geq G\left(\left(x^{*},y^{*}\right)\right)=r$,
(iii) $x'+v_{r}\left(y'\right)>x+v_{r}\left(y\right)$, and (iv) $x'+v_{0}\left(y'\right)<x+v_{0}\left(y\right)$.

Then, (i), (ii), and (iii) together give
\begin{align}
c\left(\left\{ \left(x^{*},y^{*}\right),\left(x,y\right),\left(x',y'\right)\right\} \right) & =\left\{ \left(x',y'\right)\right\} ,\label{eq:warp3_3}
\end{align}
whereas (i) and (iv) together give
\begin{align}
c\left(\left\{ \left(w,w\right),\left(x,y\right),\left(x',y'\right)\right\} \right) & =\left\{ \left(x,y\right)\right\} \text{ and}\label{eq:warp3_4}\\
c\left(\left\{ \left(w,w\right),\left(x+\epsilon,y\right),\left(x'+\epsilon,y'\right)\right\} \right) & =\left\{ \left(x+\epsilon,y\right)\right\} \,\forall\epsilon>0.\label{eq:warp3_5}
\end{align}
Note that \eqref{warp3_3} and \eqref{warp3_5} jointly violate Quasi-linearity.
Separately, by WARP, \eqref{warp3_3} and \eqref{warp3_4} imply $c\left(\left\{ \left(x,y\right),\left(x',y'\right)\right\} \right)=\left\{ \left(x',y'\right)\right\} $
and $c\left(\left\{ \left(x,y\right),\left(x',y'\right)\right\} \right)=\left\{ \left(x,y\right)\right\} $
respectively, which is also a contradiction. We conclude that if either
WARP or Quasi-linearity (or both) holds, then $c$ admits a quasi-linear
utility representation.
\end{document}